\documentclass[a4paper,10pt,twoside]{cpc-hepnp}

\usepackage{multicol}
\usepackage{graphicx}
\usepackage{booktabs}
\usepackage{amssymb,bm,mathrsfs,bbm,amscd}
\usepackage[tbtags]{amsmath}
\usepackage{lastpage}
\usepackage{amssymb, amsmath, amsthm, amsfonts, bm, mathrsfs, wasysym, txfonts, bbm, enumerate, color}

\begin{document}

\fancyhead[co]{\footnotesize Mohsen Fathi: A dynamical approach to
the exterior geometry of a perfect fluid as a relativistic star}

\footnotetext[0]{Received 27 March 2012}

\title{A dynamical approach to the exterior geometry of a perfect fluid as a relativistic star}

\author{%
      Mohsen Fathi$^{1;1)}$\email{mohsen.fathi@gmail.com}%
} \maketitle

\address{%
$^1$ Department of Physics, Islamic Azad University, Central Tehran Branch, Tehran, Iran\\
}

\begin{abstract}
In this article, we assume that a cold charged perfect fluid is
constructing a spherical relativistic star. Our purpose is the
investigation of the dynamical properties of its exterior
geometry, through simulating the geodesic motion of a charged test
particle, while moving on the star.
\end{abstract}



\begin{multicols}{2}

\section{Introduction}

Einstein field equation leads to the solutions satisfying the
definition of a perfect fluid, a homogenous isotropic mass and
energy distribution with no viscosity, such as
Friedmann-Lema\^{i}tre model (for a good review see \cite{Str}).
Moreover, It also provides the chance to investigate the interior
distributions of the perfect fluid through energy-momentum tensor;
not to forget that the energy-momentum tensor can actually be
extracted from charged bodies bearing interior pressure.

A relativistic star is supposed to consist of such a fluid either
charged or with axial rotation  \cite{Zel}. In order to find out
the relativistic, isotropic mass and charge the interior solution
has been suggested by B. Guilfoyl \cite{Guilfoyl}.

In this article, we are about to form a static charged
relativistic star using these so-called solutions; besides,
studying the dynamical properties of the exterior geometry via a
rotating test-particle is included.

As it goes without saying, in general relativity geodesic motions
are of great importance. Since, it makes precious anticipations
about the cases such as precessions of periastron in planetary
orbits and theoretical explanations for light deflection in
gravitational fields.

Furthermore, there are lots of researches which have been
dedicated to the study of geodesic motions in different kinds of
geometrical background, which could be derived from general
relativity, for example, Schwarzchild, Schwarzchild-de Sitter,
Reissner-Nordstr\"{o}m and Kerr geometries (see
\cite{Zhai}-\cite{Prasanna} and an important text book
\cite{Chandrasakhar}). The exact equations of motion have been
also derived \cite{lam1}, and lots of further theoretical
predictions provided (see \cite{lam2}-\cite{lam8}).

In this case we are considering Reissner-Nordstr\"{o}m (RN)
geometry as the exterior geometry of a static charged relativistic
star. The procedure that we are going through in this paper
concerns the process in which charged particles get trapped in
gravitational and electrical fields of relativistic stars.

The chosen approach is a Lagrangian formalism and numerical
simulations also have been involved. In order to do so, three
classes of solutions presented by Guilfoyl have been utilized to
construct a relativistic star and to compare the pertinent
geodesic behaviors of the test charged particle.

The article is organized as follows: Section two contains
dynamical preliminaries for a charged particle in RN geometry.
Section three considers a brief review on Guilfoyl's interior
solutions. In Section four, we introduce the Lagrangian formalism.
Finally, Sections five and six, consist of the simulations of
charged particle's motions on various types of relativistic stars.


\section{Charged particle in RN geometry}

 First, let us consider a spherical source which has been
schematically illustrated in Figure \ref{Star-1}. It indicates the
charged star, considered to be the gravitational source, with
$R_0$ being the radius, and the total mass and charge being $m_0$
and $q_0$ respectively. A test-particle which is moving on this
star, is assumed to travel on the
\begin{center}
{\includegraphics[height=4cm, width=4cm]{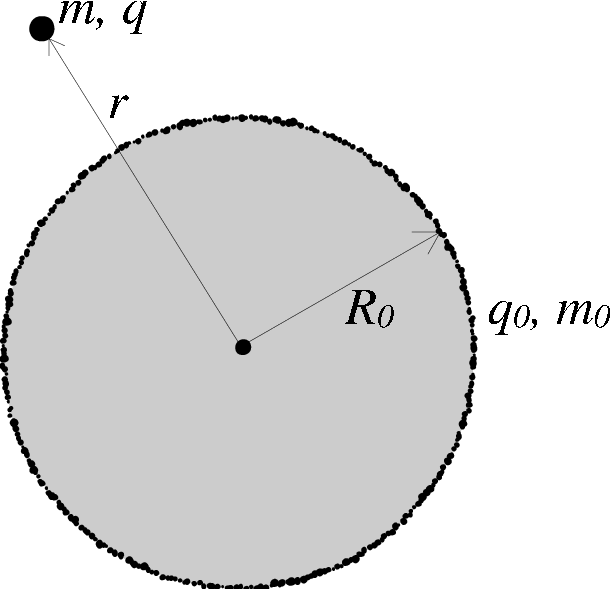}} \figcaption{\small{A
schematic illustration of a charged spherical star and rotating
test charged particle.}} \label{Star-1}
\end{center}
RN geodesics. The RN geometry is a static spherically symmetric
vacuum solution for the Einstein equations  in four dimensions
($d=4$) and is defined by the metric:
\begin{equation}
ds^2 = -c^2 A(r)dt^2 + B(r)dr^2 + r^2d\theta^2+r^2\sin^2\theta
d\phi^2, \label{pre-RN-1}
\end{equation}

in which the variable coefficients are:
$$A(r)=\frac{1}{B(r)} = 1-\frac{2G m_0}{rc^2}+\frac{q_0^2 G}{4\pi\epsilon_0 r^2c^4},$$

where $G$ stands for the gravitational constant and $c$ for the
speed of light. In order to simplify the calculations, we use
geometrical units, fixing:
$$G=c=\frac{1}{4\pi\epsilon_0}=1.$$

According to these units, one can relate length dimensions to
quantities like mass and electric charge, i.e. (see \cite{Wald}
appendix 4):
$$[\text{mass}] = [m_0] = \ell,$$
$$[\text{electric charge}] = [q_0] = \ell,$$
where $\ell$ notates the length dimension. Since $c$ and $G$ are
supposed to be 1, the remaining ratios $\frac{2m_0}{r}$ and
$\frac{q_0^2}{r^2}$, become dimensionless. These assumptions are
commonly used in general relativity. Taking these into account,
one can rewrite (\ref{pre-RN-1}) as:
\begin{equation}
ds^2=-\Big(1-\frac{2m_0}{r}+\frac{q_0^2}{r^2}\Big)dt^2+\Big(1-\frac{2m_0}{r}+\frac{q_0^2}{r^2}\Big)^{-1}dr^2+r^2d\Omega^2.
\label{RN-1}
\end{equation}
We will use this metric in further considerations. Moreover, it is
clear that (\ref{RN-1}) admits at least two time-like Killing
vectors,
\begin{equation}\label{1-1}
    X_{(1)}^\mu=\left(1,0,0,0\right)\equiv \partial_t,$$$$
 X_{(2)}^\mu=\left(0,0,0,1\right)\equiv \partial_\phi.
\end{equation}
The above Killing symmetry provides energy and angular momentum in
this spacetime.

By considering the characteristics of the test-particle ($m$
stands for the mass and $q$ for the charge), one can obtain the
potential and energy conditions, using the Hamilton-Jacobi
equation of wave crests \cite{Wheeler}:
\begin{equation}
g^{\mu\nu}(P_\mu+qA_\mu)(P_\nu+qA_\nu)+m^2=0. \label{H-J}
\end{equation}

$P_\mu$ is the momentum 4-vector\footnote{Here we use the notation
in Ref.  \cite{Wheeler}, in which in \S 25.3 the 4-momentum has
been defined like Eq. (\ref{p-1}).}
\begin{equation}
P_\mu=g_{\mu\sigma}P^\sigma=g_{\mu\sigma}\frac{dx^\sigma}{d\lambda}.
\label{p-1}
\end{equation}

Here $\lambda$ is the affine parameter of the trajectory. The
metric components $g_{\mu\nu}$ can be derived from the exterior
geometry of the source, the RN metric (\ref{RN-1}). Also the
vector potential $A_\mu$ for the static charged source will be:
\begin{equation}
A_\mu = (\varphi(r),0,0,0), \label{A-mu}
\end{equation}

where $\varphi(r)=\frac{q_0}{r}$ is the scalar electrical
potential, outside the star, producing a scalar filed within the
spacetime. One can define the two conserved quantities as:
\begin{equation}
E=-P_0, \label{Energy}
\end{equation}

the energy, and
\begin{equation}
L =P_\phi\,\,\,\,\,\,\, (L\geq0). \label{Angular Momentum}
\end{equation}

the angular momentum. Here we consider $\theta=\frac{\pi}{2}$, for
which the particle's motion is confined to the equatorial
rotations. Therefore
$$P^\theta=\frac{d\theta}{d\lambda}=0.$$

Substituting the metric (\ref{RN-1}) in (\ref{H-J}) yields:
$$-\frac{(E-\frac{qq_0}{r})^2}{1-\frac{2m_0}{r}+\frac{q_0^2}{r^2}} + (1-\frac{2m_0}{r}+\frac{q_0^2}{r^2})^{-1}(\frac{dr}{d\lambda})^2+\frac{L^2}{r^2}+m^2=0,$$

or
\begin{equation}
(\frac{dr}{d\lambda})^2=
(E-\frac{qq_0}{r})^2-(1-\frac{2m_0}{r}+\frac{q_0^2}{r^2})(m^2+\frac{L^2}{r^2}).\label{dr-1}
\end{equation}

Equation (\ref{dr-1}) can be rewritten as:
\end{multicols}

\ruleup

$$(\frac{dr}{d\lambda})^2=[E-\big(\frac{qq_0}{r}-\sqrt{(1-\frac{2m_0}{r}+\frac{q_0^2}{r^2})(m^2+\frac{L^2}{r^2})}\,\,\big)][E-\big(\frac{qq_0}{r}+\sqrt{(1-\frac{2m_0}{r}+\frac{q_0^2}{r^2})(m^2+\frac{L^2}{r^2})}\,\,\big)].$$
\\
\ruledown \vspace{0.5cm}

\begin{multicols}{2}

We introduce the potential, felt by the test-particle
as:
\begin{equation}
V(r)=\frac{qq_0}{r}+\sqrt{(1-\frac{2m_0}{r}+\frac{q_0^2}{r^2})(m^2+\frac{L^2}{r^2})},
\label{V-eff}
\end{equation}
to be assured that positive potential is available. To discuss the
motion of the test-particle also, we need the rate of variation of
rotation angle $\phi$ with respect to $r$. Previously we defined:
$$g_{\phi\phi}P^\phi=L\,\,\,\Rightarrow\,\,\,\frac{d\phi}{d\lambda}=\frac{L}{r^2}.$$

Substituting in Eq. (\ref{dr-1}) yields:
\begin{equation}
(\frac{dr}{d\phi})^2=\frac{r^4}{L^2}[(E-\frac{qq_0}{r})^2-(1-\frac{2m_0}{r}+\frac{q_0^2}{r^2})(m^2+\frac{L^2}{r^2})].
\label{dr/dphi}
\end{equation}

Solutions to equations like (\ref{dr/dphi}), have been obtained
using the Weierstrass function and elliptic integrals in
\cite{lam1}, however in this context, we use a Lagrangian
formalism instead of the geodesic equations.

For different values of $L$ and $E$, the test-particle experiences
different types of motions. The motion of the particle, depends on
its energy, for sets of solutions for $E-V(r)=0$. The RN spacetime
allows three types of orbits; for one zero, we will have scape
orbits, for two zeros, periodic bound orbits are available, and
for three zeros, we will have scape/capture orbits and periodic
bound orbits.

It is also worthy to investigate the potential that the particle
is going through. The potential may have some extremum points, as
below:
\begin{equation}
\frac{dV(r)}{dr}|_{(r=r_e)}=0. \label{dv/dr=0}
\end{equation}

We can rewrite the potential in Eq. (\ref{V-eff}) as:
\begin{equation}
V(r)=C(r)+\sqrt{g(r)(m^2+\frac{L^2}{r^2})}. \label{V-eff-1}
\end{equation}

We have $C(r)=\frac{qq_0}{r}$ and
$g(r)=1-\frac{2m_0}{r}+\frac{q_0^2}{r^2}$. Using (\ref{V-eff-1})
in (\ref{dv/dr=0}) forms the following differential equation:
$${C'\sqrt {g \left( {m}^{2}+{\frac {{L}^{2}}{{r}^{2}}} \right) }}+\frac{1}{2}\, \left[ {g'}\, \left( {m}^{2}+{\frac {{L}^{2}}{{r}^{
2}}} \right) -2\,{\frac {g{L}^{2}}{{r}^{3}}} \right] { {}{ {
 }}}=0.
$$

Solving the above equation for $L$ yields:
\end{multicols}

\ruleup
\begin{equation}
L=r \,\Big(\,\,{\frac {2\,{r}^{2}g{{ C'}}^{2}-{r}^{2}{{
g'}}^{2}{m}^{2}+ 2\,{g'}\,{m}^{2}gr\pm2\,\sqrt
{{r}^{4}{g}^{2}{{C'}}^{4}-2\,{r}^{
3}{g}^{2}{{C'}}^{2}{g'}\,{m}^{2}+4\,{g}^{3}{r}^{2}{m}^{2}{{
C'}}^{2}}}{{r}^{2}{{g'}}^{2}-4\,{g'}\,rg+4\,{g}^{2}}}\,\,\Big)^{\frac{1}{2}}
. \label{L-main}
\end{equation}
\\
\ruledown \vspace{0.5cm}

\begin{multicols}{2}

This solution, corresponds to the stable orbits of the
test-particle around the star, where the potential has its
extremum points. The expression for $L$ in Eq. (\ref{L-main})
indicates that, for stable orbits regardless of the energy of the
particle, the total angular momentum, for definite values of mass,
charge and radius of the source, can opt only definite values. As
we will see, the energy of the particle is peculiar, when we work
with a particular potential.

We are going to discuss the types of motions in Section 5,
considering different interior mass and charge distributions for
the star. We shall introduce these interior solutions in the next
section.

\section{Weyl-type interior solutions for a charged perfect fluid}

 In this article, a relativistic star, is defined to be a charged
spherically symmetric cold perfect fluid. Such a fluid is a
solution of the Einstein field equations
\begin{equation}
R_{\mu\nu}-\frac{1}{2}g_{\mu\nu}R=8\pi(T_{\mu\nu}+E_{\mu\nu}).
\label{Einstein}
\end{equation}

and the Maxwell equations
\begin{equation}
\nabla_{\nu}F^{\mu\nu}=4\pi J^\mu, \label{Maxwell}
\end{equation}

for a charged isotropic mass distribution. In Eq.
(\ref{Einstein}), $E_{\mu\nu}$ is the electromagnetic
stress-energy tensor, defined as \cite{Wheeler}:
\begin{equation}
4\pi E_{\mu\nu}=F^\rho_\mu
F_{\nu\rho}-\frac{1}{4}g_{\mu\nu}F_{\rho\sigma}F^{\rho\sigma},
\label{stress-energy-E}
\end{equation}

in which
$$F_{\mu\nu}=\nabla_\mu A_\nu-\nabla_\nu A_\mu,$$

is the antisymmetric electromagnetic field strength tensor. The
energy-momentum tensor for an isotropic mass distribution with
pressure is:
\begin{equation}
T_{\mu\nu}= \begin{pmatrix} \rho &0 &0 &
0\\0&p&0&0\\0&0&p&0\\0&0&0&p
\end{pmatrix},\label{energy-momentum-T}
\end{equation}
where $\rho$ is the matter density and $p$ is the fluid pressure.
The following spherically symmetric metric is assumed to explain
the interior geometry of the perfect fluid: \begin{equation} ds^2
= -\omega^2dt^2+\xi(r)^2dr^2+r^2d\Omega^2, \label{Weyl-1}
\end{equation}

in which the metric potential
\begin{equation}
\omega(\varphi)^2 = a + b\varphi+\varphi^2, \label{omega(phi)}
\end{equation}
by Weyl's assumption \cite{Weyl}, only depends on the electric
potential $\varphi(r)$, and $a$ and $b$ in (\ref{omega(phi)}) are
constants.

The non-zero components of Maxwell equations (\ref{Maxwell}) for
metric (\ref{Weyl-1}), determine the total charge inside a volume
with radius $r$. We have \cite{Lemos}:
\begin{equation}
Q(r)=\frac{r^2\varphi'(r)}{\omega(\varphi)\xi(r)}=4\pi\int_0^r\sigma(r)r^2\xi(r)dr,
\label{Q(r)-1}
\end{equation}

in which $\sigma(r)$ is the volume charge density. Since outside
the star, the geometry obeys the RN metric (\ref{RN-1}), Equation
(\ref{Q(r)-1}) can be rewritten as
$$Q(r)=r^2\varphi'(r)[A(r)B(r)]^{-\frac{1}{2}}.$$

According to the Weyl's assumption in (\ref{omega(phi)}), this
gives:
$$\varphi'(r)=\mp\frac{A'(r)}{\sqrt{b^2+4A(r)-4a}}.$$

Therefore
\begin{equation}
Q(r)=\mp A'(r)\{A(r)B(r)(b^2+4A(r)-4a)\}^{-\frac{1}{2}}.
\label{Q(r)-2}
\end{equation}

Now to realize what the interior solutions of a relativistic star
really are, let us consider another star, with radius $r_0$
($r_0=R_0+\delta r$), charge $Q_0$ and mass $M$ which are
considered to be known values. This star (Figure \ref{Boundary
Condition}) is not the one that will be used as the massive source
for the test particle to move on. It is only used for discussing
the interior solutions and it is in fact, a boundary condition to
derive the desired constants in the interior solutions. However,
the massive source in Figure \ref{Star-1} has the same
distributions but in lower dimensions.
\begin{center}
{
\includegraphics[height=4cm,width=4cm]{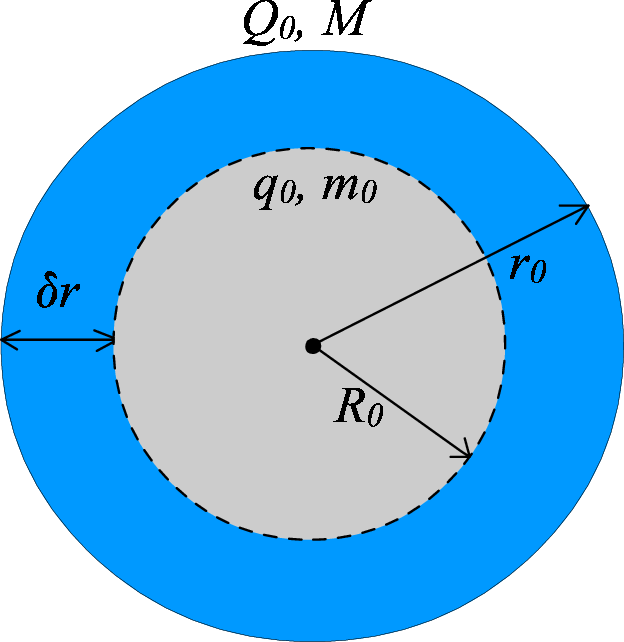}}
\figcaption{\small{A star with higher dimensions, having the same
distributions as Figure \ref{Star-1}, but with known values for
charge and mass.}} \label{Boundary Condition}
\end{center}
The radius $r_0$ will be considered to be the same for all cases,
therefore the stars under discussion, will have the same
geometrical appearance and obviously the amounts $Q_0$ and $M$
will differ for different interior solutions. Now for the star in
Figure \ref{Boundary Condition}, the continuity condition at
$r=r_0$ implies that:
$$A(r_0)=\frac{1}{B(r_0)}=1-\frac{2M}{r_0}+\frac{Q_0^2}{r_0^2},$$

in which $Q_0=Q(r_0)$. Therefore, from Eq. (\ref{Q(r)-2}) we
obtain:
\begin{equation}
\frac{M}{Q_0}=\pm\frac{1}{2}\sqrt{b^2+4(1-aQ_0^2)}, \label{M/Q0-1}
\end{equation}

which implies that for $a=b=0$ we have:
$$M=|Q_0|.$$

This is the condition concerning a Majumdar-Papapetrou star for a
Weyl-type metric potential. The pressure for these stars is zero;
in this case the relation between the metric potential $\omega^2$
in (\ref{omega(phi)}) and the electrical potential $\varphi(r)$ is
always a perfect square \cite{Majumdar, Papapetrou}. Also we will
discuss the motion of the test-particle around these stars in
Section 5.

In Ref. \cite{Guilfoyl}, Guilfoyl has presented some interesting
Weyl-type interior solutions for a spherically symmetric charged
perfect fluid. He based his solutions on two presuppositions:
\begin{itemize}
\item{The Schwarzchild condition which implies:
$$8\pi T_{00} - 8\pi E_{00}=\frac{3}{R^2} = \text{constant},$$
or
\begin{equation}
8\pi\rho(r)+\frac{Q(r)^2}{r^4}=\frac{3}{R^2}=\text{constant}.
\label{Sch-1}
\end{equation}
} \item{The expression for the radial part of metric
(\ref{Weyl-1}):
\begin{equation}
\xi(r)^2=(1-\frac{r^2}{R^2})^{-1} \label{xi}
\end{equation}
}
\end{itemize}

In these two presuppositions, the quantity $\frac{1}{R^2}$, is a
constant that must be determined using the boundary conditions at
$r=r_0$ (in Figure \ref{Boundary Condition}). We now bring up two
classes of Guilfoyl's solutions which we will use later to form
the test-particle's potential \cite{Guilfoyl}:

\subsection{The case of Class I}
 In this case, the metric potential is considered to have the
spherical form:
\begin{equation}
a\omega^2=b+(c+\varphi)^2, \label{w-Guilfoyl-I}
\end{equation}
where $c$ is a constant. We call this, the Weyl-Guilfoyl metric
potential as it is in \cite{Lemos}. Both using the RN limit in
(\ref{Weyl-1}) and also considering (\ref{w-Guilfoyl-I}) we have:
\begin{equation}
\frac{M}{Q_0}=\frac{Q_0}{ar_0}(a-1)\pm\frac{1}{ar_0}\sqrt{r_0^2(a-b)+Q_0^2(1-a)}.
\label{M/Q0-2}
\end{equation}

As we can see, while $a=1$ and $b=0$ we have $M=|Q_0|$ which is
the Majumdar condition. For the expression (\ref{w-Guilfoyl-I}),
two sets of solutions have been
introduced:\\\\

\underline{Class Ia}: $a=\frac{1}{2}$, $b=0$\\
\begin{equation}
\left\{
\begin{array}{c}
\omega^2=e^{\sqrt{2}\psi(r)}\\\\
Q(r)=\pm\frac{kr^3}{2}\\\\
8\pi\rho(r)=\frac{3}{R^2}-\frac{k^2r^2}{4}
\end{array}
 \right.
 \label{Class-Ia-a=1/2}
\end{equation}

The function $\psi(r)$ is defined as:
\begin{equation}
\psi(r)=l-kR^2(1-\frac{r^2}{R^2})^{\frac{1}{2}}. \label{psi}
\end{equation}

Also the constants $\frac{1}{R^2}$, $k$ and $l$ can be determined,
using the boundary conditions at $r=r_0$ (Figure \ref{Boundary
Condition}) as below:
\begin{equation}
k=\frac{2|Q_0|}{r_0^3}, \label{k-Ia-1/2}
\end{equation}
\begin{equation}
l=(\frac{M}{Q_0}-\frac{Q_0}{2r_0})^{-1}(1-\frac{2M}{r_0}+\frac{Q_0^2}{r_0^2})^{\frac{1}{2}}+\sqrt{2}\ln(1-\frac{2M}{r_0}+\frac{Q_0^2}{r_0^2}),
\label{l-Ia-1/2}
\end{equation}
\begin{equation}
\frac{1}{R^2}=\frac{2}{r_0^3}(M-\frac{Q_0^2}{2r_0}). \label{1/R^2}
\end{equation}\\\\

\underline{Class Ib}: $a=1$, $b<0$\\
\begin{equation}
\left\{
\begin{array}{c}
\omega^2=-b\tan^2[\sqrt{-b}\psi(r)]\\\\
Q(r)=\pm\sqrt{-b}kr^3\sec[\sqrt{-b}\psi(r)]\\\\
8\pi\rho(r)=\frac{3}{R^2}+bk^2r^2\sec^2[\sqrt{-b}\psi(r)]
\end{array}
 \right.
 \label{Class-Ib}
\end{equation}

The solution set Class Ib, are confirmed for
$$b\leq0\,\,\,\Leftrightarrow\,\,\, M\geq|Q_0|.$$

The corresponding coefficients are:
\begin{equation}
b=1-(\frac{M}{Q_0})^2, \label{b-Ib}
\end{equation}
\begin{equation}
k=\frac{Q_0^2}{r_0^3}(M-\frac{Q_0^2}{r_0})^{-1}, \label{k-Ib}
\end{equation}
\begin{equation}
\begin{array}{l}
l=[(\frac{M}{Q_0})^2-1]^{-\frac{1}{2}}\sec^{-1}[(\frac{M}{Q_0}-\frac{Q_0}{r_0})\{(\frac{M}{Q_0})^2-1\}^{-\frac{1}{2}}]\\
+(1-\frac{2M}{r_0}
+\frac{Q_0^2}{r_0^2})^{\frac{1}{2}}(\frac{M}{Q_0}-\frac{Q_0}{r_0})^{-1}(\frac{2M}{Q_0}-\frac{Q_0}{r_0})^{-1}.
\end{array}
\label{l-Ib}
\end{equation}

\subsection{The case of Class II}
 In this case Guilfoyl assumed that the metric potential $\omega^2$
has the following form:
\begin{equation}
\frac{\sqrt{2}}{3}a\omega^{\frac{3}{2}}=c+\varphi(r).
\label{w-Guilfoyl-II}
\end{equation}

The interior solutions for this case are:
\begin{equation}
\left\{
\begin{array}{c}
\omega^2=\frac{1}{a^4}[\ln\big(-a^2\psi(r)\big)]^2\\\\
Q(r)=\pm\frac{kr^3}{\sqrt{2}\psi(r)}[-\ln\big(-a^2\psi(r)\big)]^{-\frac{1}{2}}\\\\
8\pi\rho(r)=\frac{3}{R^2}-\frac{k^2r^2}{2\psi(r)^2}[-\ln\big(-a^2\psi(r)\big)]^{-1}
\end{array}
\right. \label{Class-II}
\end{equation}

And also the constant coefficients:
\begin{equation}
a=\sqrt{2}(\frac{M}{Q_0}-\frac{Q_0}{r_0})^{-1}(1-\frac{2M}{r_0}
+\frac{Q_0^2}{r_0^2})^{\frac{1}{4}}, \label{a-II}
\end{equation}
\begin{equation}
k=\frac{1}{r_0^3}(M-\frac{Q_0^2}{r_0})\exp[-2(\frac{M}{Q_0}-\frac{Q_0}{r_0})^{-2}(1-\frac{2M}{r_0}
+\frac{Q_0^2}{r_0^2})], \label{k-II}
\end{equation}
\begin{equation}
\begin{array}{l}
l=\frac{1}{2}(\frac{M}{Q_0}-\frac{Q_0}{r_0})(1-\frac{2M}{r_0}
+\frac{Q_0^2}{r_0^2})^{\frac{1}{2}}[(\frac{M}{Q_0}-\frac{Q_0}{2r_0})^{-1}\\
-(\frac{M}{Q_0}-\frac{Q_0}{r_0})(1-\frac{2M}{r_0}
+\frac{Q_0^2}{r_0^2})^{-1}]\\
.\exp[-2(\frac{M}{Q_0}-\frac{Q_0}{r_0})^{-2}(1-\frac{2M}{r_0}
+\frac{Q_0^2}{r_0^2})].
\end{array}
\label{l-II}
\end{equation}

We use these interior solutions in Section 5, to form different
types of relativistic stars.

\section{Lagrangian formalism for a charged particle moving in RN spacetime}
 Now let us consider a Lagrangian like:
\begin{equation}
\mathcal{L}=\frac{1}{2}mu_i u^i-V(r), \label{Lagrangian1}
\end{equation}

where $u^i=\frac{dx^i}{d\lambda}$ is the velocity $4$-vector for
the test-particle and $V(r)$ is the potential in (\ref{V-eff}).
One can write down this equation for the RN spacetime as:
\begin{equation}
\mathcal{L}=\frac{1}{2}mg_{ii}(u^i)^2-V(r), \label{Lagrangian2}
\end{equation}

in which the diagonal components of the metric (\ref{RN-1}) are
used. Therefore the Lagrangian is a function like:
\begin{equation}
\mathcal{L}\equiv\mathcal{L}(t,r,\theta,\phi,\dot t, \dot r, \dot
\theta, \dot \phi). \label{Lagrangian3}
\end{equation}

Here the dot stands for differentiation with respect to affine
parameter $\lambda$ in geodesic motion. Using the RN metric
(\ref{RN-1}) yields:
\end{multicols}

\ruleup
\begin{equation}
\begin{array}{l}
\mathcal{L}=\frac{1}{2}\,m\, \left( -1+2\,{\frac {m_0}{r}}-{\frac
{{q_0}^{2}}{{r}^{2}}}
 \right) {\dot t}^{2}+\frac{1}{2}\,m\,{\dot r}^{2} \left( 1-2\,{
\frac {m_0}{r}}+{\frac {{q_0}^{2}}{{r}^{2}}} \right)
^{-1}+\frac{1}{2}\,m\,{r}^{ 2}{\dot \theta}^{2}
+\frac{1}{2}\,m\,{r}^{2} \left( \sin \left( \theta
 \right)  \right) ^{2}{\dot\phi}^{2}\\- {\frac {q\,q_0}{r}}-\sqrt { \left( 1-2\,{\frac
{m_0}{r}}+{\frac {{q_0}^{2}}{{r}^{2}}} \right)
 \left( {m}^{2}+{\frac {{L}^{2}}{{r}^{2}}} \right) }.
\end{array}
\label{Lagrangian4}
\end{equation}
\\
\ruledown \vspace{0.5cm}

\begin{multicols}{2}

Since we take the geometrical units, we have $d\lambda=d\tau$,
where $\tau$ is the proper time. The action in this space time,
therefore is defined by \cite{Greiner}:
\begin{equation}
S=\int\mathcal{L}(t,r,\theta,\phi,\dot t,\dot r,\dot \theta, \dot
\phi)d\tau, \label{action1}
\end{equation}

turning to the following one for equatorial rotations:
\begin{equation}
S=\int\mathcal{L}(t,r,\phi,\dot t,\dot r,\dot
\phi)d\tau.\label{action2}
\end{equation}

Varying this action, we can obtain the Euler-Lagrange equation of
motion in RN spacetime:
\begin{equation}
\frac{\partial\mathcal{L}}{\partial
x^i}-\frac{d}{d\tau}\Big(\frac{\partial\mathcal{L}}{\partial\dot
x^i}\Big)=0. \label{E-L1}
\end{equation}

For $i=0,1,3$, this leads to three equations which together are
the equations of equatorial motion ($\theta=\frac{\pi}{2}$) for
the test-particle, moving in RN geometry. Using the definition of
the potential in (\ref{V-eff}), and the Lagrangian
(\ref{Lagrangian4}) in (\ref{E-L1}) we have:

\end{multicols}

\ruleup
\begin{eqnarray}
\begin{array}{l}
\left( -2\,{\frac {m_{{0}}{\frac {d}{d\tau}}r \left( \tau \right)
}{
 \left( r \left( \tau \right)  \right) ^{2}}}+2\,{\frac {{q_{{0}}}^{2}
{\frac {d}{d\tau}}r \left( \tau \right) }{ \left( r \left( \tau
 \right)  \right) ^{3}}} \right) {\frac {d}{d\tau}}t \left( \tau
 \right)
  + \left( -1+2\,{\frac {m_{{0}}}{r \left( \tau \right) }}-{
\frac {{q_{{0}}}^{2}}{ \left( r \left( \tau \right)  \right)
^{2}}}
 \right) {\frac {d^{2}}{d{\tau}^{2}}}t \left( \tau \right) =0
,
\\\\

-m \left( {\frac {d}{d\tau}}r \left( \tau \right)  \right)  \left(
2\, {\frac {m_{{0}}{\frac {d}{d\tau}}r \left( \tau \right) }{
\left( r
 \left( \tau \right)  \right) ^{2}}}-2\,{\frac {{q_{{0}}}^{2}{\frac {d
}{d\tau}}r \left( \tau \right) }{ \left( r \left( \tau \right)
 \right) ^{3}}} \right)\left( 1-2\,{\frac {m_{{0}}}{r \left( \tau
 \right) }}+{\frac {{q_{{0}}}^{2}}{ \left( r \left( \tau \right)
 \right) ^{2}}} \right) ^{-2}+m{\frac {d^{2}}{d{\tau}^{2}}}r \left(
\tau \right)  \left( 1-2\,{\frac {m_{{0}}}{r \left( \tau \right)
}}+{ \frac {{q_{{0}}}^{2}}{ \left( r \left( \tau \right)  \right)
^{2}}}
 \right) ^{-1}\\-\frac{1}{2}\,m \left( -2\,{\frac {m_{{0}}}{ \left( r \left(
\tau \right)  \right) ^{2}}}+2\,{\frac {{q_{{0}}}^{2}}{ \left( r
 \left( \tau \right)  \right) ^{3}}} \right)  \left( {\frac {d}{d\tau}
}t \left( \tau \right)  \right) ^{2} +\frac{1}{2}\,m \left( {\frac
{d}{d\tau}}r
 \left( \tau \right)  \right) ^{2} \left( 2\,{\frac {m_{{0}}}{ \left(
r \left( \tau \right)  \right) ^{2}}}-2\,{\frac {{q_{{0}}}^{2}}{
 \left( r \left( \tau \right)  \right) ^{3}}} \right) \\
 \times \left( 1-2\,{
\frac {m_{{0}}}{r \left( \tau \right) }}+{\frac {{q_{{0}}}^{2}}{
 \left( r \left( \tau \right)  \right) ^{2}}} \right) ^{-2}-mr \left(
\tau \right)  \left( {\frac {d}{d\tau}}\phi \left( \tau \right)
 \right) ^{2}-{\frac {qq_{{0}}}{ \left( r \left( \tau \right)
 \right) ^{2}}}+\frac{1}{2}\Big{[}\sqrt { \left( 1-2\,{ \frac{m_{{0}}}{r \left(
\tau \right) } }+{\frac {{q_{{0}}}^{2}}{ \left( r \left( \tau
\right) \right) ^{2}}}
 \right)  \left( {m}^{2}+{\frac {{L}^{2}}{ \left( r \left( \tau
 \right)  \right) ^{2}}} \right) }\Big{]}\,\\
 \times  \Big[\left( 2\,\frac {m_{{0}}}{ \left( r
 \left( \tau \right)  \right) ^{2}}-2\,\frac {{q_{{0}}}^{2}}{
 \left( r \left( \tau \right)  \right) ^{3}} \right)  \left( m^2+
\frac {L^2}{ \left( r \left( \tau \right)  \right)^2}
 \right)
-2 \left( 1-2\,\frac {m_{{0}}}{r \left( \tau \right) }+\frac
{{q_{{0}}}^{2}}{ \left( r \left( \tau \right)  \right) ^{2}}
 \right) {L}^{2} \left( r \left( \tau \right)  \right) ^{-3}\Big]=0
,\\\\

2\,r \left( \tau \right)  \left( {\frac {d}{d\tau}}\phi \left(
\tau
 \right)  \right) {\frac {d}{d\tau}}r \left( \tau \right) + \left( r
 \left( \tau \right)  \right) ^{2}{\frac {d^{2}}{d{\tau}^{2}}}\phi
 \left( \tau \right) =0.
\end{array}
\label{E-L2}
\end{eqnarray}
\\
\ruledown \vspace{0.5cm}

\begin{multicols}{2}

Using various interior solutions, in the next section we solve
these equations of motion numerically, for a particle moving on a
relativistic star (Figure \ref{Star-1}), and obtain the shape of
possible orbits due to different available potentials.

\section{Moving charge on a relativistic star: various interior solutions}
 Now let us come back to our original source, i.e. the star with
radius $r=R_0$ (Figure \ref{Star-1}). We will illustrate the
orbits of a massive charged particle, moving around this star. The
interior distributions in the star, obey the three cases of
interior solutions discussed in Section 3. Since we have discussed
the interior solutions for a star with radius $r_0$, for the total
charge amount of the star in Figure \ref{Star-1} we have:
\begin{equation}
q_0=Q(R_0)=Q(r_0-\delta r), \label{q0}
\end{equation}

in which $Q(r)$ is one of the interior solutions introduced in
Section 3. Also for this star's total mass we have
\cite{Guilfoyl}:
$$m_0=\frac{1}{2}\int_0^{R_0}r^2dr\big(8\pi\rho(r)+\frac{Q(r)^2}{r^4}\big)+\frac{q_0^2}{2R_0}.$$

Using Eq. (\ref{Sch-1}) this can be summarized as:
\begin{equation}
m_0=\frac{(r_0-\delta r)^3}{2R^2}+\frac{q_0^2}{2(r_0-\delta r)},
\label{m0}
\end{equation}

where $\frac{1}{R^2}$ was determined in (\ref{1/R^2}). In the rest
of this section we will discuss different sources.

\subsection{For Class Ia}
 In this case the total charge $q_0$ can be derived from the
general formula (\ref{q0}) using Class Ia solution, namely Eqs.
(\ref{Class-Ia-a=1/2}-\ref{1/R^2}):
\begin{equation}
q_0={\frac {Q_{{0}} \left(r_0-\delta r \right) ^{3}}{{r_0}^{3}}}.
\label{q0-Ia}
\end{equation}

The total mass $m_0$ has the same relation as (\ref{m0}). We
should notice that, in our investigations we suppose that the
final amounts of charge and mass, $Q_0$ and $M$, have the same
ratio ($\frac{Q_0}{M}$) in all considered cases, to facilitate the
comparison of interior solutions. The potential in (\ref{V-eff}),
for an intermediate angular momentum, is shown in Figure
\ref{V-eff-Ia-plot}. Falling in this potential, the test-particle
can access three types of orbits, due to the total energy $E$
(notated in Figure \ref{V-eff-Ia-plot}).
\begin{center}
{\includegraphics[height=6.5cm]{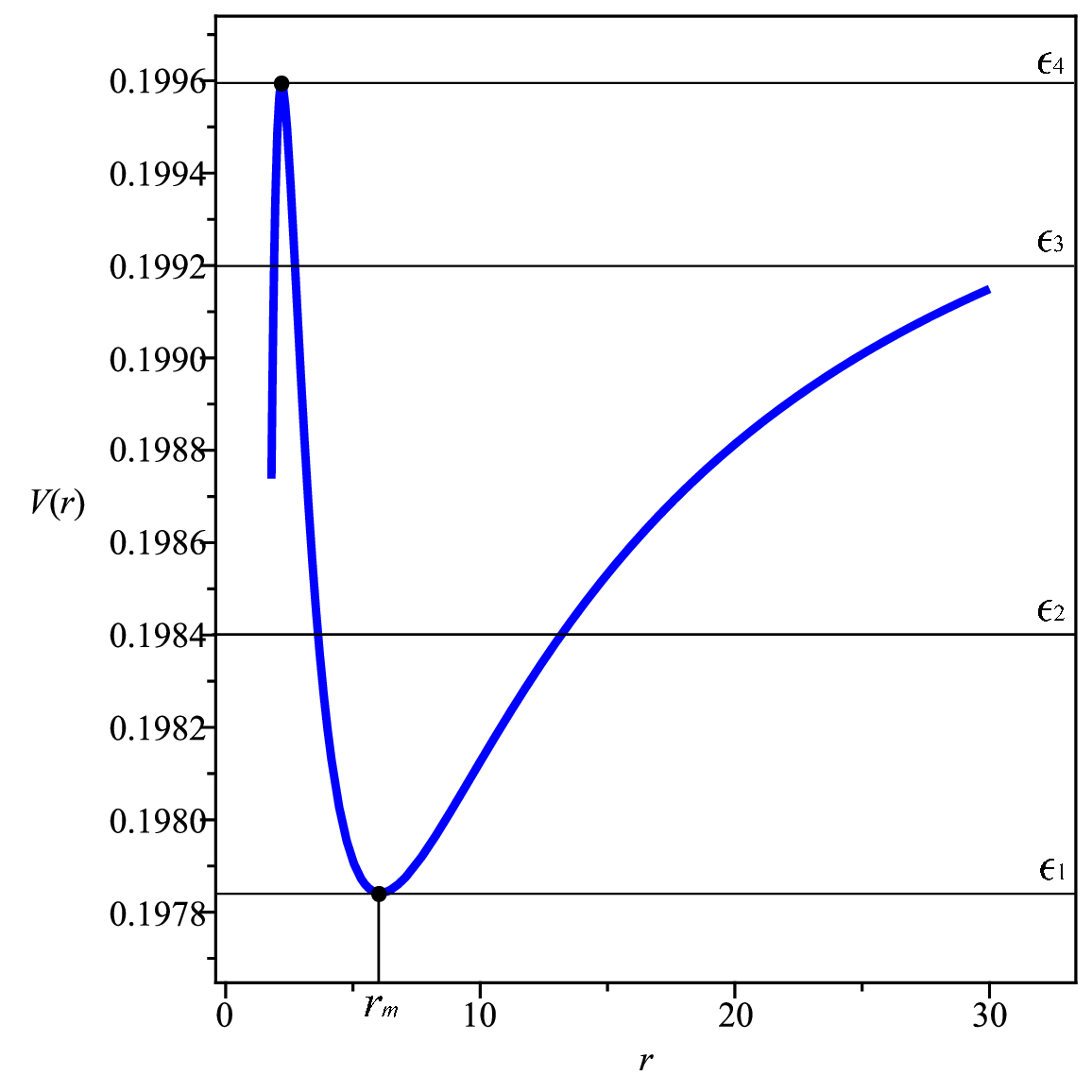}} \figcaption{\small{The
potential for a test-particle moving on a relativistic star, with
the interior solution Class Ia. The illustration is plotted for
$Q_0=0.85$, $q=0.18$ ,$m=0.2$, $L=0.225$, $r_0=2$, $\delta r=0.1\,
r_0$. The unit of length along the coordinate axis is $M$.}}
\label{V-eff-Ia-plot}
\end{center}

1) {\it{Periodic bound orbits}}\\\\

This type of orbit, is often called planetary orbit, specially
when the shape of trajectories, is elliptical. Figure
\ref{bound-classIa} shows some possible bound orbits for different
values of initial energy $E$.
\end{multicols}
\ruleup
\begin{center}
\includegraphics[width=40mm,height=40mm]{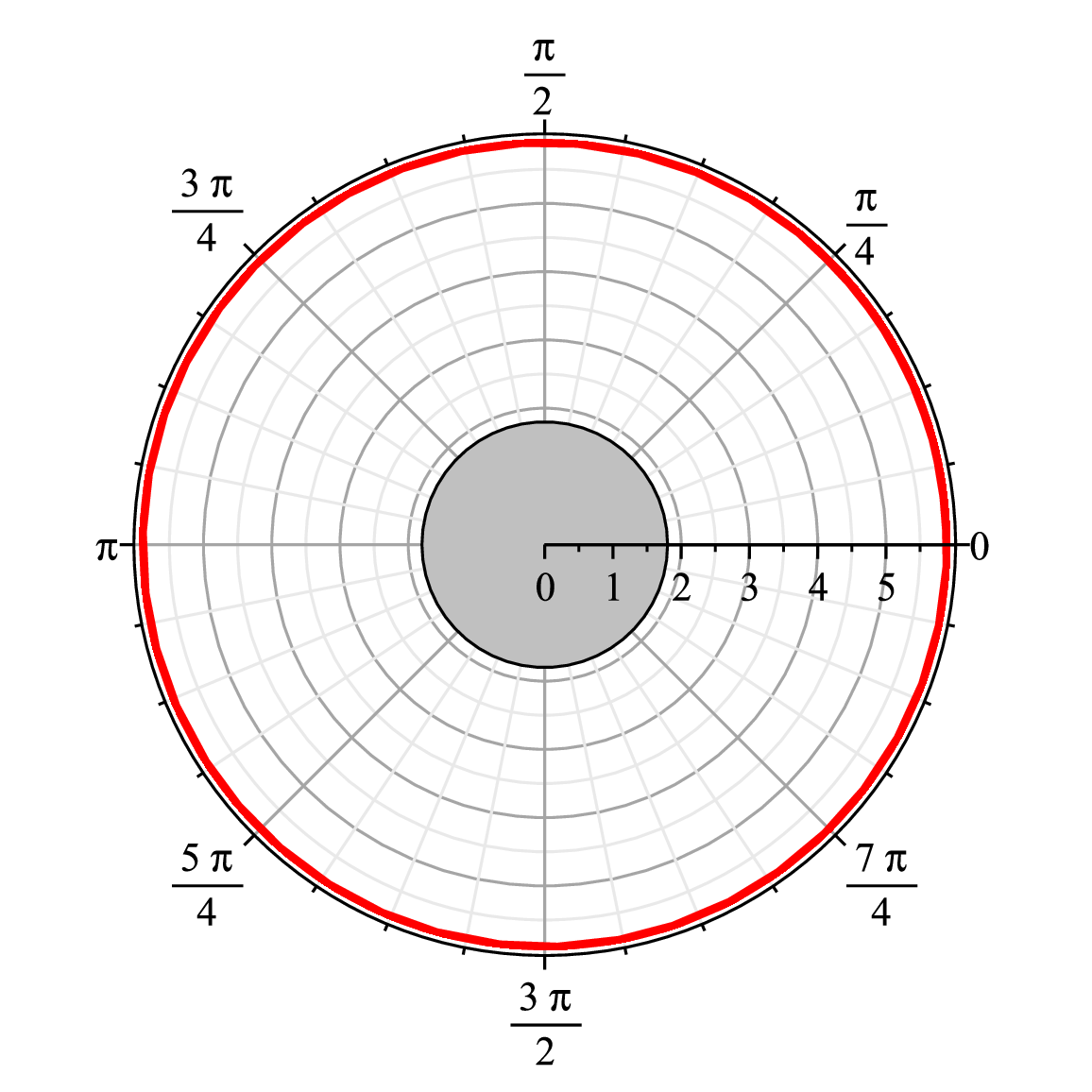}~a)
\hfil
\includegraphics[width=40mm,height=40mm]{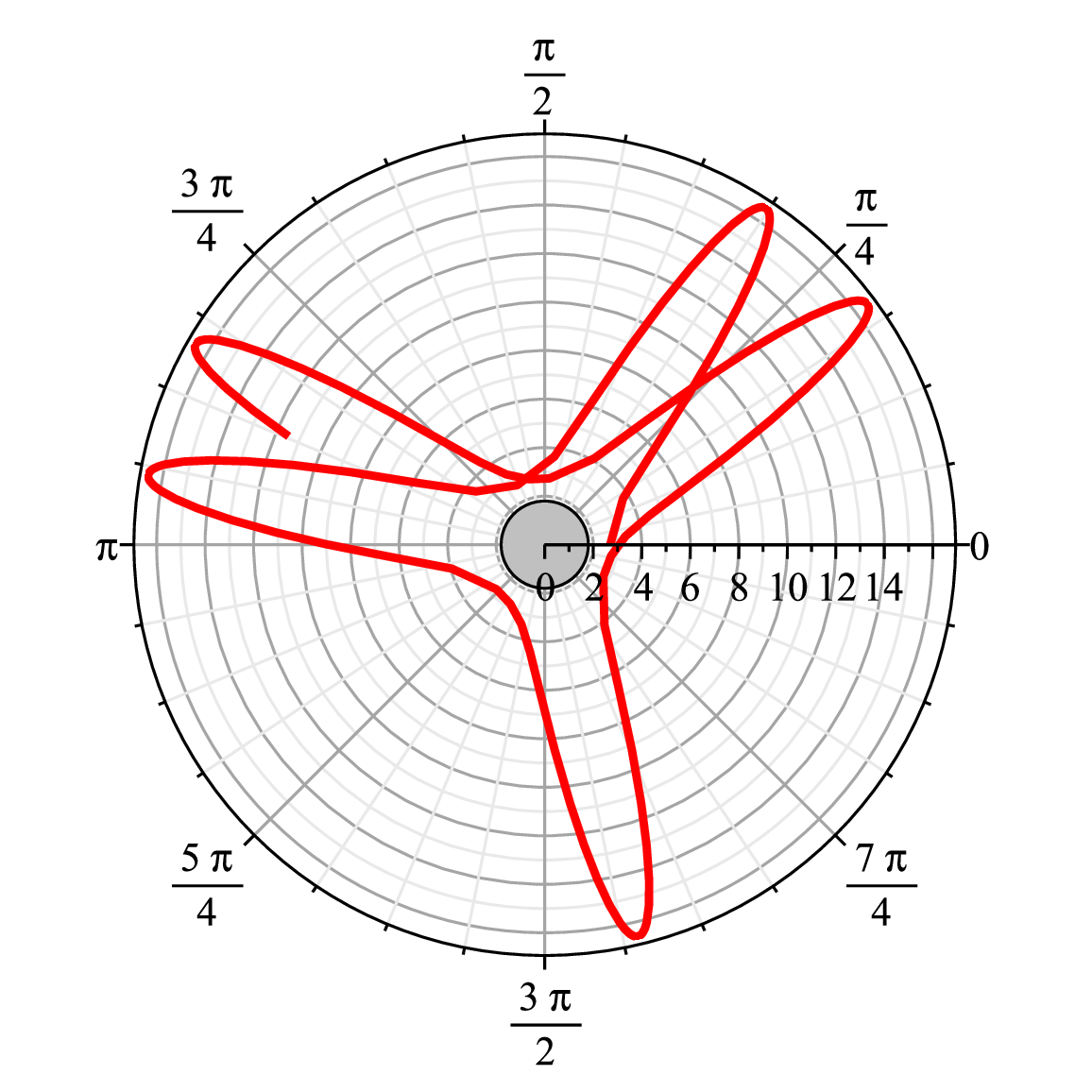}~b)
\hfil
\includegraphics[width=40mm,height=40mm]{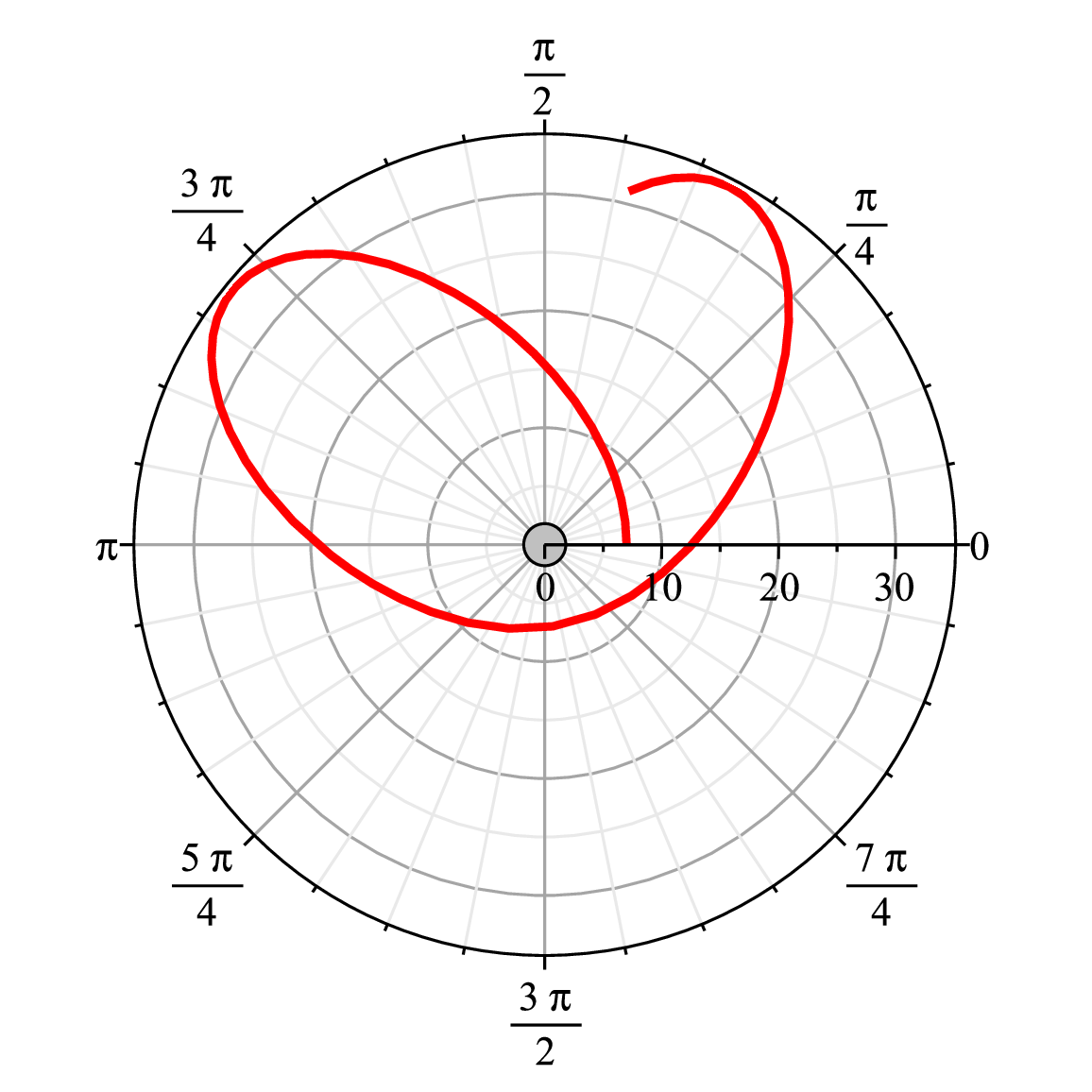}~c)
\hfil
\includegraphics[width=40mm,height=40mm]{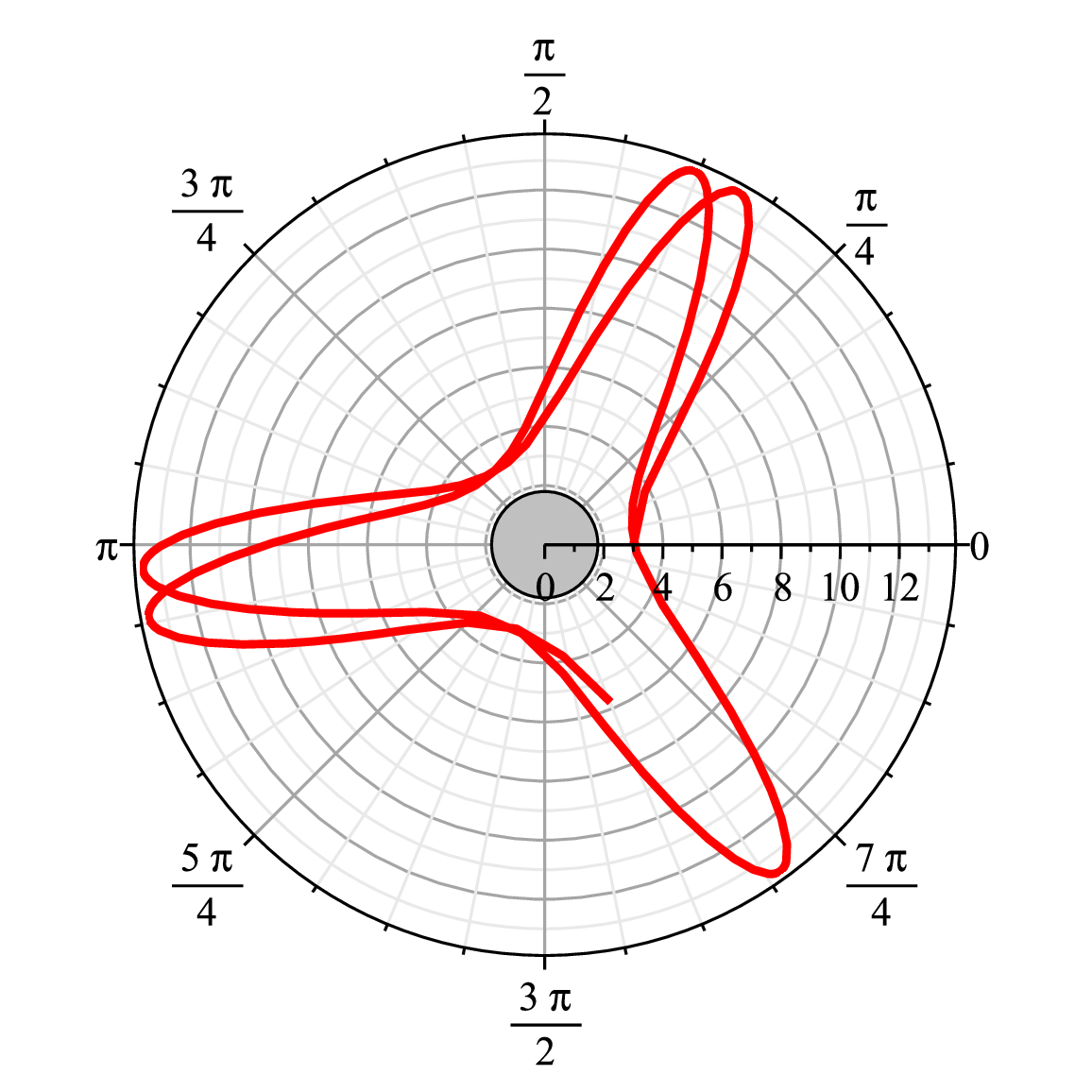}~d)
\hfil
\includegraphics[width=40mm,height=40mm]{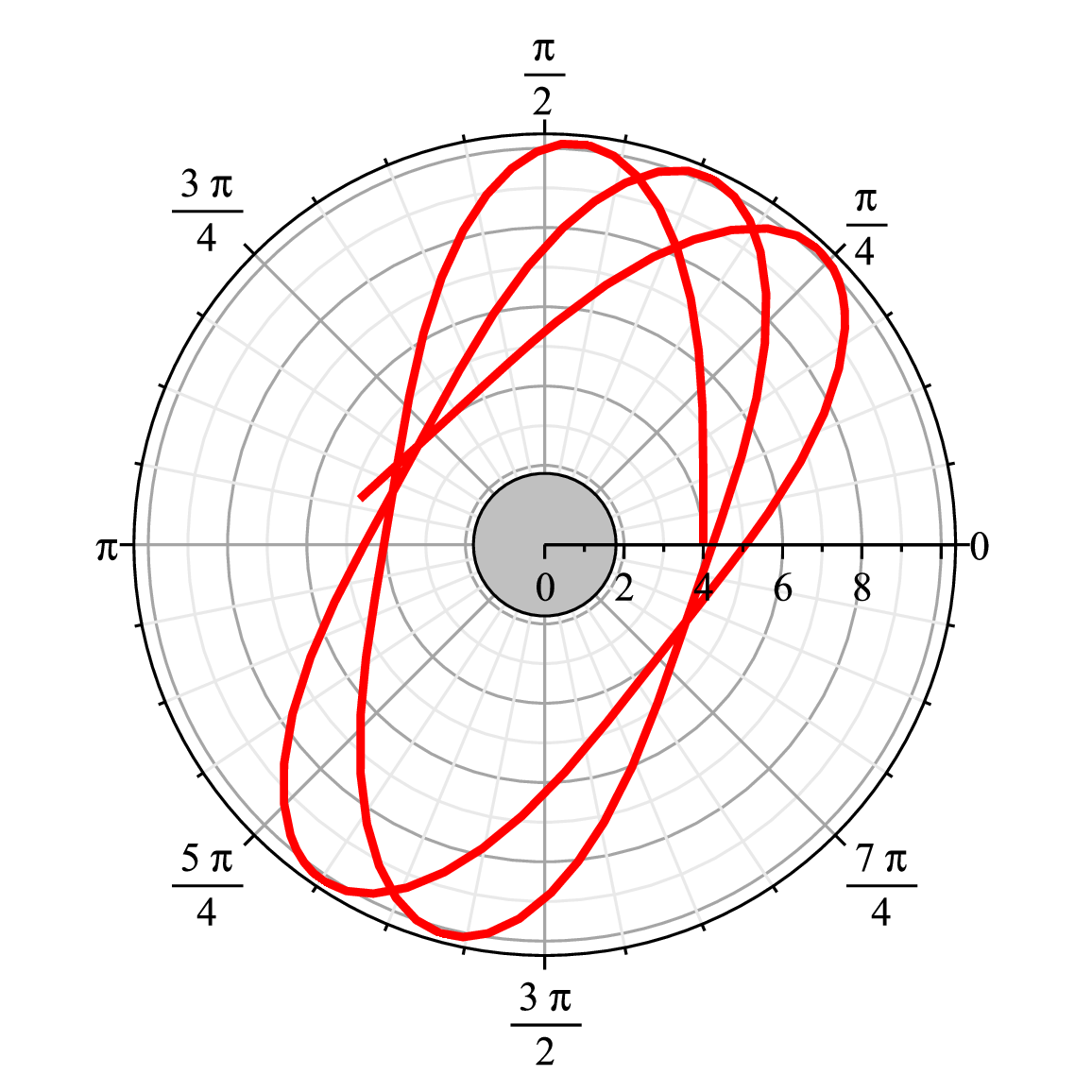}~e)
\hfil
\includegraphics[width=40mm,height=40mm]{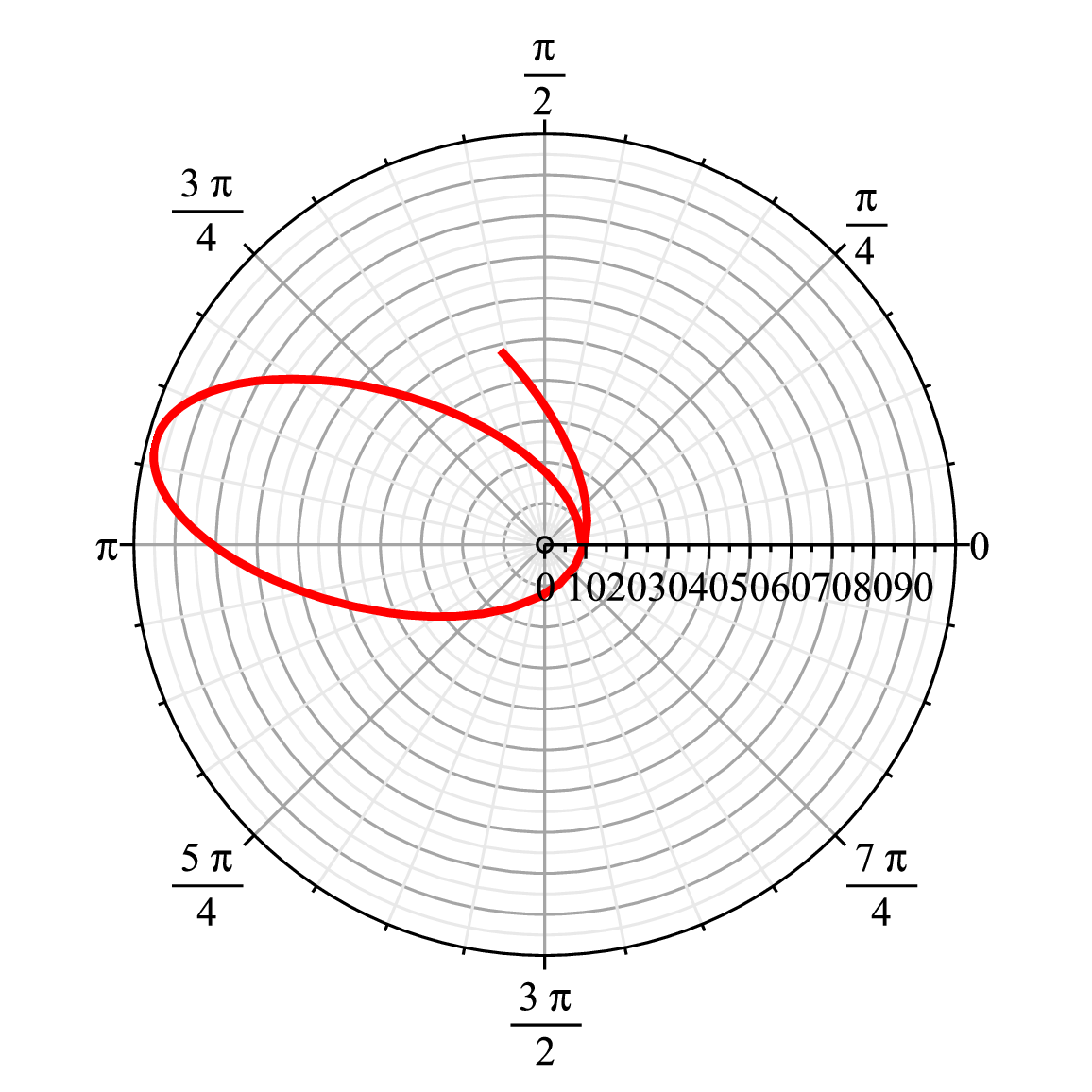}~f)
\hfil
\includegraphics[width=40mm,height=40mm]{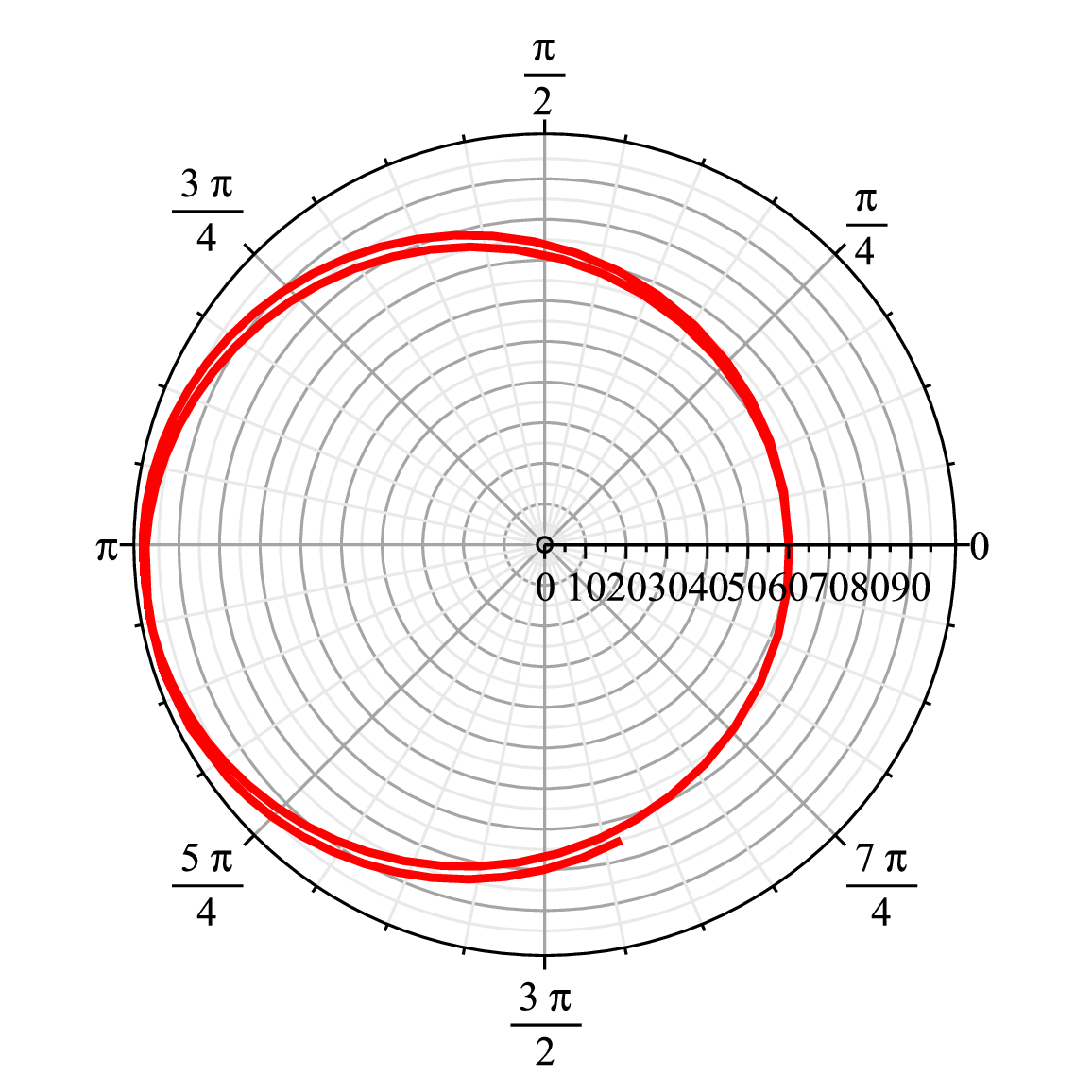}~g)
\figcaption{The possible bound orbits (or planetary orbits) for a
test-particle moving on a star with interior solution Class Ia,
using different initial points of approach and different initial
energies: \textbf{a}) $r=r_m$, $E=\epsilon_1$; \textbf{b})
$r=2.7$, $E=\epsilon_1$; \textbf{c}) $r=7$, $E=\epsilon_2$;
\textbf{d}) $r=3$, $E=\epsilon_3$; \textbf{e}) $r=4$,
$E=\epsilon_4$; \textbf{f}) $r=9$, $E=0.36$; \textbf{g}) $r=60$,
$E=10$.} \label{bound-classIa}
\end{center}
\ruledown

\begin{multicols}{2}

2) {\it{Hyperbolic motion}}\\\\

For particles coming from infinity, hyperbolic motions or escape
orbits are available. That is when $E-V(r)=0$ possesses only one
zero. Figure \ref{escape-classIa} shows some possible types of
escape orbits, for which the particle comes to the vicinity of the
star, having a definite minimum distance, and then repels again to
the infinity.
\end{multicols}
\ruleup
\begin{center}
\includegraphics[width=40mm,height=40mm]{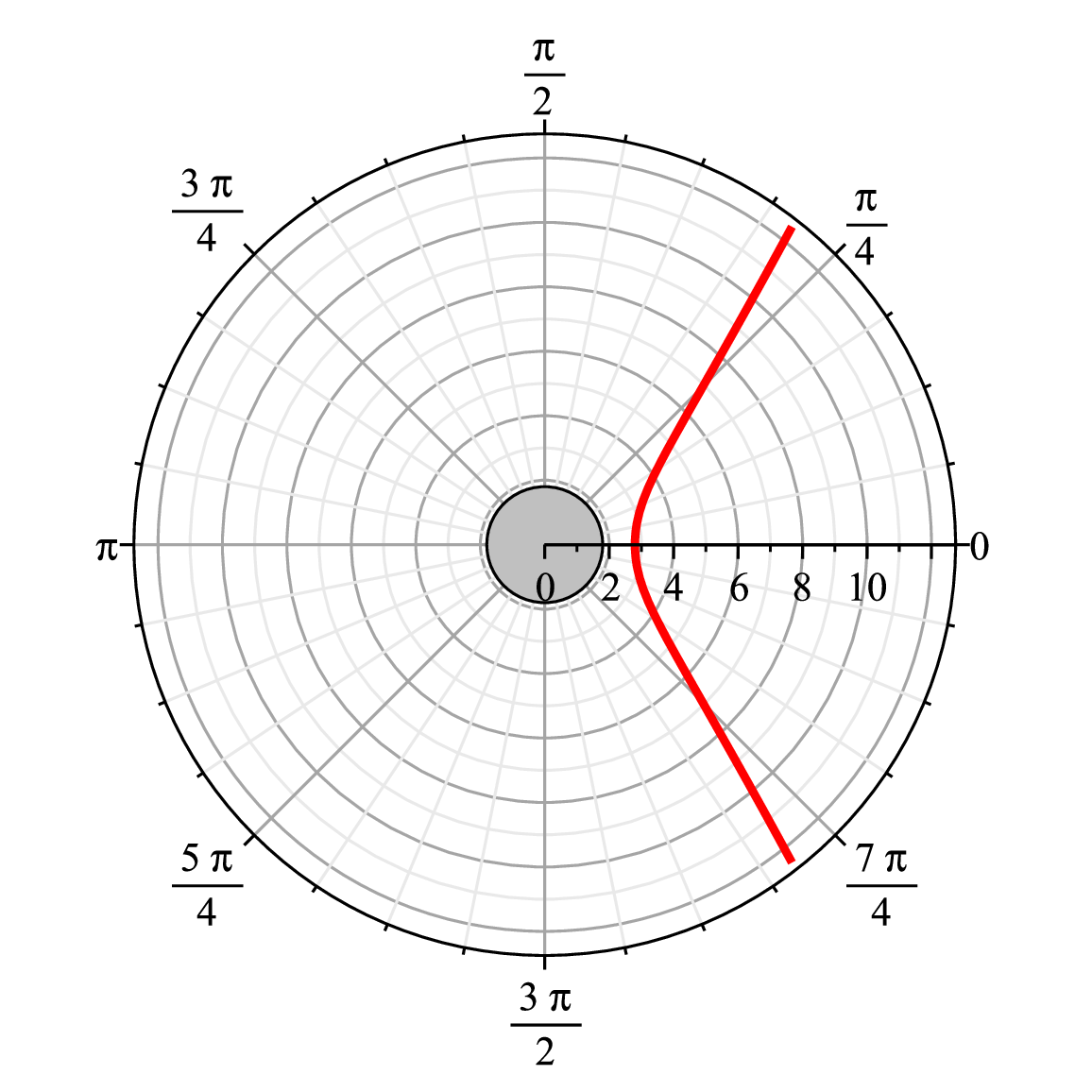}~a)
\hfil
\includegraphics[width=40mm,height=40mm]{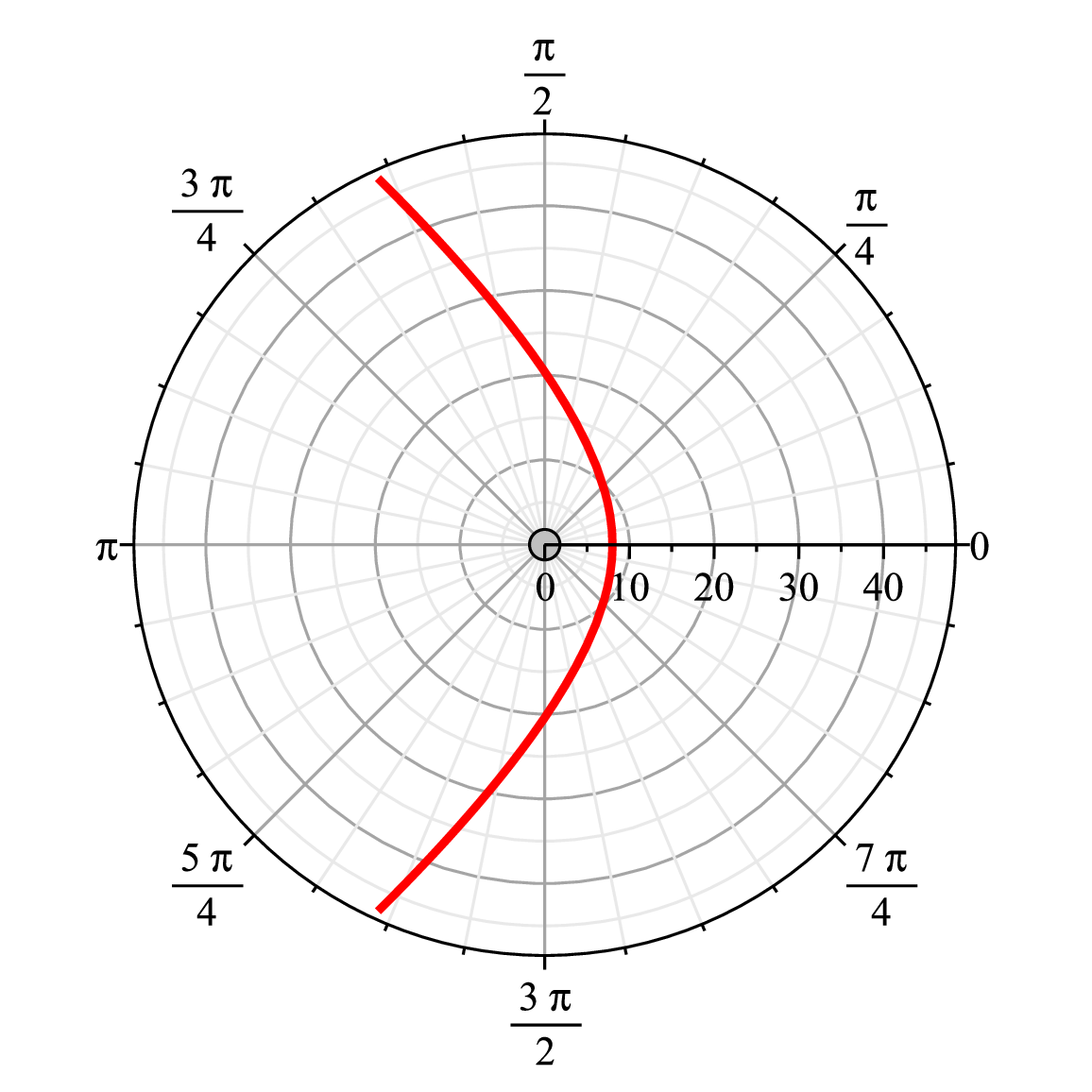}~b)
\hfil
\includegraphics[width=40mm,height=40mm]{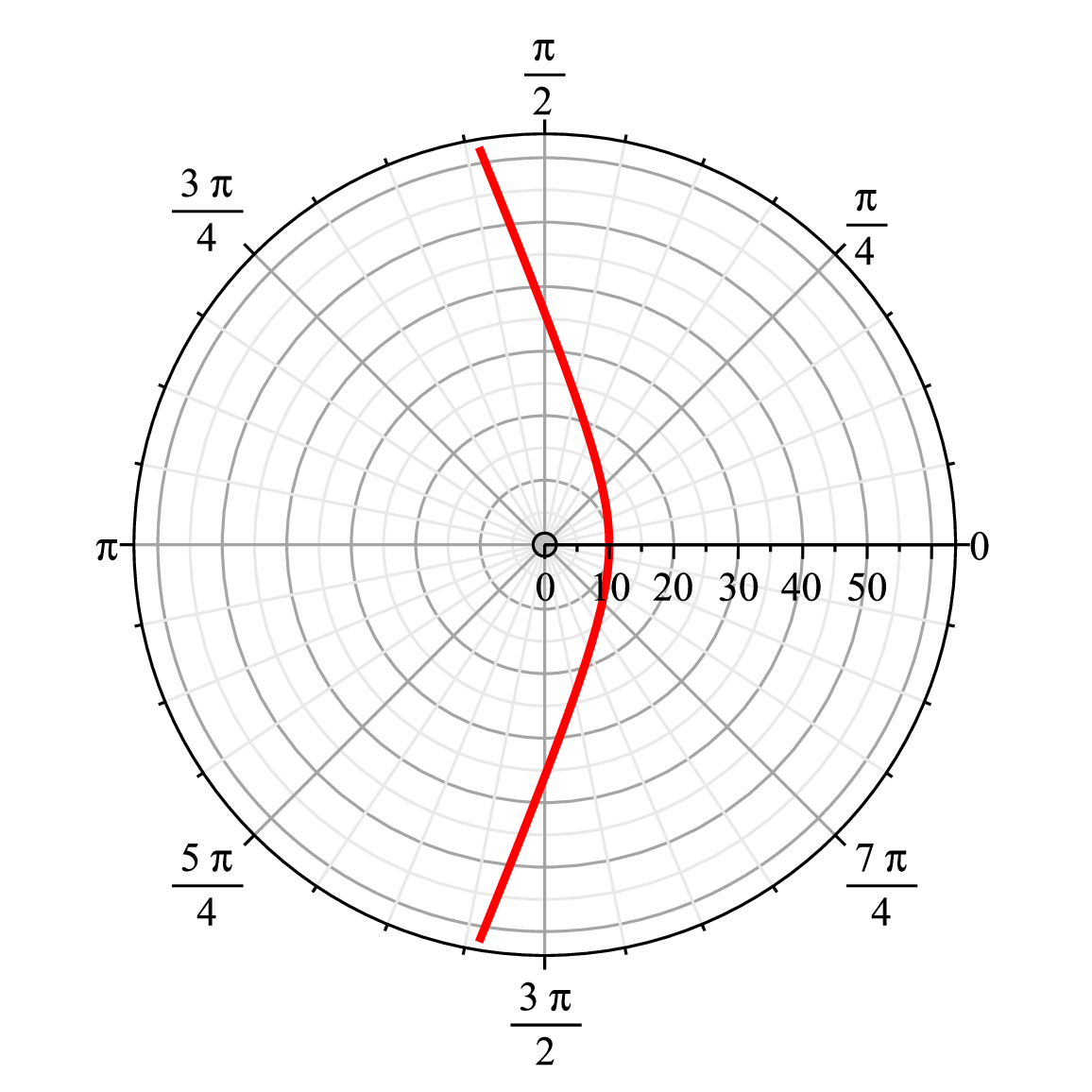}~c)
\hfil
\includegraphics[width=40mm,height=40mm]{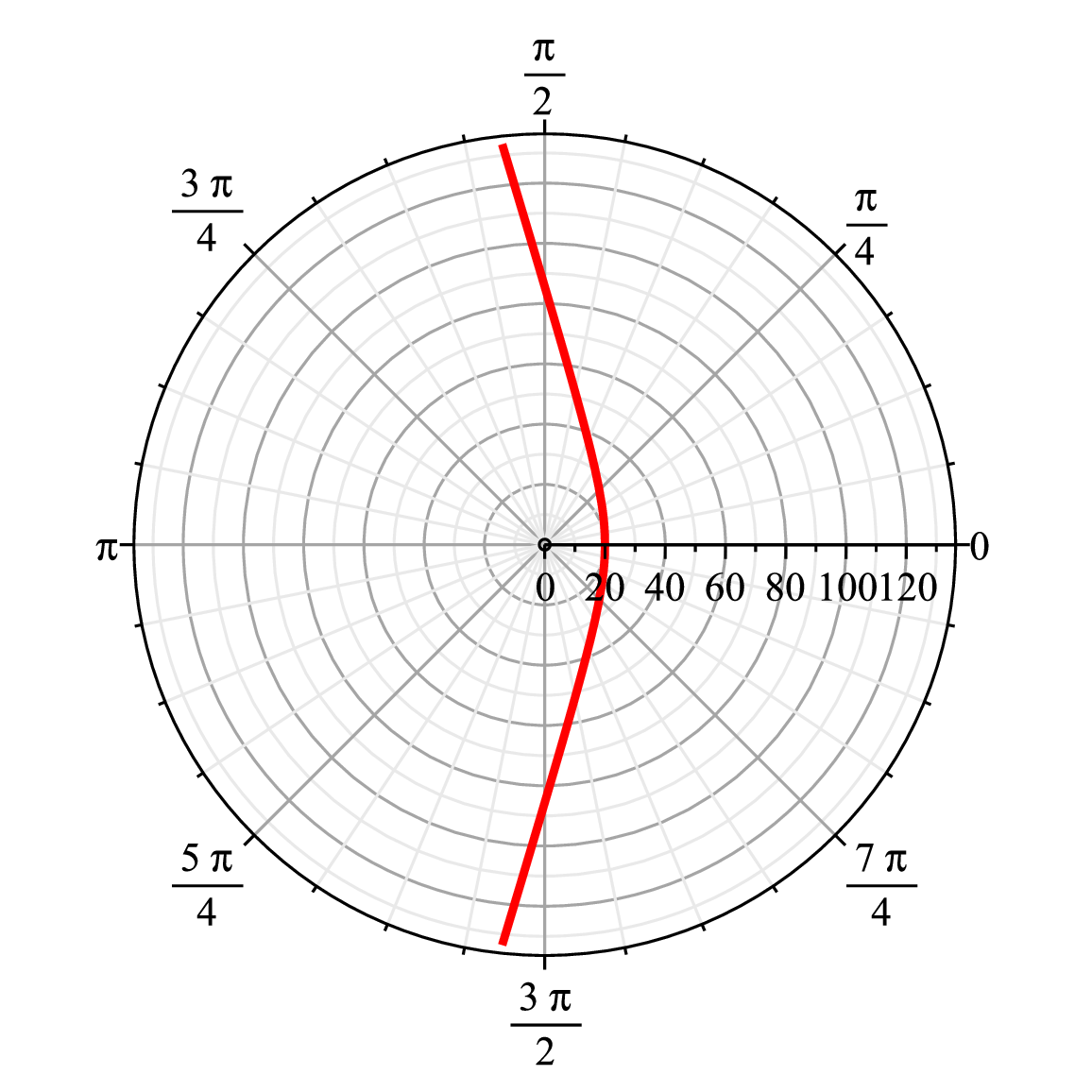}~d)
\figcaption{The possible escape orbits (or hyperbolic motions) for
a test-particle approaching to a star with interior solution Class
Ia, using different initial points of approach and different
initial energies: \textbf{a}) $r=2.8$, $E=\epsilon_3$; \textbf{b})
$r=8$, $E=\epsilon_4$; \textbf{c}) $r=10$, $E=\epsilon_2$;
\textbf{d}) $r=20$, $E=0.9$.} \label{escape-classIa}
\end{center}
\ruledown

\begin{multicols}{2}

2) {\it{Terminating orbits}}\\\\

Having high energies, when the particle moves in the vicinity of
the star, terminating orbits are also possible. This type of orbit
is unstable and terminates when the particle falls on the star's
surface (being captured). Figure \ref{capture-classIa} shows this
type of motion for different energies.
\end{multicols}
\ruleup
\begin{center}
\includegraphics[width=40mm,height=40mm]{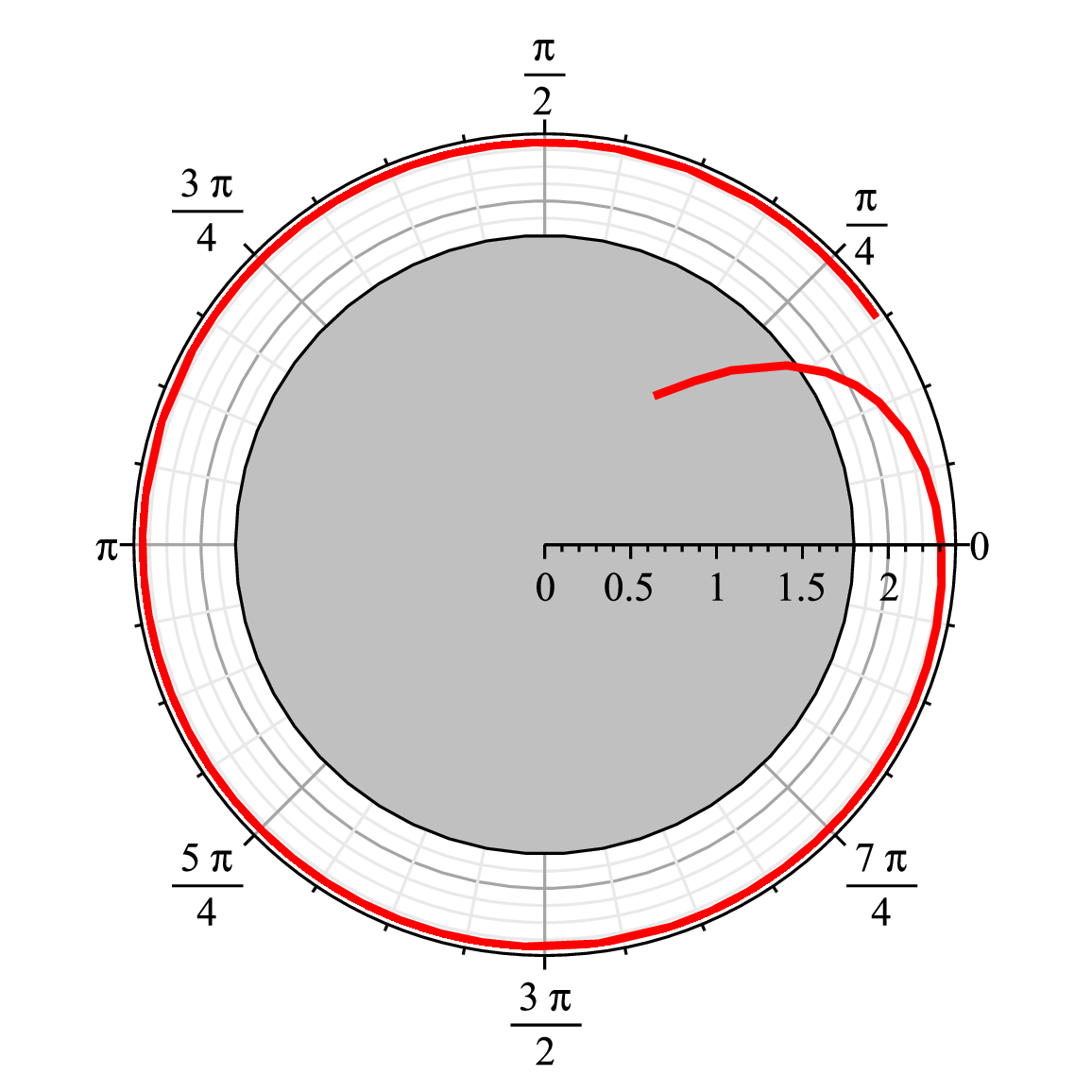}~a)
\hfil
\includegraphics[width=40mm,height=40mm]{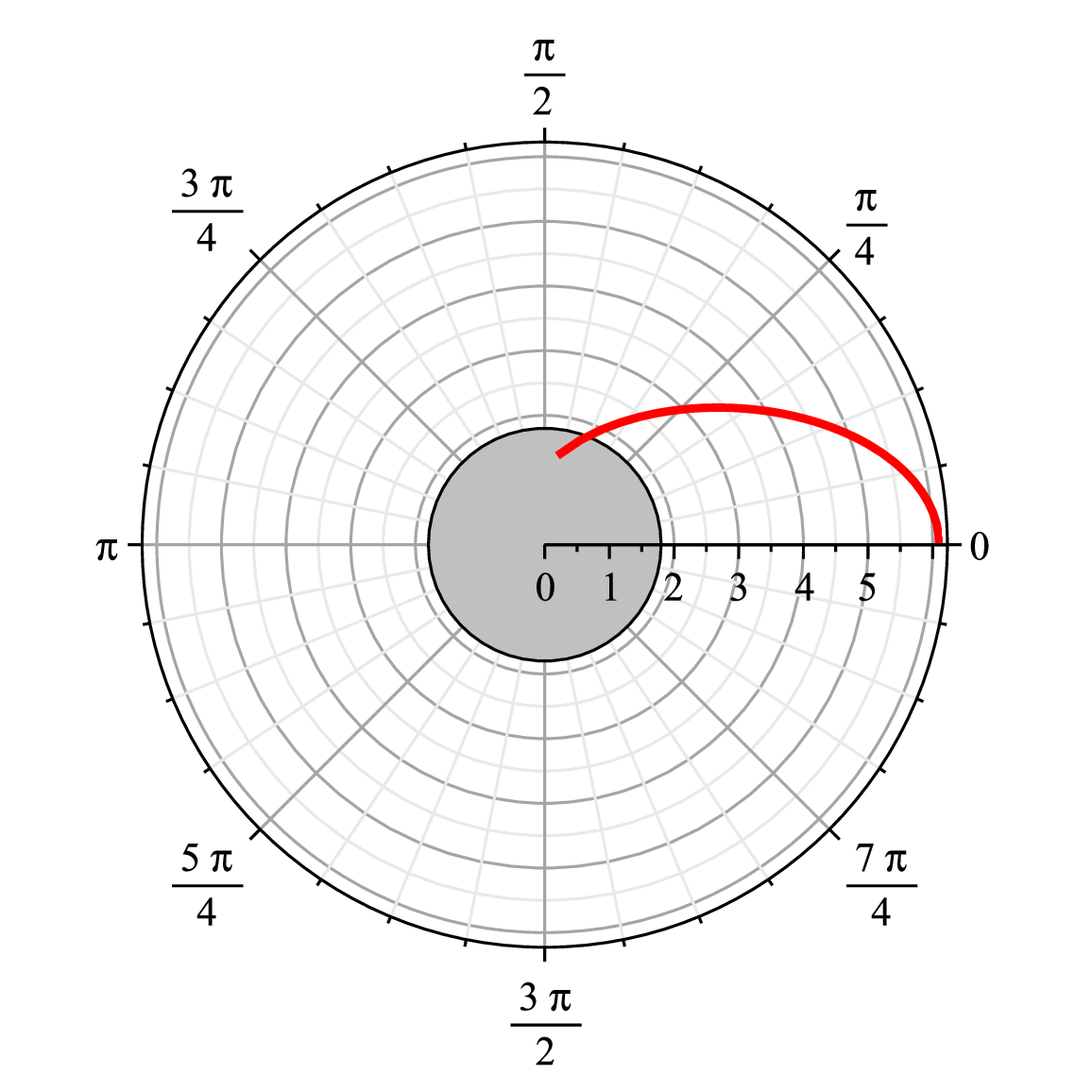}~b)
\figcaption{The terminating orbits (or capture) for a
test-particle approaching to a star with interior solution Class
Ia, using different initial points of approach and different
initial energies: \textbf{a}) $r=2.197$, $E=\epsilon_4$;
\textbf{b}) $r=r_m$, $E=0.9$.} \label{capture-classIa}
\end{center}
\ruledown

\begin{multicols}{2}

\subsection{For Class Ib}
 The total charge of the star, using Eqs.
(\ref{Class-Ib}-\ref{l-Ib}) is:

\end{multicols}
\ruleup

\begin{equation}
\begin{array}{l}
q_0={r_0^{-3}}{} [ M-{\frac {{Q_{{0}}}^{2}}{r_0}}] ^{-1}\sqrt
{{\frac {{M}^{2}}{{Q_{{0}}}^{2}}}-1}\,{Q_{{0}}}^{2} ( r_0- \delta
r ) ^{3}\sec \Big( \sqrt {{\frac {{M}^{2}}{{Q_{{0}}}^{2}}} -1} \{
{\sec^{-1}} \big( [ {\frac {M}{Q_{{0}}}}-{\frac {Q_ {{0}}}{r_0}} ]
{\frac {1}{\sqrt {{\frac {{M}^{2}}{{Q_{{0}}}^{2}} }-1}}} \big)
{\frac {1}{\sqrt {{\frac {{M}^{2}}{{Q_{{0}}}^{2}}}-1}}}
\\+\sqrt {1-2\,{\frac {M}{r_0}}+{\frac
{{Q_{{0}}}^{2}}{{r_0^2}{}}}} ( {\frac {M}{Q_{{0}}}}-{\frac
{Q_{{0}}}{r_0}} ) ^{-1} ( 2\,{\frac {M}{Q_{{0}}}}-{\frac
{Q_{{0}}}{r_0}} ) ^{-1}-Q _{{0}} \sqrt {1- ( r_0-\delta r )
^{2}Q_{{0}} ( 2\,{ \frac {M}{Q_{{0}}}}-{\frac {Q_{{0}}}{r_0}} )
{r_0^{-3}}{}} ( M-{\frac {{Q_{{0}}}^{2}}{r_0}} ) ^{-1} ( 2\,{\frac
{M}{Q_{{0 }}}}-{\frac {Q_{{0}}}{r_0}} ) ^{-1} \} \Big).\\
\end{array}
\label{q0-Ib}
\end{equation}
\\
\ruledown \vspace{0.5cm}

\begin{multicols}{2}

Utilizing the same data which were used to plot the potential in
Figure \ref{V-eff-Ia-plot}, can lead us to the potential for Class
Ib, shown in Figure \ref{V-eff-Ib-plot}.
\begin{center}
\center{\includegraphics[height=7cm]{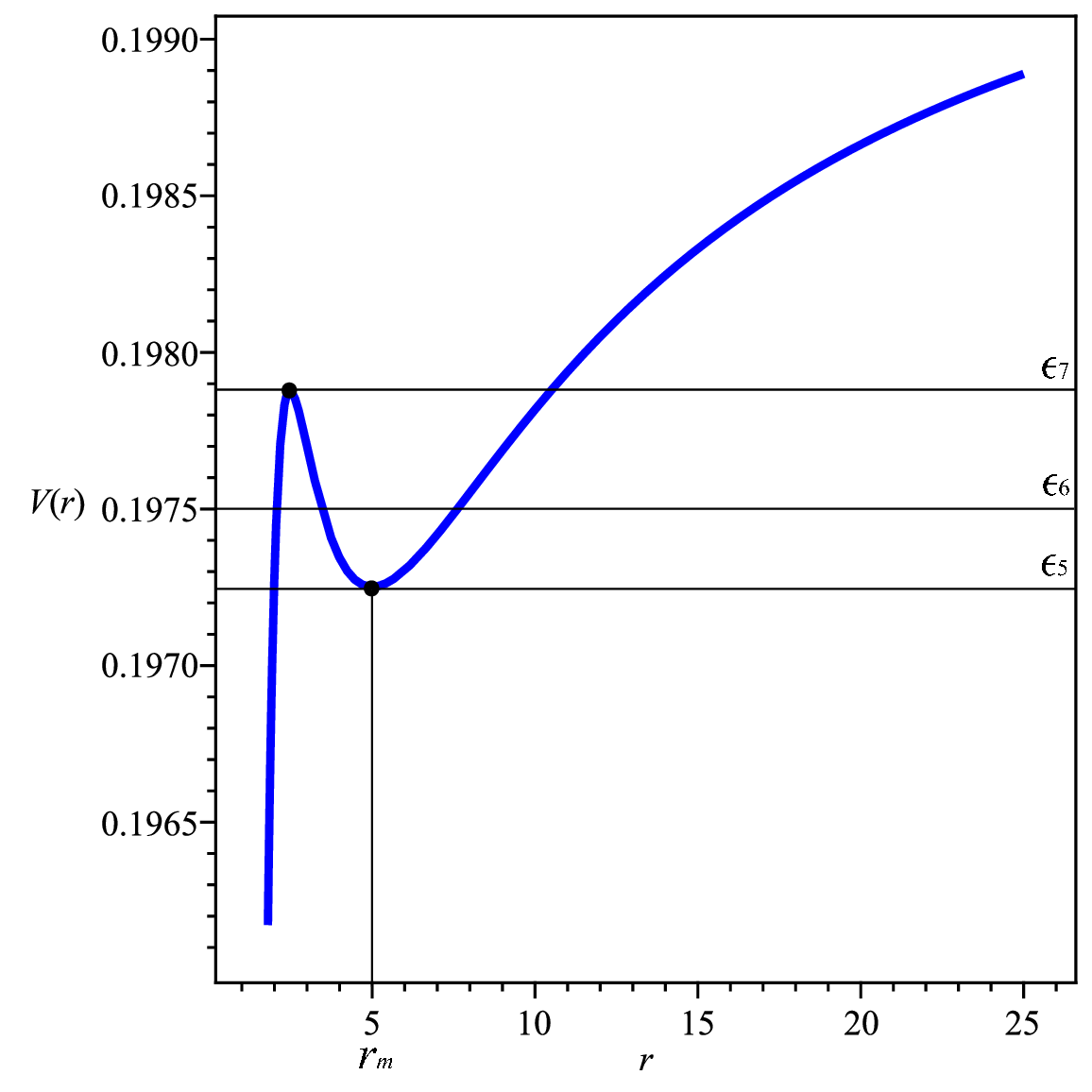}} \figcaption{\small{The
potential for a test-particle moving on a relativistic star, with
the interior solution Class Ib. The illustration is plotted for
$Q_0=0.85$, $q=0.18$, $m=0.2$, $L=0.225$, $r_0=2$, $\delta r=0.1\,
r_0$. The unit of length along the coordinate axis is $M$.}}
\label{V-eff-Ib-plot}
\end{center}
For a particle moving on a star with Class Ib as the interior
solutions, also the so-called three types of orbits are available.
The periodic bound orbits (are shown in Figure
\ref{bound-classIb}), the scape orbit (is shown Figure
\ref{escape-classIb}) and the capture (is shown in Figure
\ref{capture-classIb}).

\end{multicols}
\ruleup
\begin{center}
\includegraphics[width=40mm,height=40mm]{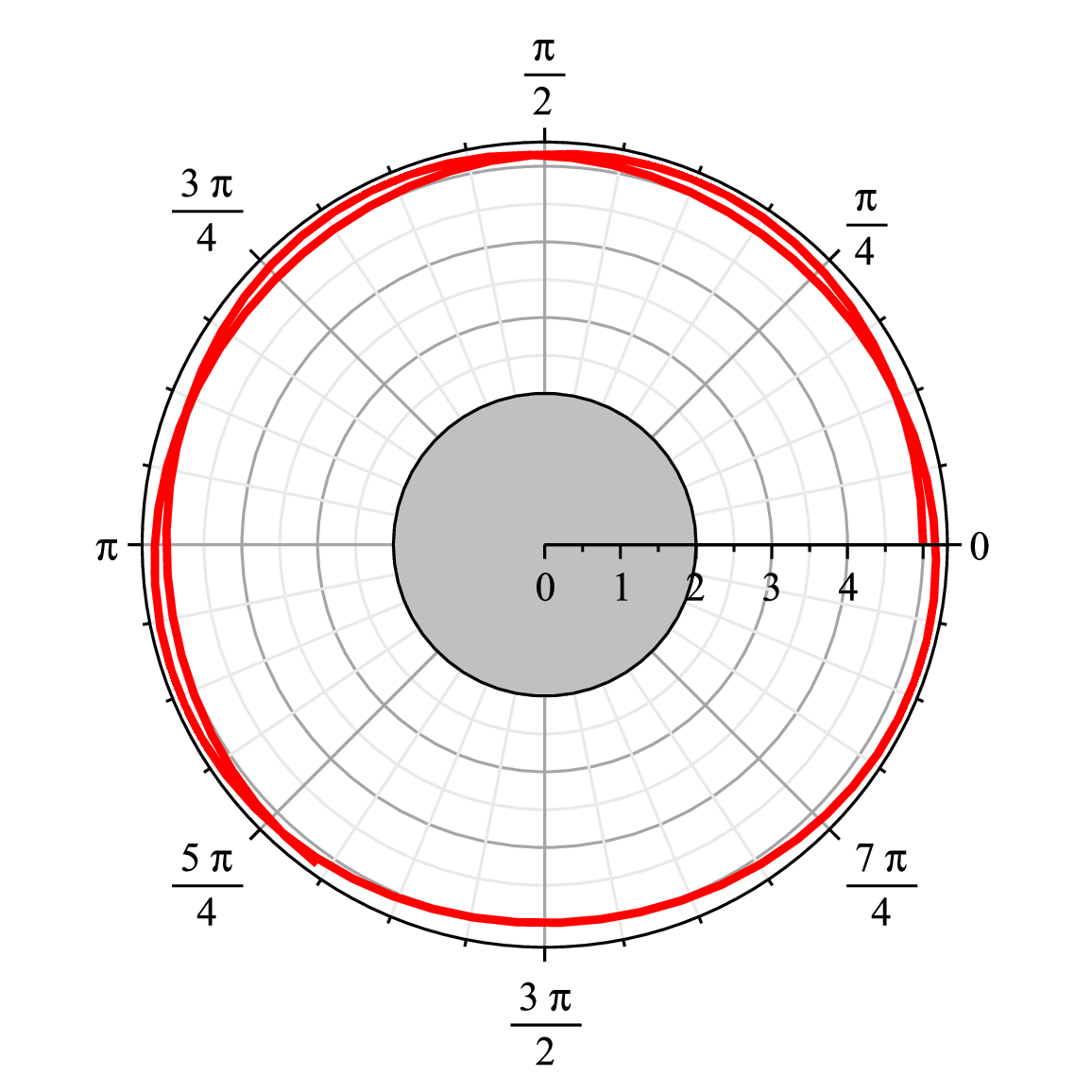}~a)
\hfil
\includegraphics[width=40mm,height=40mm]{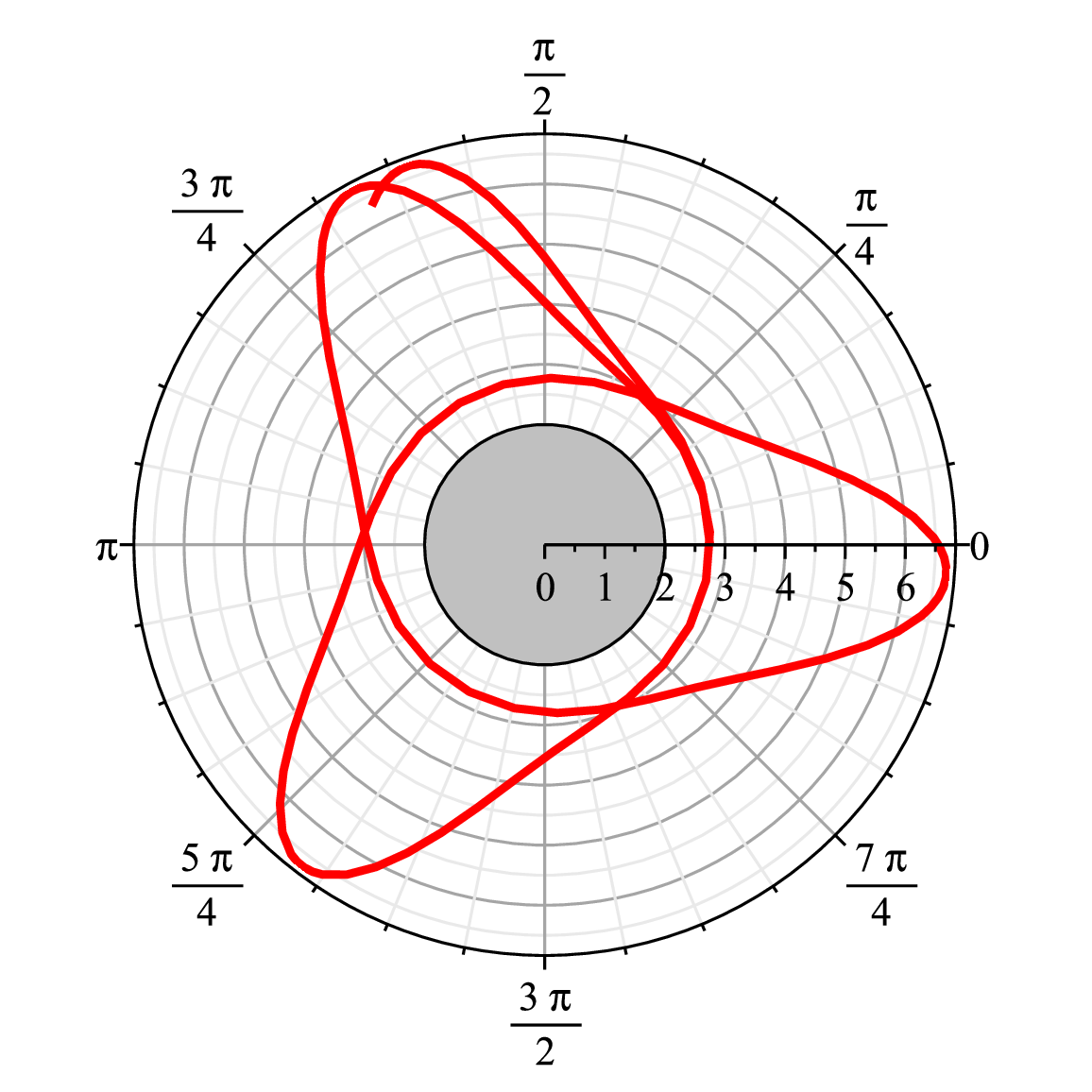}~b)
\hfil
\includegraphics[width=40mm,height=40mm]{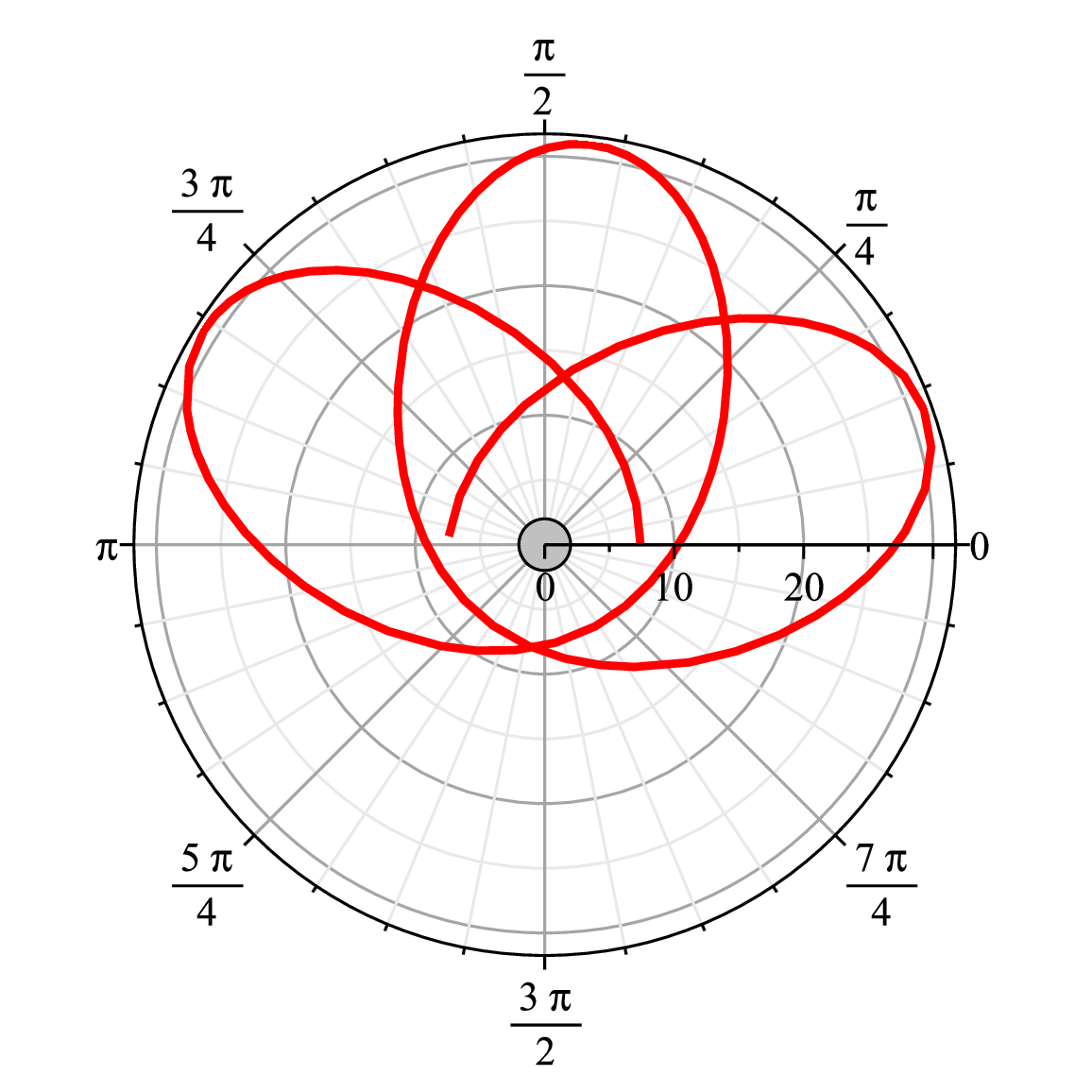}~c)
\hfil
\includegraphics[width=40mm,height=40mm]{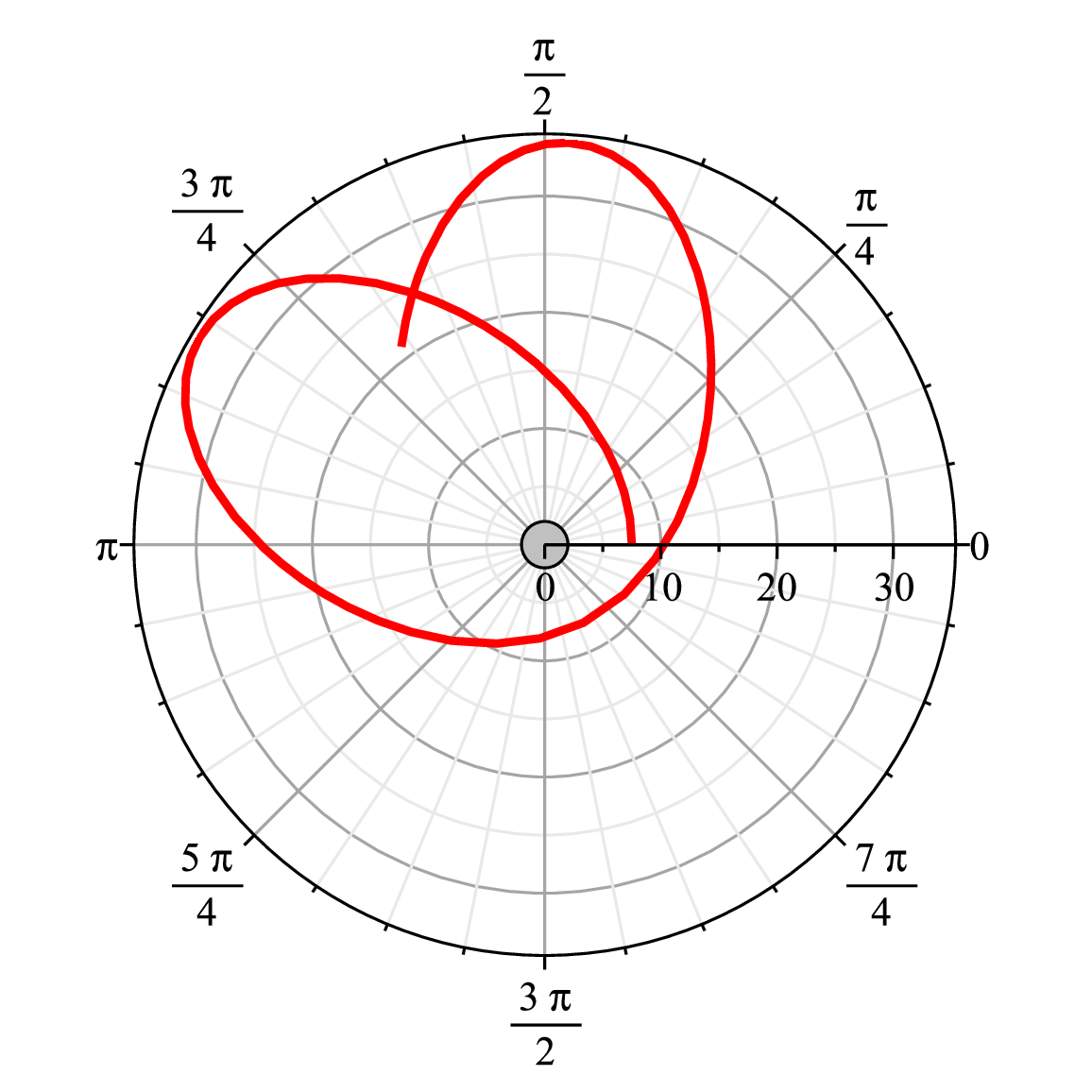}~d)
\hfil
\includegraphics[width=40mm,height=40mm]{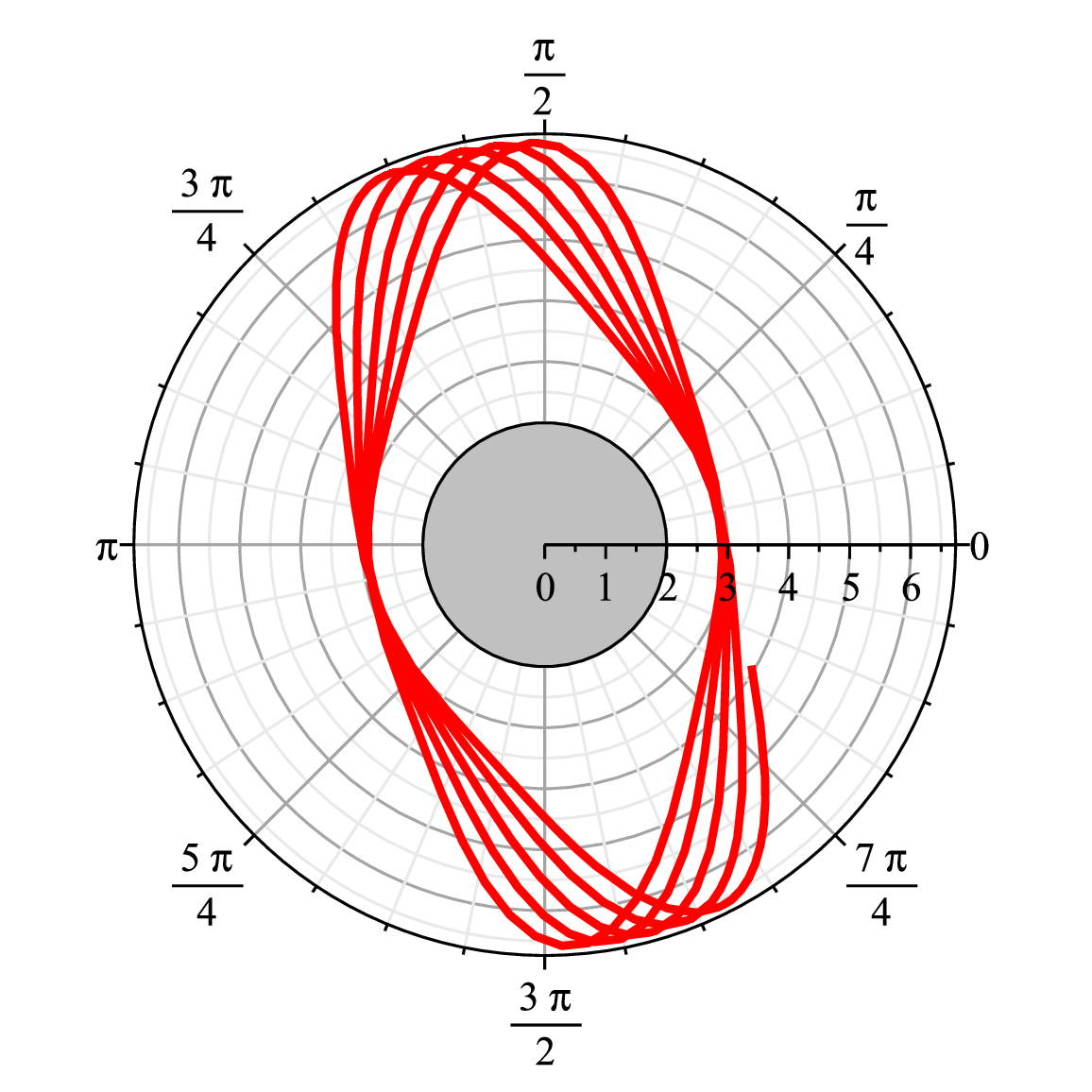}~e)
\hfil
\includegraphics[width=40mm,height=40mm]{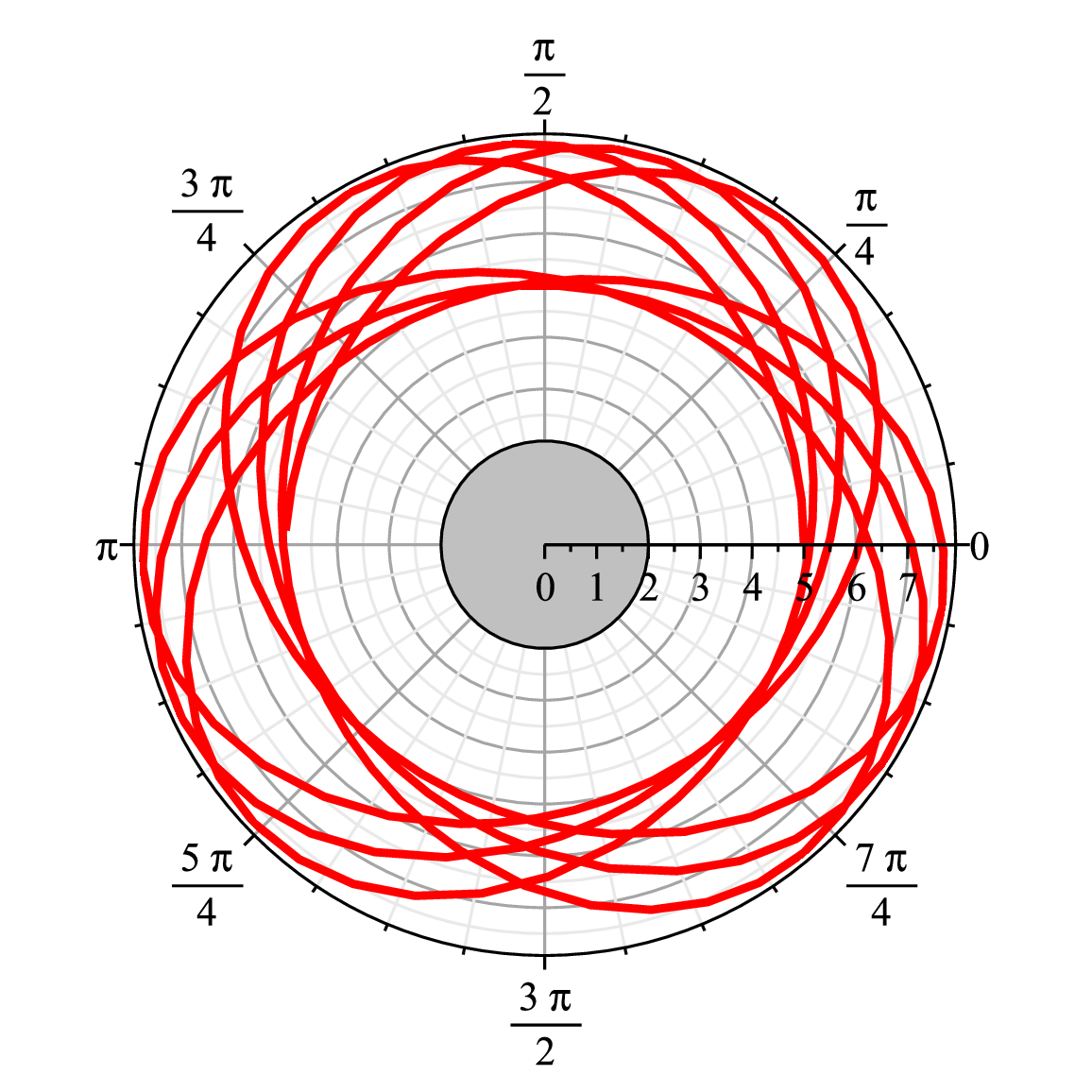}~f)
\hfil
\includegraphics[width=40mm,height=40mm]{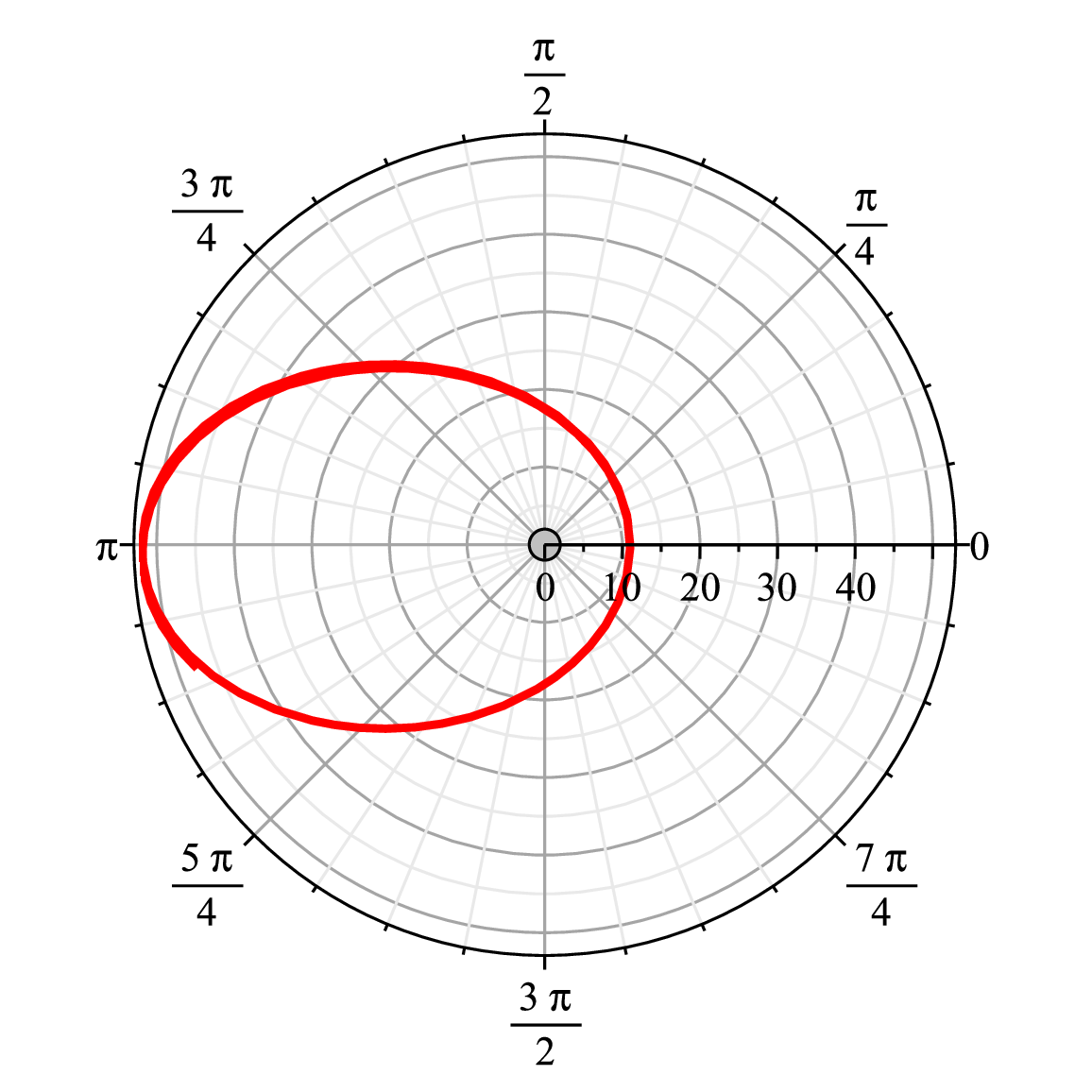}~g)
\figcaption{The possible bound orbits (or planetary orbits) for a
test-particle moving on a star with interior solution Class Ib,
using different initial points of approach and different initial
energies: \textbf{a}) $r=r_m$, $E=\epsilon_5$; \textbf{b})
$r=2.75$, $E=\epsilon_5$; \textbf{c}) $r=7.4$, $E=\epsilon_5$;
\textbf{d}) $r=7.5$, $E=\epsilon_6$; \textbf{e}) $r=2.9$,
$E=\epsilon_7$; \textbf{f}) $r=r_m$, $E=\epsilon_7$; \textbf{g})
$r=11$, $E=0.6$.} \label{bound-classIb}
\end{center}
\ruledown

\begin{multicols}{2}

\end{multicols}
\ruleup
\begin{center}
\includegraphics[width=40mm,height=40mm]{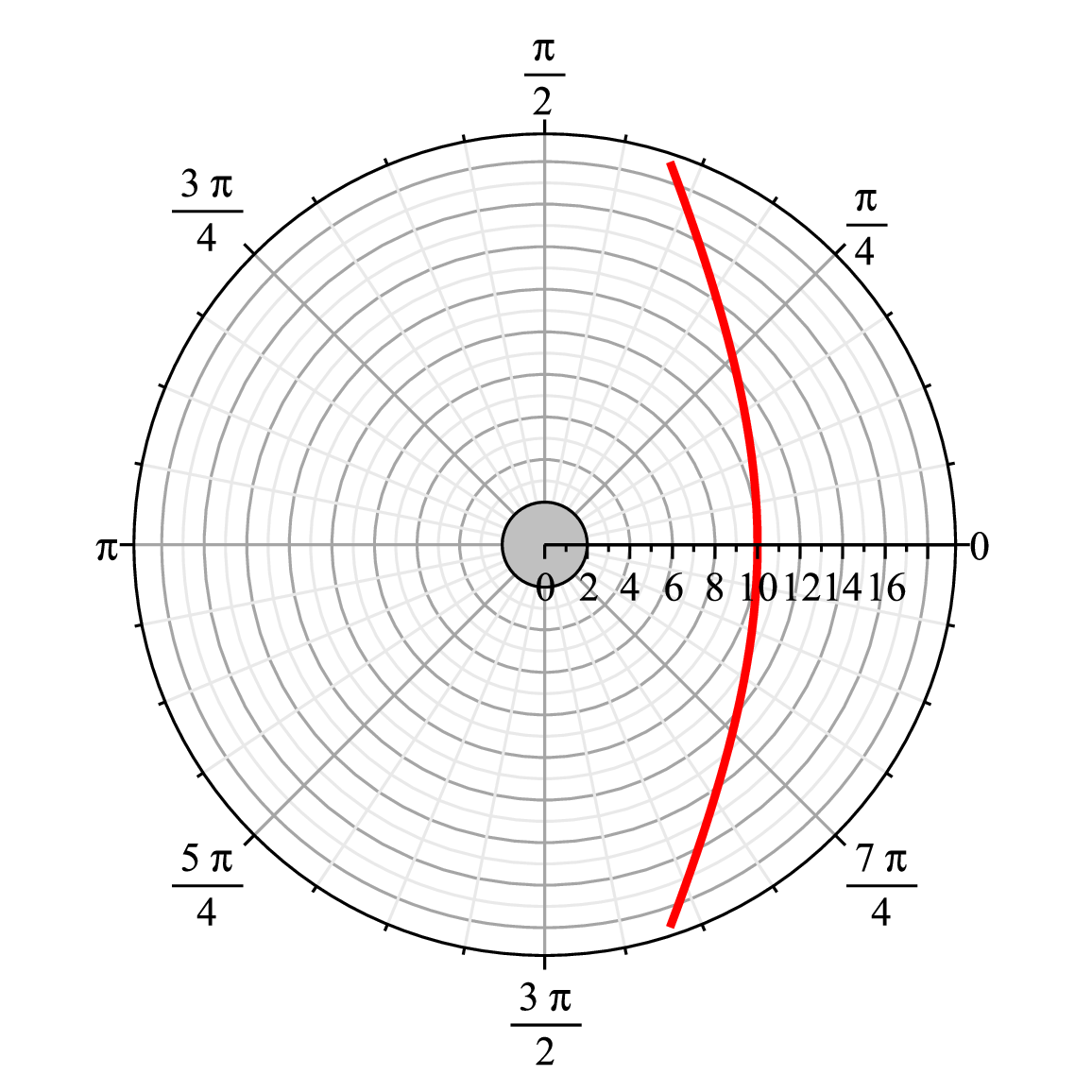}~a)
\hfil
\includegraphics[width=40mm,height=40mm]{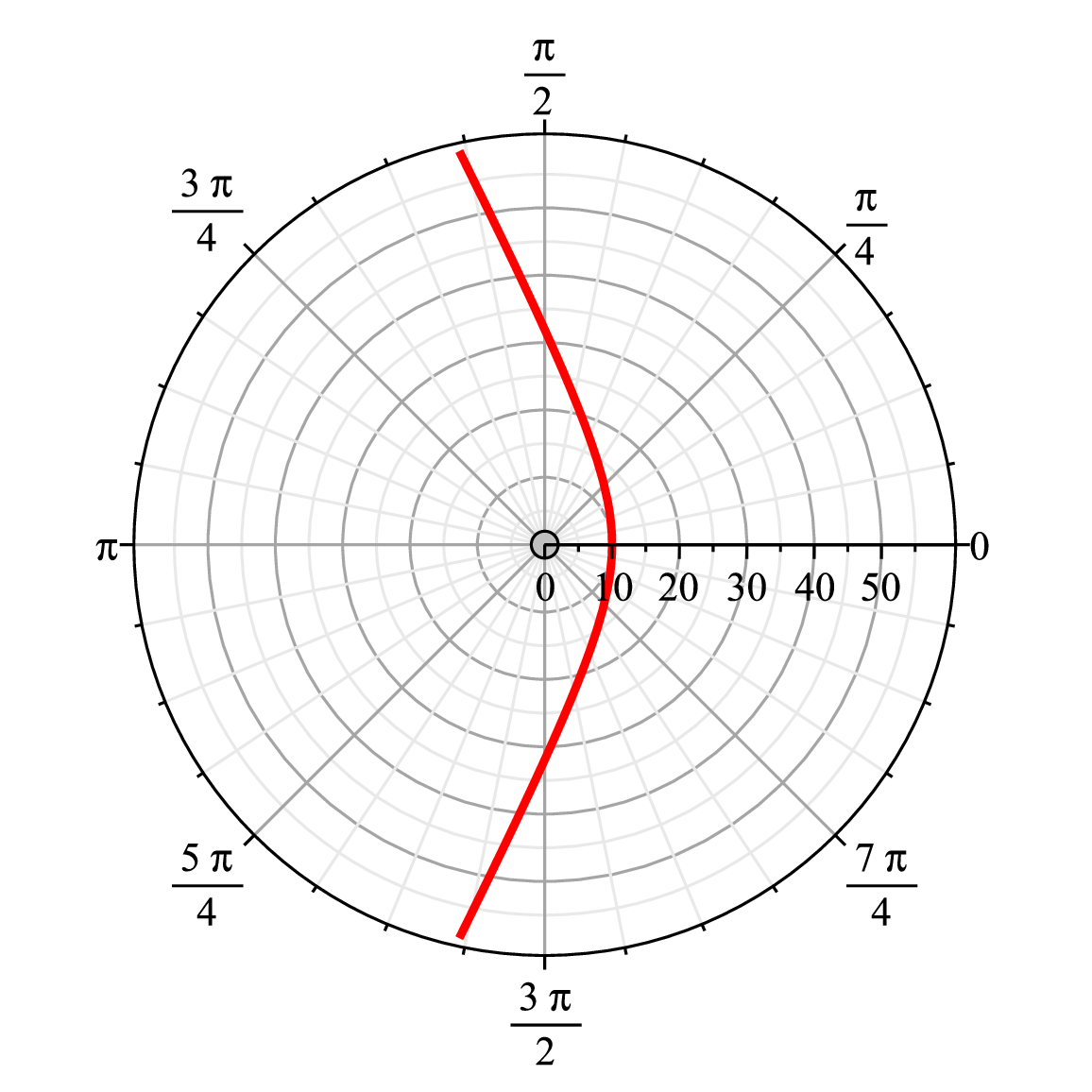}~b)
\figcaption{The possible escape orbits (or hyperbolic motions) for
a test-particle approaching to a star with interior solution Class
Ib, using different initial points of approach and different
initial energies: \textbf{a}) $r=10$, $E=\epsilon_5$; \textbf{b})
$r=10$, $E=\epsilon_7$.} \label{escape-classIb}
\end{center}
\ruledown

\begin{multicols}{2}

\end{multicols}
\ruleup
\begin{center}
\includegraphics[width=40mm,height=40mm]{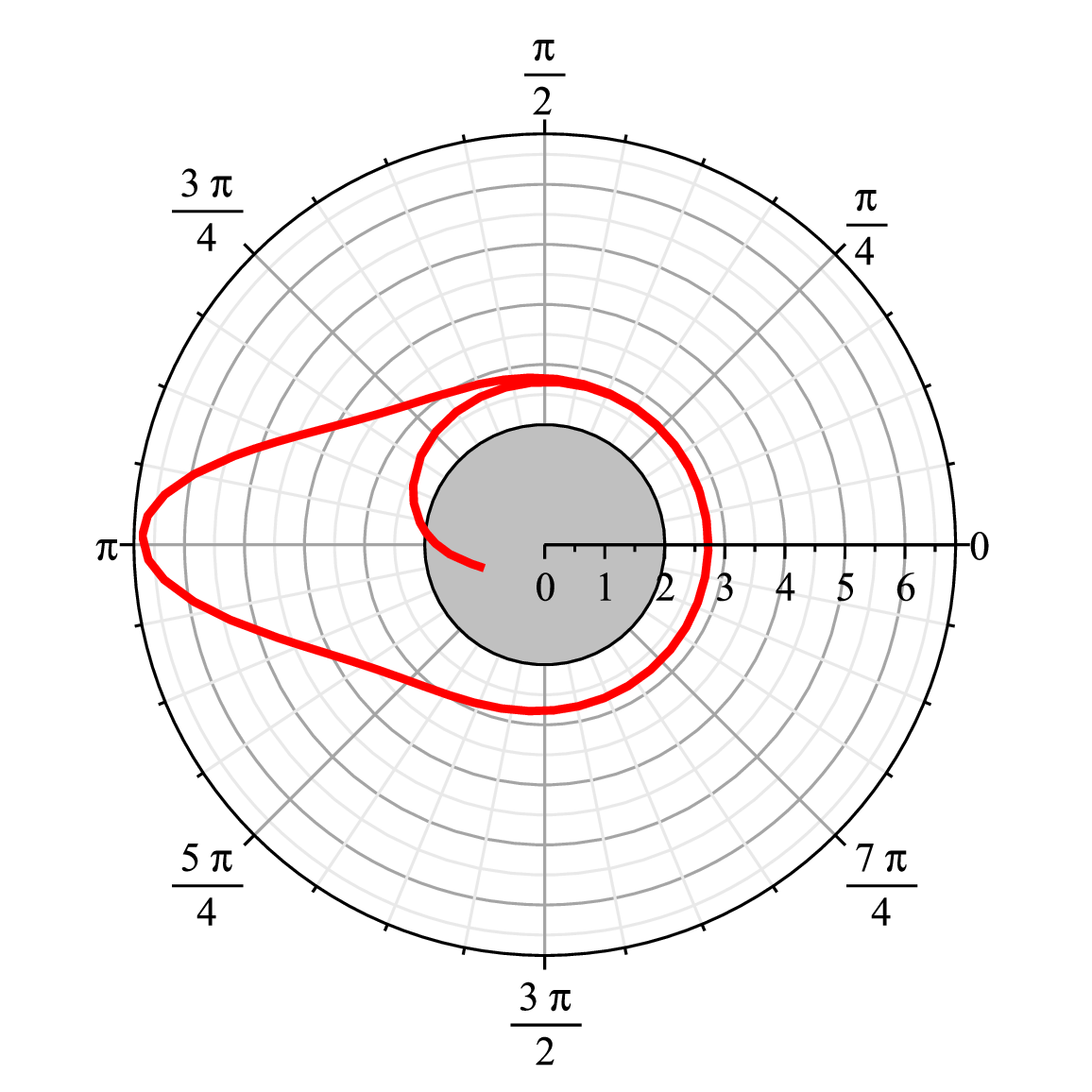}~a)
\hfil
\includegraphics[width=40mm,height=40mm]{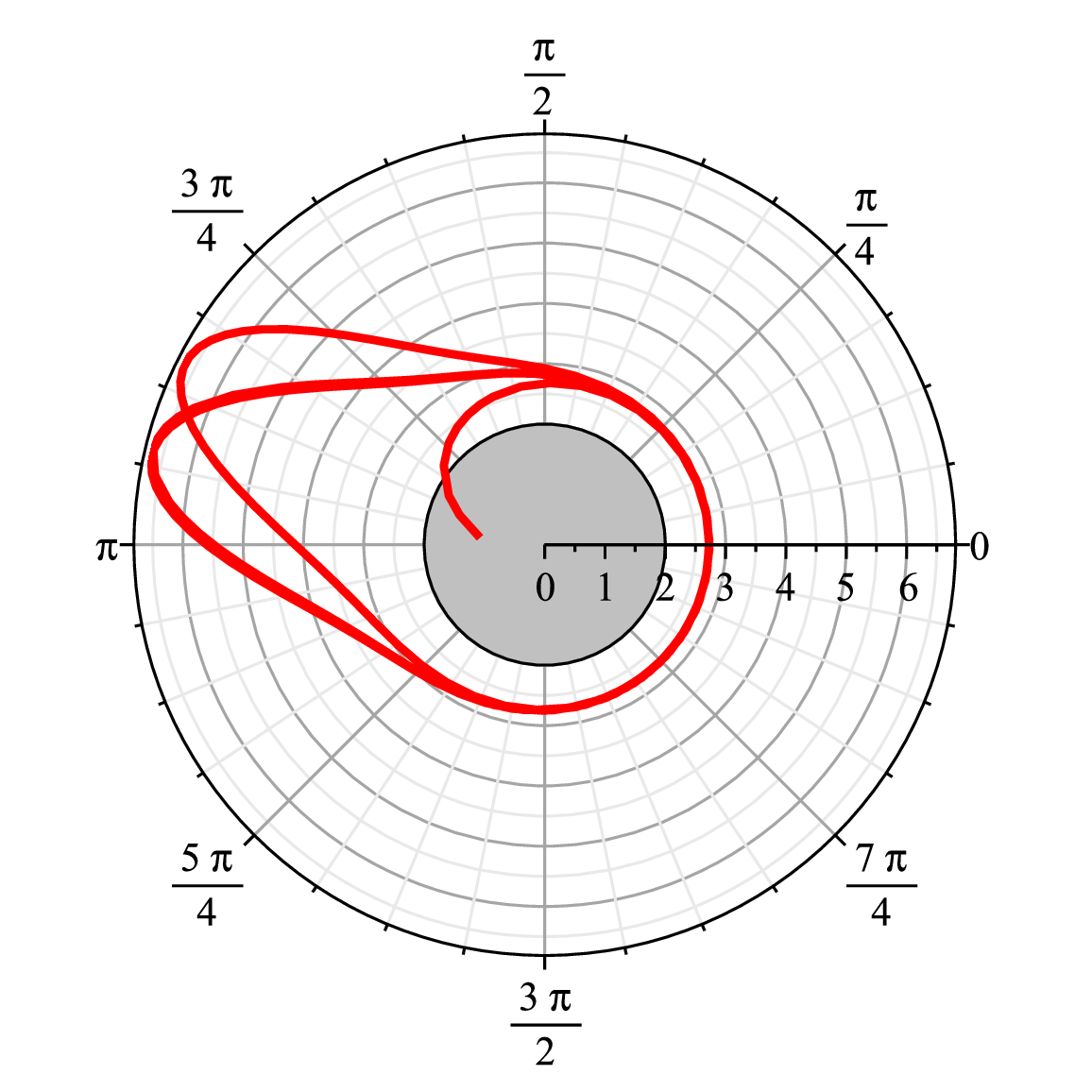}~b)
\hfil
\includegraphics[width=40mm,height=40mm]{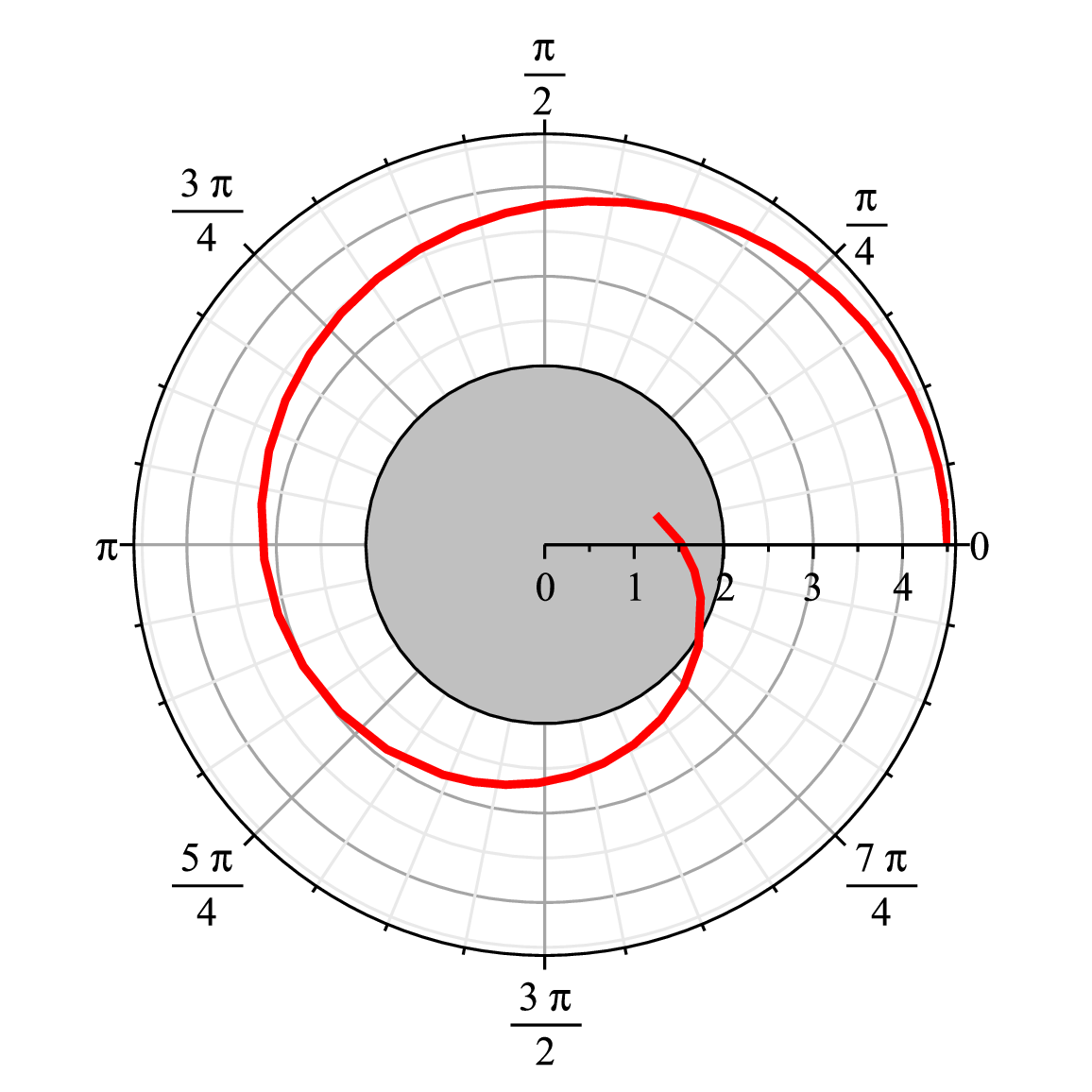}~c)
\figcaption{The terminating orbits (or capture) for a
test-particle approaching to a star with interior solution Class
Ib, using different initial points of approach and different
initial energies: \textbf{a}) $r=2.725$, $E=\epsilon_5$;
\textbf{b}) $r=2.729$, $E=\epsilon_7$; \textbf{c}) $r=4.5$,
$E=0.3$.} \label{capture-classIb}
\ruledown
\end{center}

\begin{multicols}{2}

\subsection{For Class II}
 The total charge in this case, using Eqs.
(\ref{Class-II}-\ref{l-II}) is:

\end{multicols}
\ruleup
\begin{equation}
\begin{array}{l}
q_0=(-r_0^2+2Mr_0-Q_0^2)(2Mr_0-Q_0^2)Q_0^2\sqrt{2}(r_0-\delta
r)^3\\\\
\times\Big(2-Xr_0^2Q_0^2+XQ_0^2Mr_0-XQ_0^4+2XM^2r_0^2+2YQ_0^2r_0^2-4YQ_0^2Mr_0+2YQ_0^4\Big)^{-1}\\\\
\times\Big[-r_0^8\ln\big(-\frac{r_0^2X}{(Mr_0-Q_0^2)(2Mr_0-Q_0^2)(-r_0^2+2Mr_0-Q_0^2)}\{-2Xr_0^2Q_0^2\\\\
+XQ_0^2Mr_0-XQ_0^4+2XM^2r_0^2+2YQ_0^2r_0^2
-4YQ_0^2Mr_0+2YQ_0^4\}\rm
e^{\frac{2Q_0^2(-r_0^2+2Mr_0-Q_0^2)}{(Mr_0-Q_0^2)^2}}
\big)\Big]^{-\frac{1}{2}}, \label{q0-II}
\end{array}
\end{equation}
\\
\ruledown \vspace{0.5cm}

\begin{multicols}{2}

in which
\end{multicols}
\ruleup
$$X = \sqrt{-\frac{2Mr_0-Q_0^2-r_0^2}{r_0^2}},\,\,\,\,Y=\sqrt{ -\frac{2r_0^3M-r_0^4-r_0^2Q_0^2-4r_0^2\delta rM+2r_0\delta rQ_0^2+2\delta r^2Mr_0-\delta r^2Q_0^2}{r_0^4} }.$$
\\
\ruledown \vspace{0.5cm}

\begin{multicols}{2}

The corresponding potential for the same data used before, has
been illustrated in Figure \ref{V-eff-II-plot}.
\begin{center}
{\includegraphics[height=7cm]{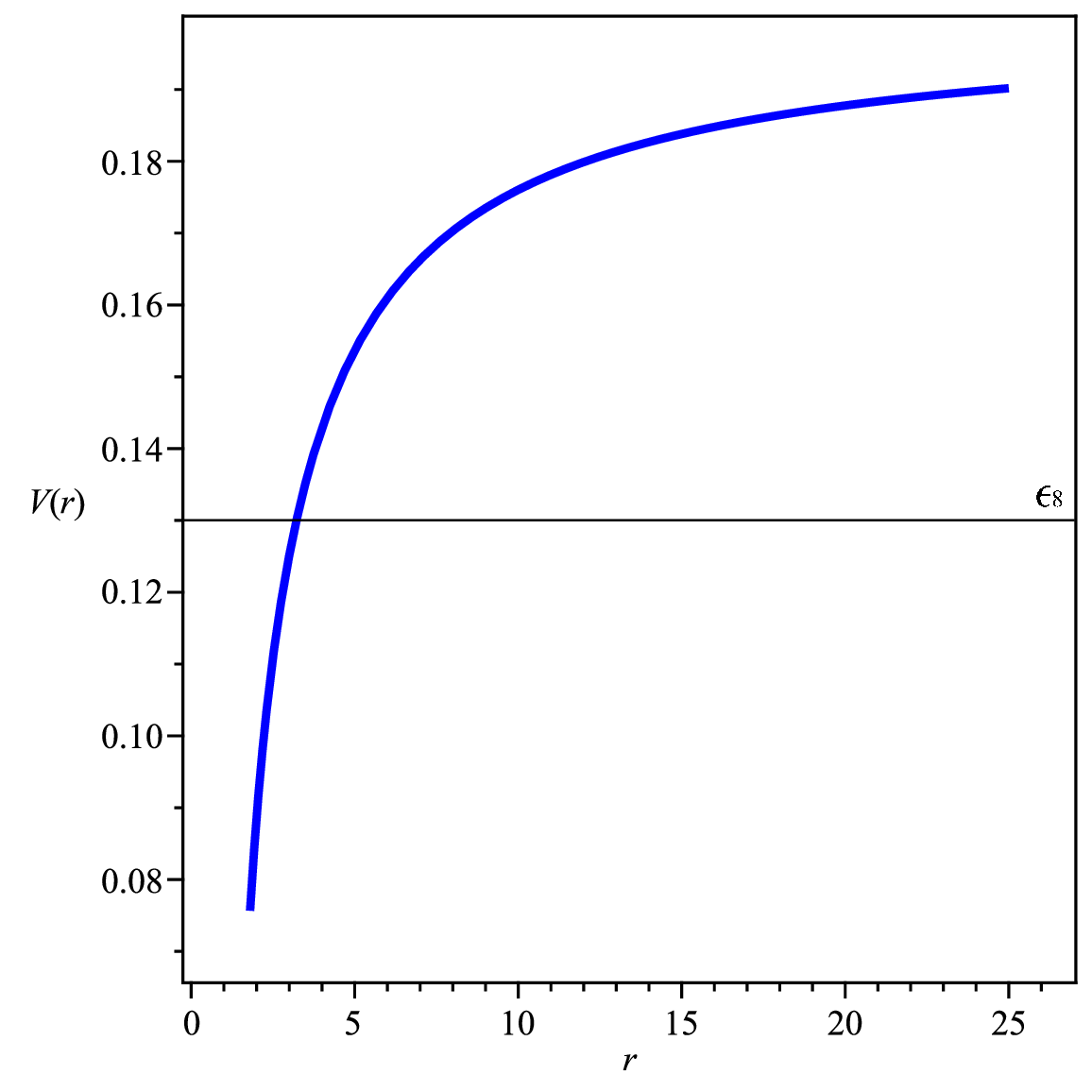}} \figcaption{\small{The
potential for a test-particle moving on a relativistic star, with
the interior solution Class II. The illustration is plotted for
$Q_0=0.85$, $q=0.18$, $m=0.2$, $L=0.225$, $r_0=2$, $\delta r=0.1
\,r_0$. The unit of length along the coordinate axis is $M$.}}
\label{V-eff-II-plot}
\end{center}
According to this potential, the equation $E-V(r)=0$, possesses
only one zero. Therefore, it is expected that the particle
exhibits only unstable orbits (escape and terminating orbits).
However, for medium and high energies, when the particle comes
from infinity, the periodic bound orbits are also available. All
types of possible orbits have been indicated in Figure
\ref{motion-classII}.
\end{multicols}
\ruleup
\begin{center}
\includegraphics[width=40mm,height=40mm]{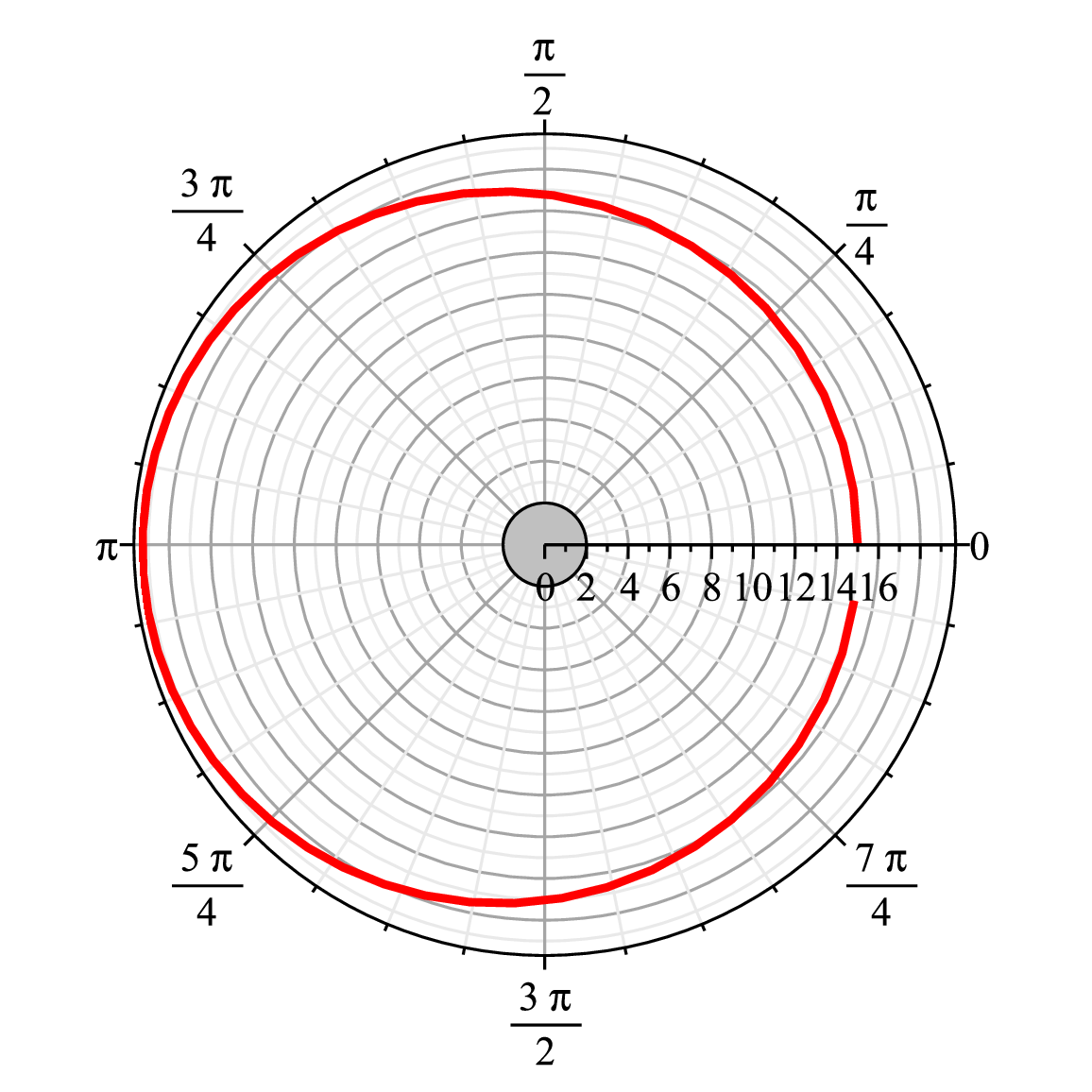}~a)
\hfil
\includegraphics[width=40mm,height=40mm]{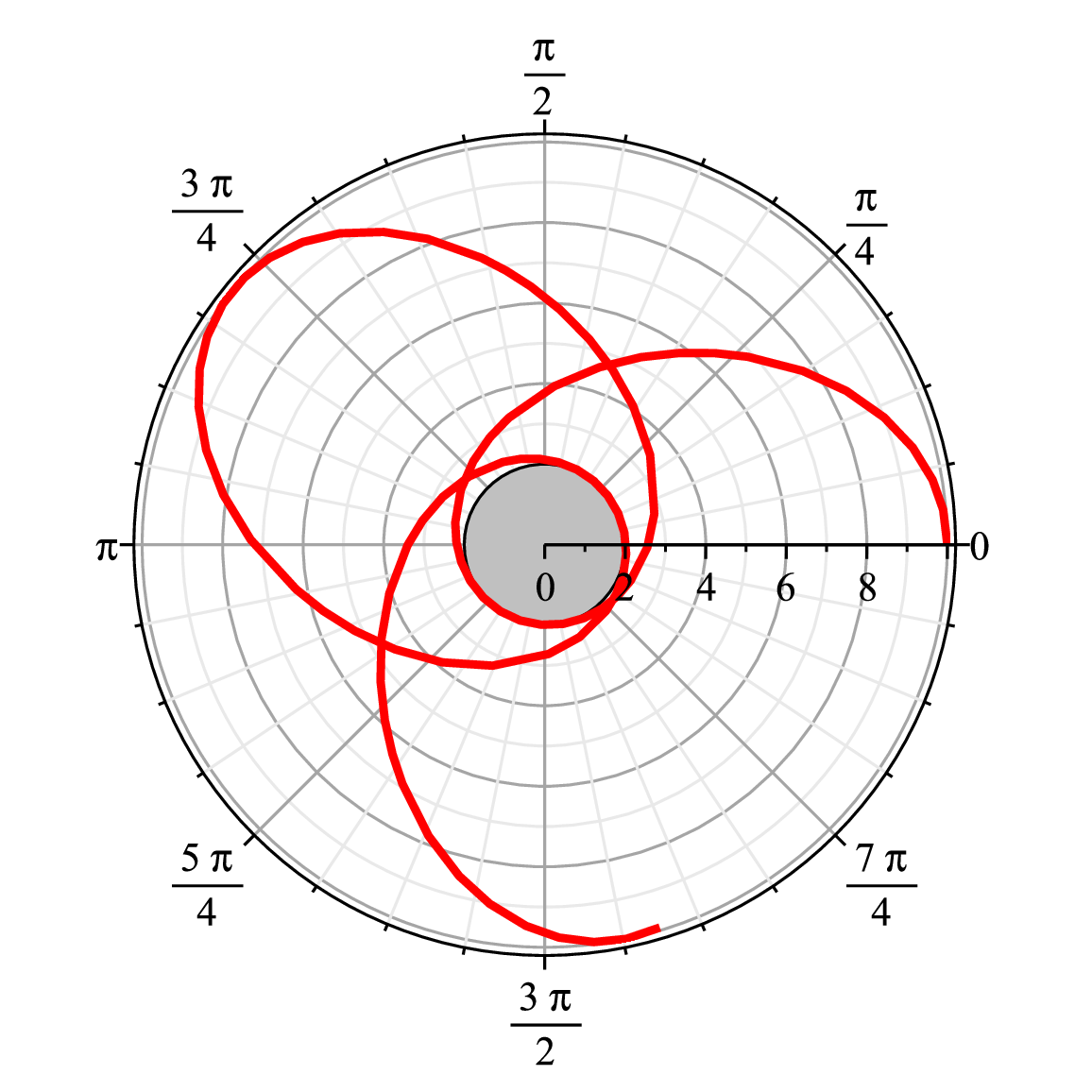}~b)
\hfil
\includegraphics[width=40mm,height=40mm]{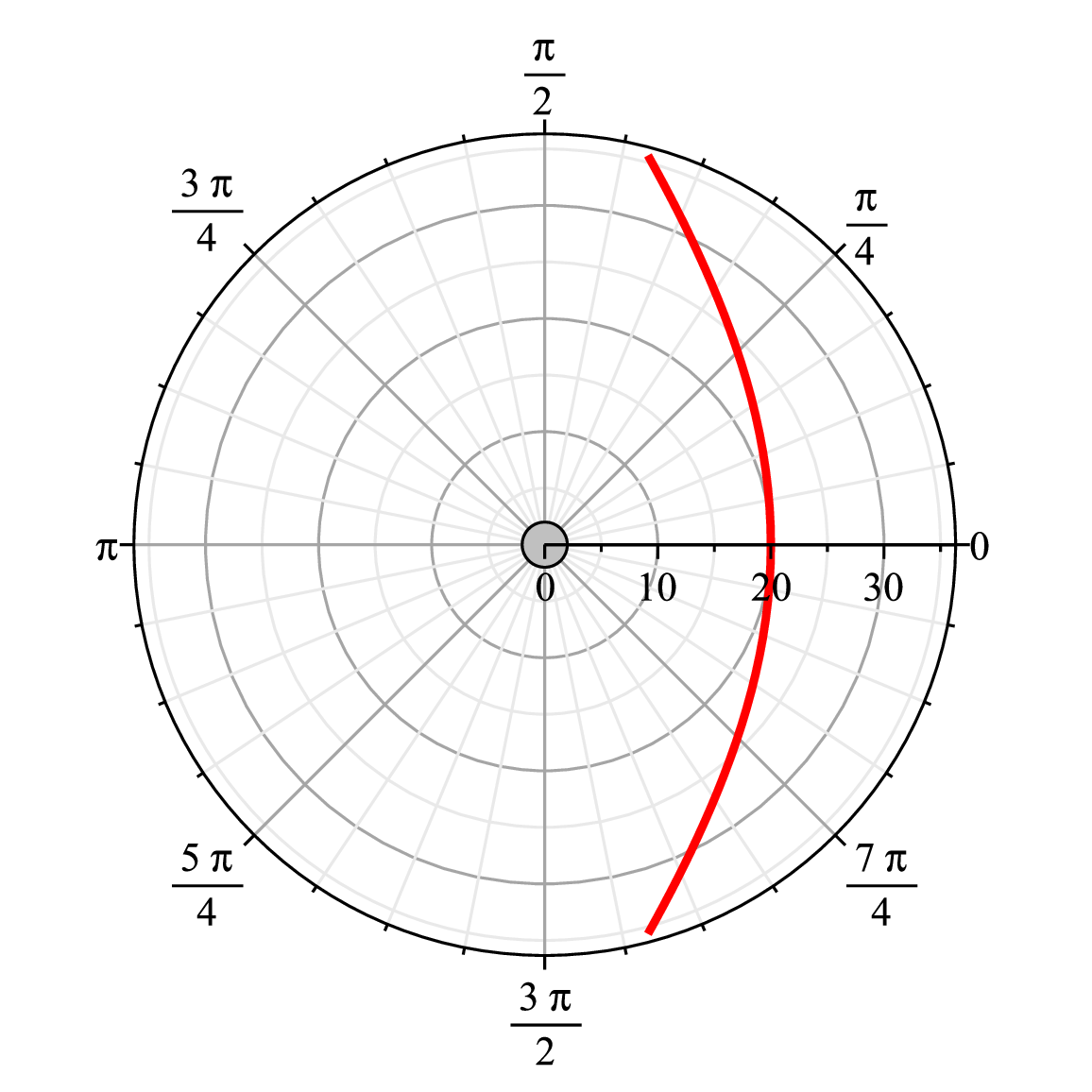}~c)
\hfil
\includegraphics[width=40mm,height=40mm]{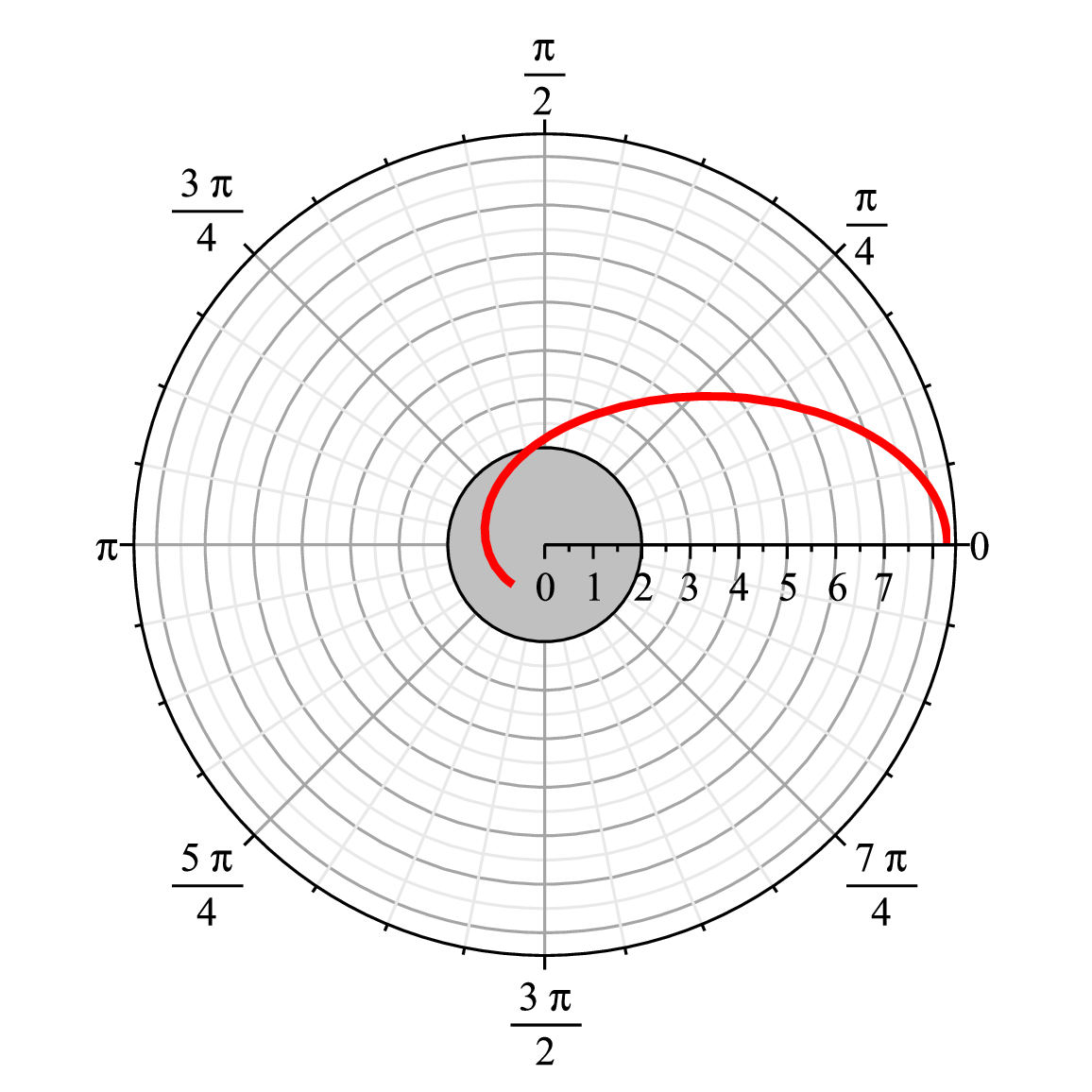}~d)
\figcaption{The possible orbits for a test-particle moving on a
star with interior solution Class II, using different initial
points of approach and different initial energies: \textbf{a},
\textbf{b}) the periodic bound orbits for $r=15$, $E=\epsilon_8$
and $r=10$, $E=0.3$; \textbf{c}) the escape orbit for $r=20$,
$E=\epsilon_8$ and \textbf{d}) the terminating orbit for $r=8.3$,
$E=\epsilon_8$.} \label{motion-classII}
\end{center}
\ruledown

\begin{multicols}{2}

\section{Motion of a charged particle on a Majumdar-Papapetrou star}
 Our last discussion, belongs to the stars with a special case of
interior solutions Class Ia, corresponding to the Weyl-Guilfoyle
spherically symmetric metric potential in Eq.
(\ref{w-Guilfoyl-I}). From Equation (\ref{M/Q0-2}) one can confirm
that for such metric, the conditions $a=1$ and $b=0$ form a star
with equal total mass and charge, i.e.
$$M=|Q_0|.$$

This condition for a pressure-less charged perfect fluid, is the
Majumdar condition \cite{Majumdar, Papapetrou}. In this case, the
relation between the metric potential $\omega(\varphi)^2$ and the
electric potential $\varphi(r)$ is a perfect square:
$$\omega(\varphi)^2=(c+\varphi)^2,$$

where $c$ is a constant. In this section we generally investigate
the potential for a massive charged particle moving on a
Majumdar-Papapetrou star. In order to do so, we consider the
shrunk star (the inner circle in Figure \ref{Star-1}), to apply
the interior solution. We use also, the Class Ia solution with
$a=1, b=0$
from Ref. \cite{Guilfoyl}. The solution is:\\

\underline{Class Ia}: $a=1$, $b=0$\\
\begin{equation}
\left\{
\begin{array}{c}
\omega^2=-[\psi(r)]^{-2}\\\\
Q(r)=\mp\frac{kr^3}{\psi(r)}\\\\
8\pi\rho(r)=\frac{3}{R^2}-\frac{k^2r^2}{\psi(r)^2}
\end{array}
 \right.
 \label{Class-Ia-a=1}
\end{equation}

in which the function $\psi(r)$ was defined in (\ref{psi}). Since
$M=|Q_0|$, for this class of solutions, the constant coefficients
are as follows:
\begin{equation}
k=\frac{|Q_0|}{r_0^3}(1-\frac{|Q_0|}{r_0})^{-1}, \label{k-Ia-1}
\end{equation}
\begin{equation}
l^2=k^2R^4, \label{l-Ia-1}
\end{equation}

in which
\begin{equation}
\frac{1}{R^2}=\frac{2|Q_0|}{r_0^3}(1-\frac{|Q_0|}{2r_0}).
\label{1/R^2-M=Q0}
\end{equation}

As before we use the default solution of $Q(r)$. Here the default
solution is the negative one. There would be two cases:\\

I) The case $l=+kR^2$ (Type 1):\\\\

In this case the total charge of the star is:
\begin{equation}
q_0=Q(r_0-\delta r)=-\frac{k(r_0-\delta
r)^3}{(l-kR^2\sqrt{1-\frac{(r_0-\delta r)^2}{R^2}})}.
\label{q0-majumdar-main}
\end{equation}

Using Eqs. (\ref{Class-Ia-a=1}-\ref{l-Ia-1}) we obtain:
\end{multicols}
\ruleup
\begin{equation}
\begin{array}{l}
q_0=-\frac{(-2r_0+|Q_0|)(r_0-\delta r)^3|Q_0|}{r_0^4}
\Big[\sqrt{\frac{r_0^4-2|Q_0|r_0^3+r_0^2Q_0^2+4|Q_0|r_0^2\delta
r-2r_0\delta rQ_0^2-2|Q_0|\delta r^2r_0+\delta
r^2Q_0^2}{r_0^4}}-1\Big]^{-1}.
\end{array}
\label{q0-majumdar-1}
\end{equation}
\\
\ruledown \vspace{0.5cm}

\begin{multicols}{2}

The total mass also, can be calculated using Eq. (\ref{m0}). This
leads us to conclude:
$$m_0=|q_0|,$$

which is predictable. The corresponding potential is illustrated
in Figure \ref{V-eff-Majumdar-1-plot}.
\begin{center}
{\includegraphics[height=7cm]{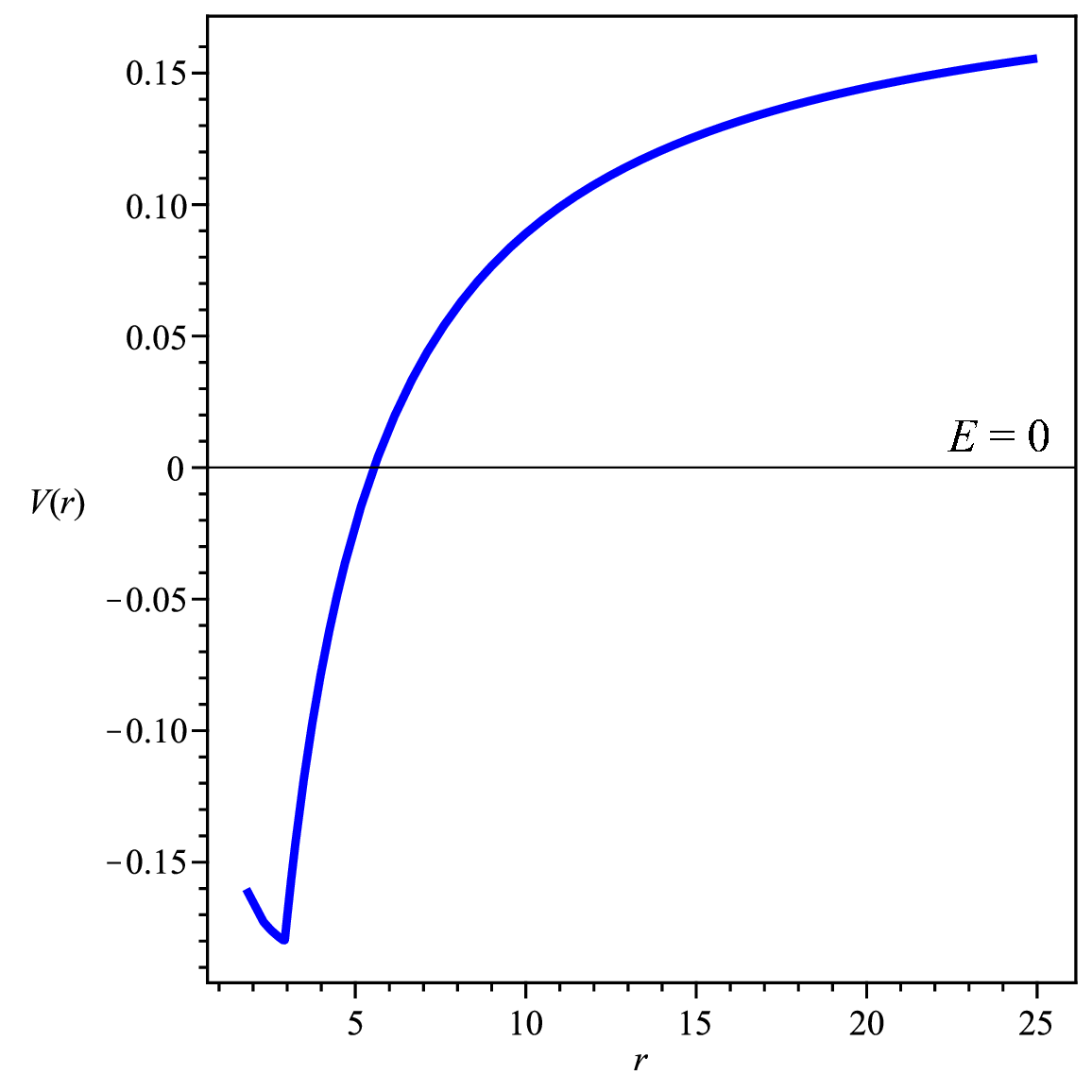}} \figcaption{\small{The
potential for a test-particle moving on a Type 1
Majumdar-Papapetrou star. The interior solution is Class Ia
considering the first case $l=+kR^2$. The illustration is plotted
for $Q_0=-1$, $q=0.18$, $m=0.2$, $L=0.124$, $r_0=2$, $\delta r=0.1
\,r_0$. The unit of length along the coordinate axis is $M$.}}
\label{V-eff-Majumdar-1-plot}
\end{center}
The minus part of the potential is of no importance, because we
previously decided to consider only the positive energies.

For $E=0$ in Figure \ref{V-eff-Majumdar-1-plot}, the escape orbits
and periodic bound orbits are available, which are illustrated in
Figures \ref{bound-Majumdar1} and \ref{escape-Majumdar1}.
\end{multicols}
\ruleup
\begin{center}
\includegraphics[width=40mm,height=40mm]{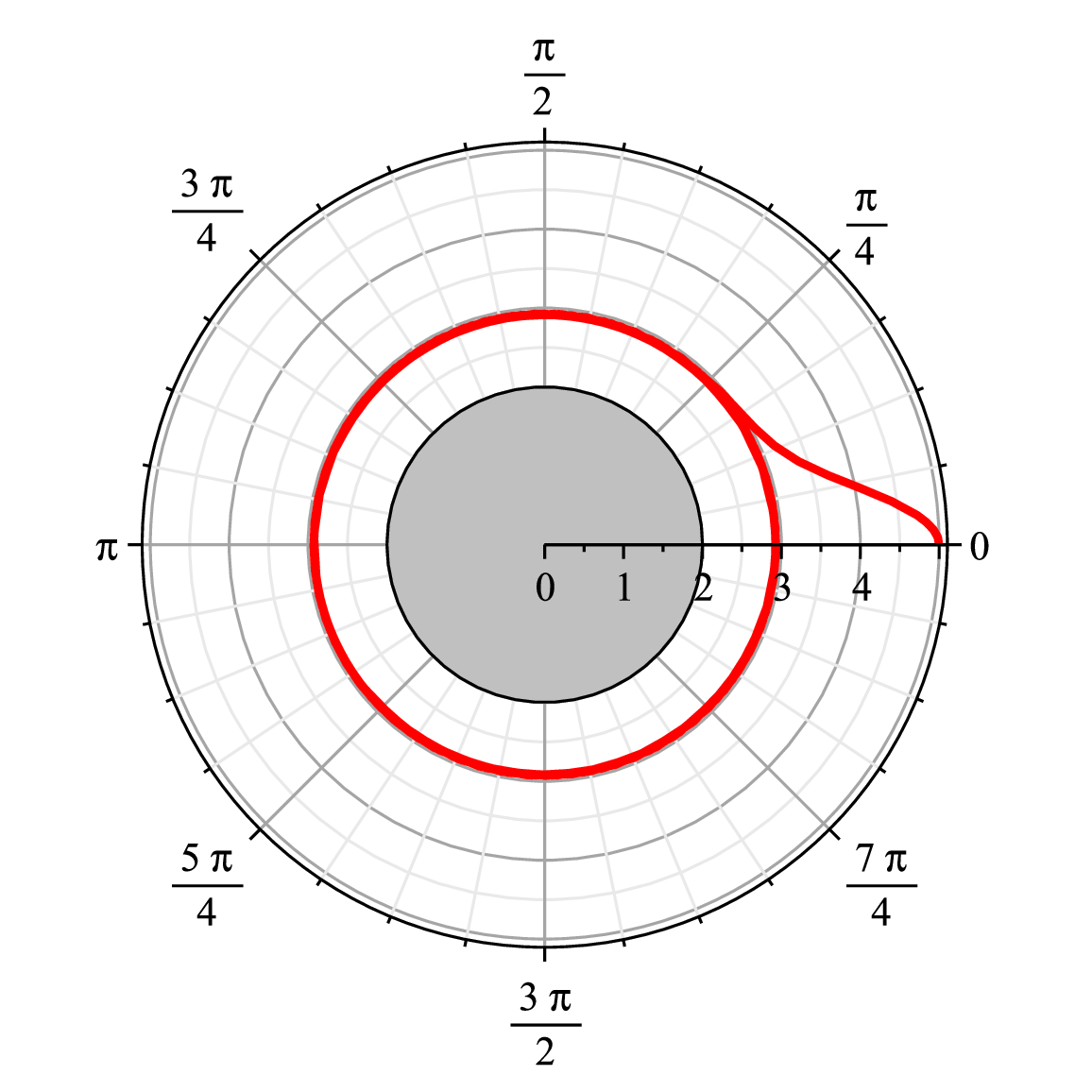}~a)
\hfil
\includegraphics[width=40mm,height=40mm]{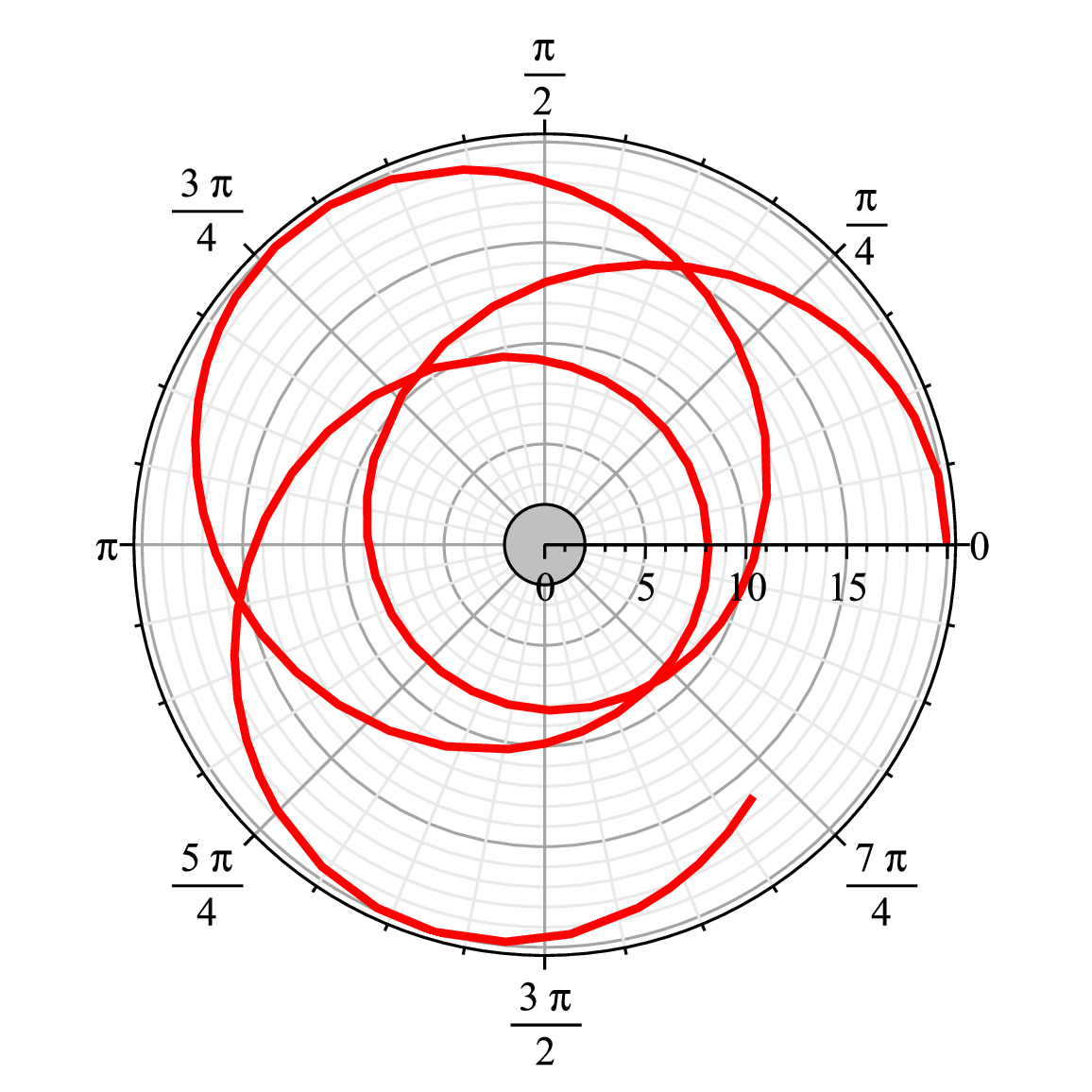}~b)
\figcaption{Periodic bound orbits for a test-particle moving on a
type 1 Majumdar-Papapetrou star, using different initial points of
approach when $E=0$: \textbf{a}) circular orbits for $r=5$, $E=0$;
\textbf{b}) $r=20$, $E=0$.} \label{bound-Majumdar1}
\end{center}
\ruledown

\begin{multicols}{2}

\end{multicols}
\ruleup
\begin{center}
\center{\includegraphics[width=40mm,height=40mm]{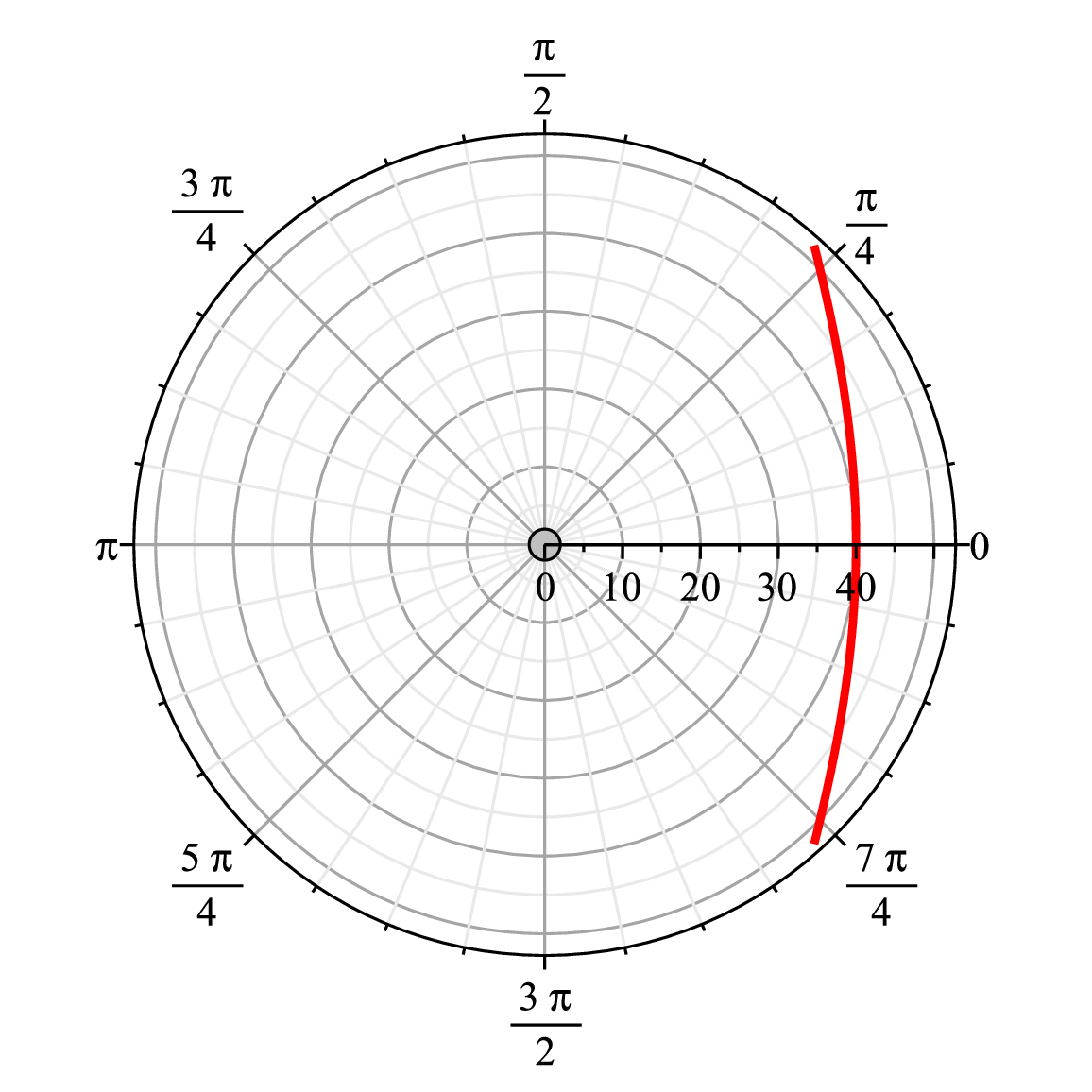}}
\figcaption{The escape orbits for a test-particle moving on a Type
1 Majumdar-Papapetrou star for: $r=40$, $E=0$.}
\label{escape-Majumdar1}
\end{center}
\ruledown

\begin{multicols}{2}

Also for higher energies, the periodic bound orbits are possible
(Figure \ref{bound2-Majumdar1}).\\\\
\end{multicols}
\ruleup
\begin{center}
\center{\includegraphics[width=40mm,height=40mm]{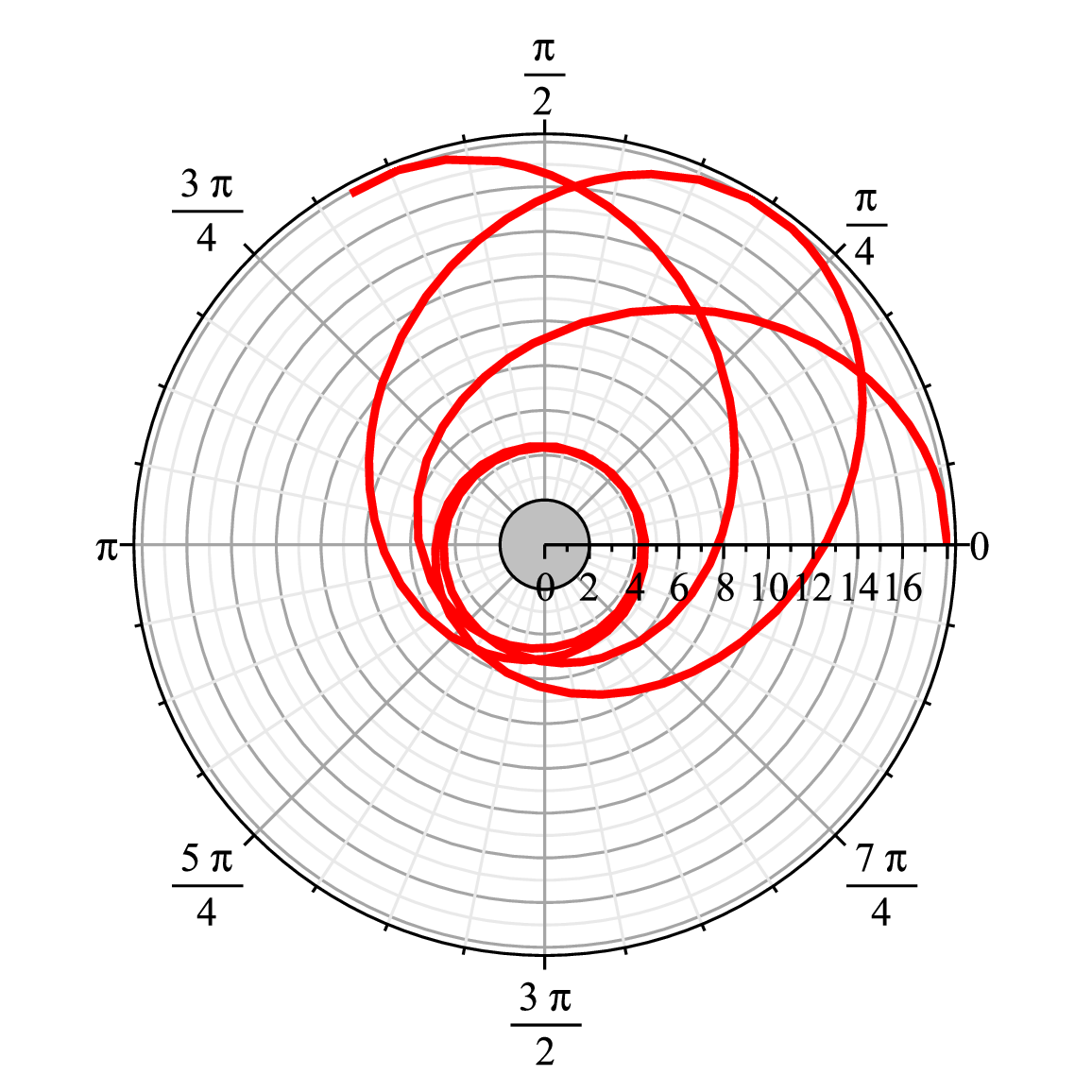}}
\figcaption{The periodic bound orbits for higher energies for a
test-particle moving on a Type 1 Majumdar-Papapetrou star taking:
$r=18$, $E=0.2$.} \label{bound2-Majumdar1}
\end{center}
\ruledown

\begin{multicols}{2}

II) The case $l=-kR^2$ (Type 2):\\\\

According to Eq. (\ref{q0-majumdar-main}), the total charge is
positive, therefore using Eqs. (\ref{Class-Ia-a=1}-\ref{l-Ia-1})
we have:
\end{multicols}
\ruleup
\begin{equation}
\begin{array}{l}
q_0=m_0= \frac{(2r_0-|Q_0|)(r_0-\delta r)^3|Q_0|}{r_0^4}
\Big[\sqrt{\frac{r_0^4-2|Q_0|r_0^3+r_0^2Q_0^2+4|Q_0|r_0^2\delta
r-2r_0\delta rQ_0^2-2|Q_0|\delta r^2r_0+\delta
r^2Q_0^2}{r_0^4}}+1\Big]^{-1}.
\end{array}
\label{q0-majumdar-2}
\end{equation}
\\
\ruledown \vspace{0.5cm}

\begin{multicols}{2}

The potential in this case for the same data is plotted in Figure
\ref{V-eff-Majumdar-2-plot} as the previous case.
\begin{center}
\center{\includegraphics[height=7cm]{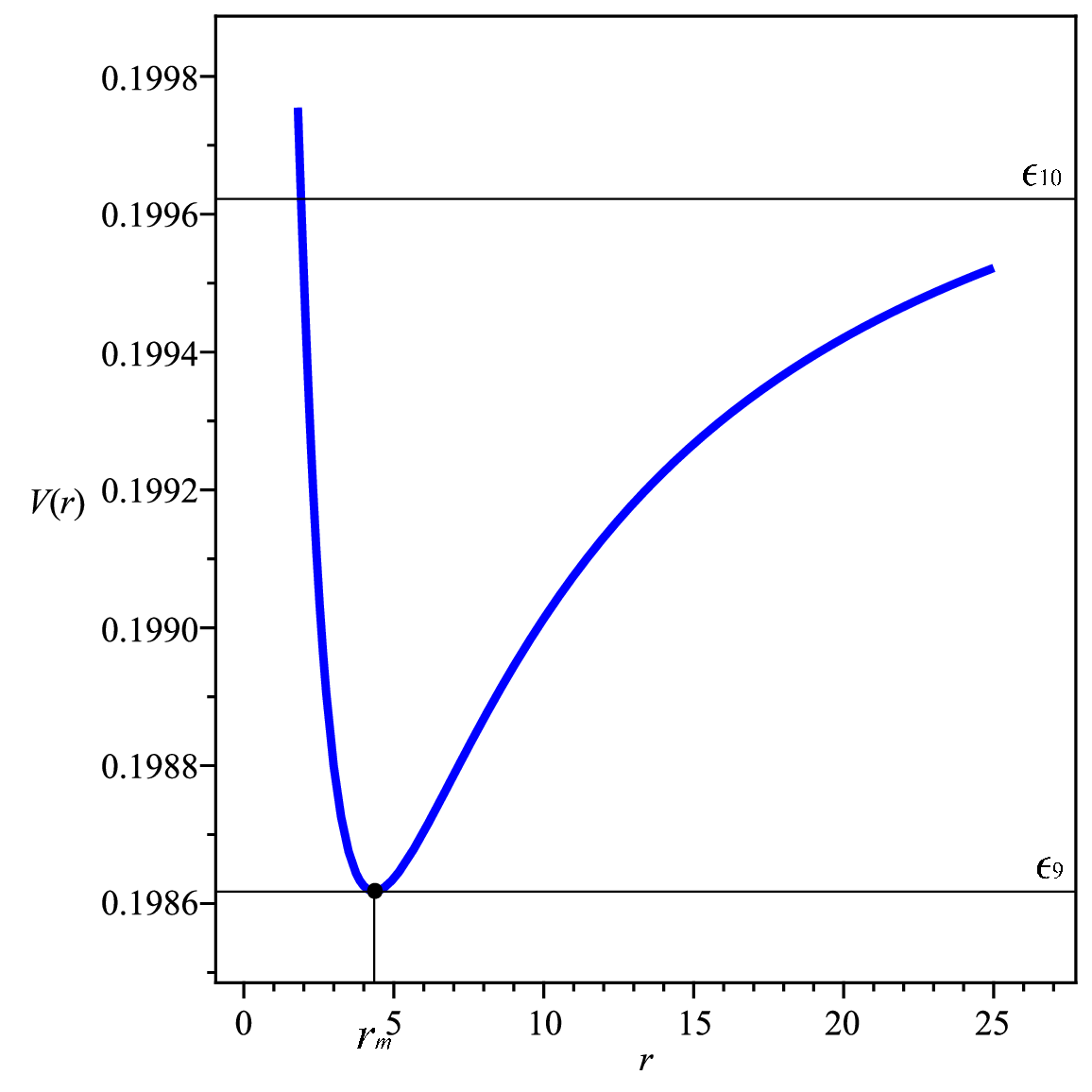}} \figcaption{\small{The
potential for a test-particle moving on a Type 2
Majumdar-Papapetrou star. The interior solution is Class Ia
considering the first case $l=-kR^2$. The illustration is plotted
for $Q_0=1$, $q=0.18$, $m=0.2$, $L=0.124$, $r_0=2$, $\delta r=0.1
\,r_0$. The unit of length along the coordinate axis is $M$.}}
\label{V-eff-Majumdar-2-plot}
\end{center}
This potential allows periodic bound orbits and escape orbit for
the test-particle (Figure \ref{bound-Majumdar2} and
\ref{escape-Majumdar2}).
\end{multicols}
\ruleup
\begin{center}
\includegraphics[width=40mm,height=40mm]{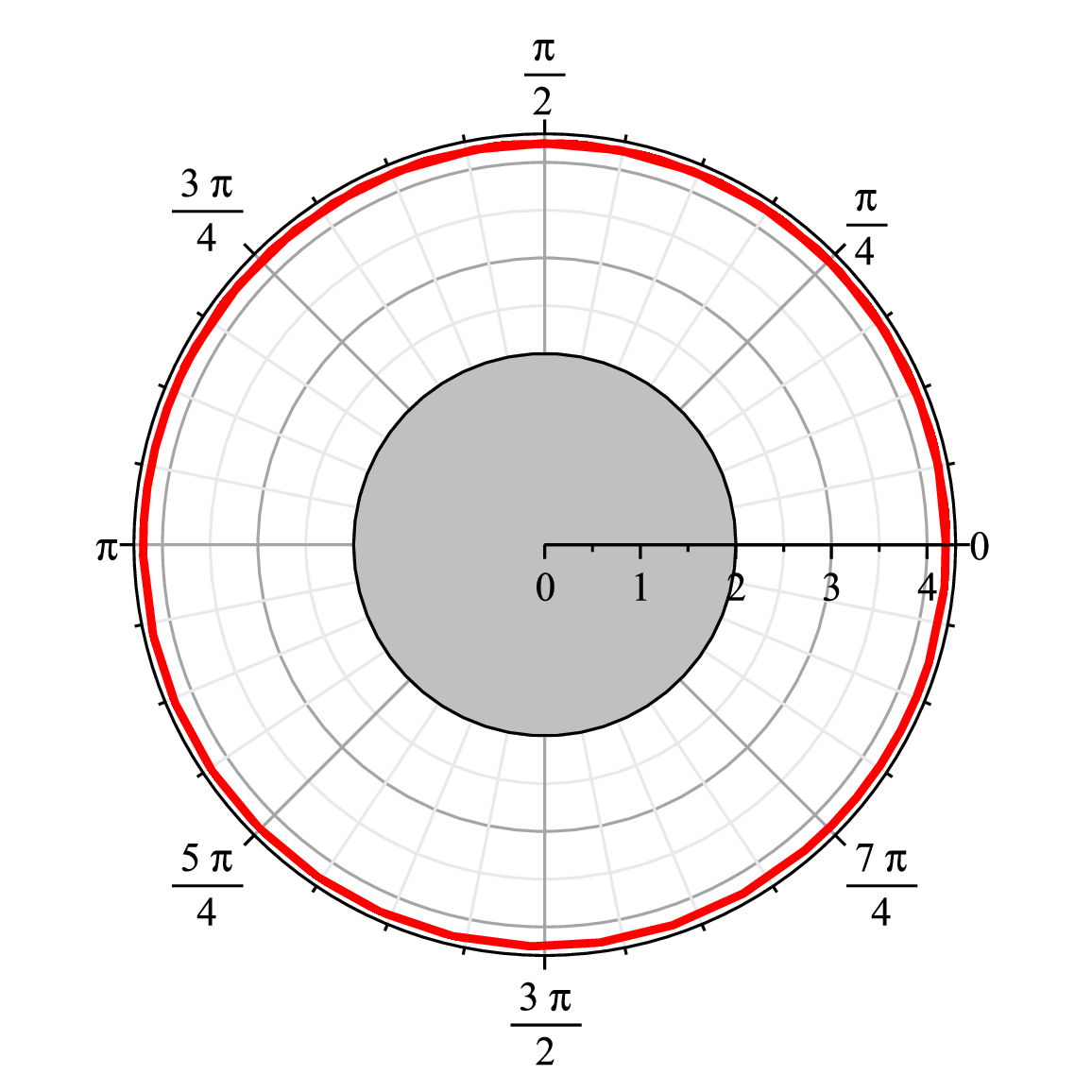}~a)
\hfil
\includegraphics[width=40mm,height=40mm]{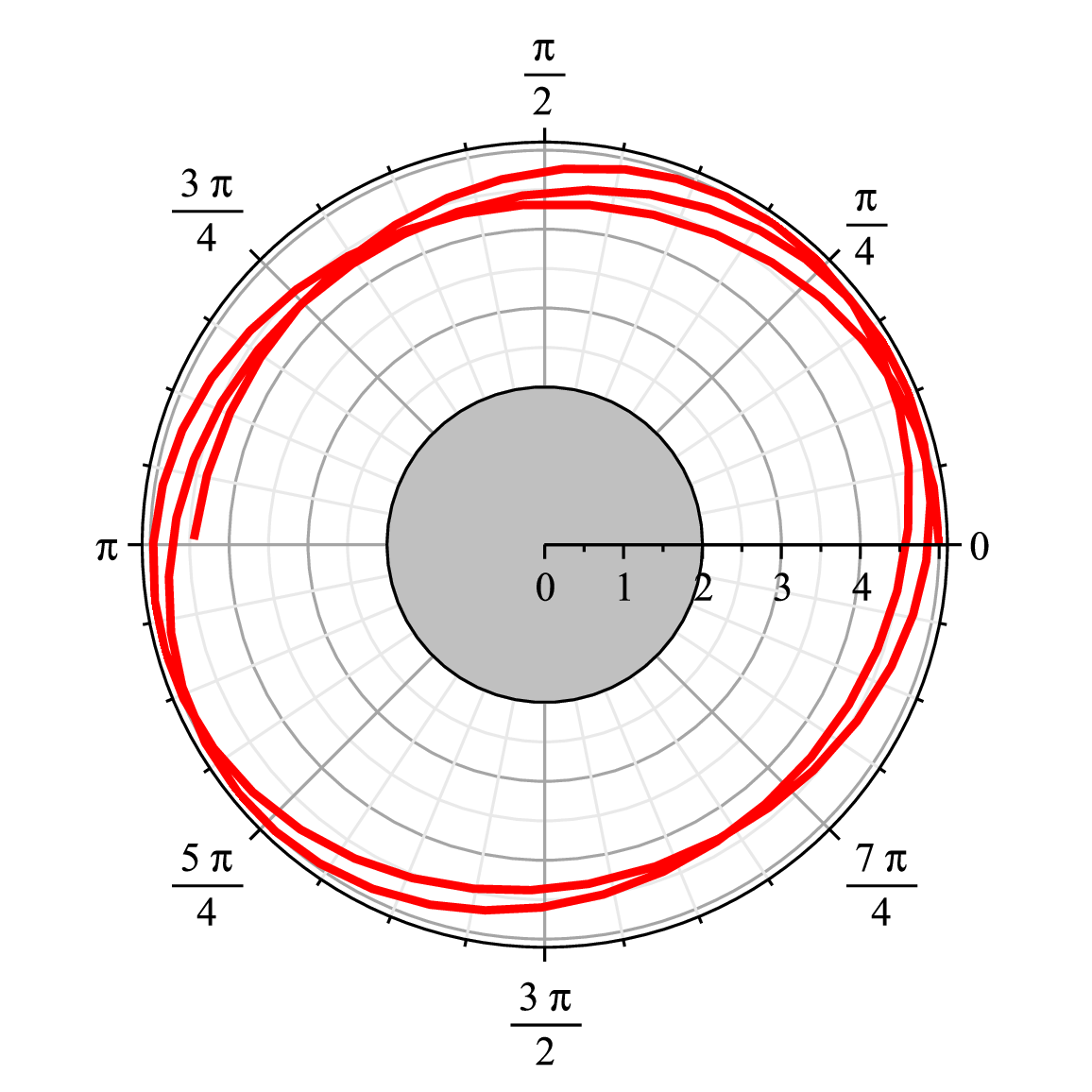}~b)
\figcaption{The periodic bound orbits for a test-particle moving
on a Type 2 Majumdar-Papapetrou star, using different initial
points of approach: \textbf{a}) $r=r_m$, $E=\epsilon_9$;
\textbf{b}) $r=5$, $E=0.1990$.} \label{bound-Majumdar2}
\end{center}
\ruledown

\begin{multicols}{2}

\end{multicols}
\ruleup
\begin{center}
\center{\includegraphics[width=40mm,height=40mm]{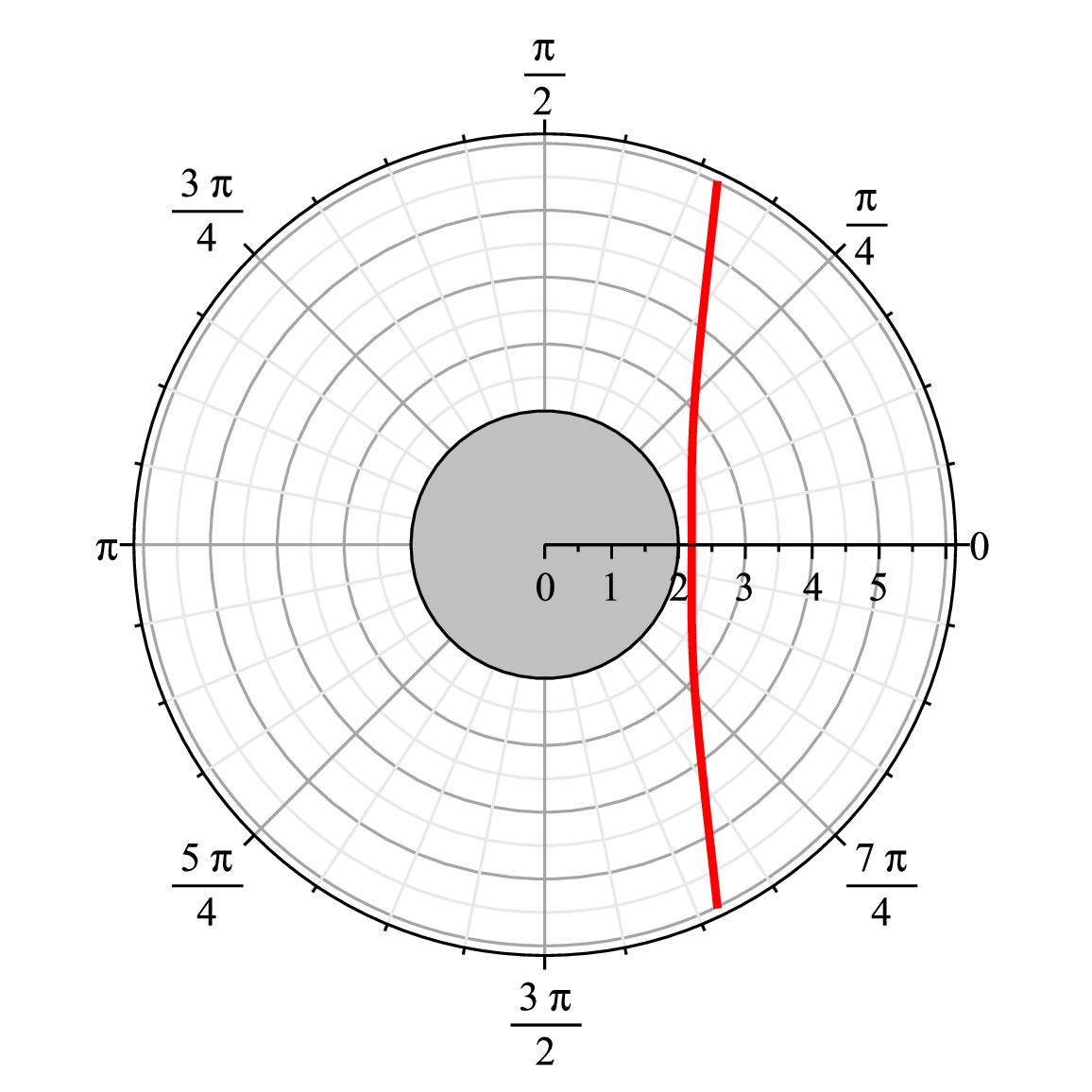}}
\figcaption{The escape orbits for a test-particle moving on a Type
2 Majumdar-Papapetrou star for: $r=2.2$, $E=\epsilon_{10}$.}
\label{escape-Majumdar2}
\end{center}
\ruledown

\begin{multicols}{2}

One can discover that for the default relation for $Q(r)$ in Eq.
(\ref{Class-Ia-a=1}), the total charge of the star, has the
possibility to be negative or positive. This depends somewhat on
the mathematical relationship between the constant coefficients
$l$ and $R^2$. As we could see, this affects the test-particle
trajectories and change its possibilities to have various kinds of
motions in each case.

\section{Conclusion and discussion}
 Beside the differences between the shapes of orbits on different
types of relativistic stars, there are other items, which can be
used for more analytic comparisons. One of them is the period of
stable orbits corresponding to the minimum energy $E_{min}$,
coinciding the minimum of the potential curve at $r_m$. From
Equation (\ref{Energy}) and (\ref{Angular Momentum}), for example,
the period of stable circular orbits in Class Ia, will be (see
appendix A):
\begin{equation}
T_{\textmd{Ia}}=2\pi\frac{r_{m}^2[C(r_{m})+\sqrt{g(r_{m})(m^2+\frac{\ell_{\textmd{Ia}}^2}{r_{m}^2})}]}{g(r_{m})\ell_{\textmd{Ia}}^2},
\label{period-Ia}
\end{equation}

in which
$$\ell_{\textmd{Ia}}=L|_{{r=r_m}},$$

and $r_m$ is indicated in Figure \ref{V-eff-Ia-plot}. We can
notice that:
$$r_m|_{\textmd{Ib}}<r_m|_{\textmd{Ia}},$$

and therefore
$$\ell_{\textmd{II}}>\ell_{\textmd{Ib}}>\ell_{\textmd{Ia}}.$$

Note that, for the current data we have:
$$T_{\textmd{II}}<T_{\textmd{Ib}}<T_{\textmd{Ia}}.$$

Another important item, useful to be compared between the cases,
is the precession of periastron/apastron in periodic bound
(planetary) orbits. For stable orbits, one can derive this
precession as (see appendix B):
\begin{equation}
\Delta\phi=\frac{2\pi}{\alpha}=2\pi[-6q_0^2u_{m}^2+\frac{q^2q_0^2}{\ell_{\textmd{Ia}}^2}+6m_0u_{m}^2-1-\frac{m^2q_0^2}{\ell_{\textmd{Ia}}^2}]^{\frac{1}{2}},
\label{precession of periastron}
\end{equation}

which has been derived for known angular momentum, energy and the
radius of stable orbits in Class Ia. We can conclude that:
$$\Delta\phi|_{\textmd{Ib}}<\Delta\phi_{\textmd{Ia}}.$$

In this article, we considered a relativistic charged perfect
fluid as a relativistic spherical star, and reviewed the
Guilfoyl's interior solutions, through comparing the motion of a
massive charged particle on such a star. We investigated the
effects of these interior solutions on the exterior geometry, for
various types of relativistic stars, by studying the geodesic
motions of a massive charged test-particle. We showed that, how
changes in distributed charge and mass in a certain volume, could
affect the dynamical properties of the exterior geometry. We also
plotted the potentials, and possible orbits, confirming that each
case, may exhibit peculiar shapes of orbits. We also discussed the
special case of a pressure-less Majumdar-Papapetrou star, and
illustrated the potentials and corresponding types of possible
orbits.

\acknowledgments{I would like to thank Zahra Gh. Moghaddam, for
her interest in this work.}

\subsection*{Appendices A}

\noindent{\bf Derivation of the period for stable orbits}

In Equation (\ref{V-eff-1}) we substitute:
\begin{equation}
E_{min}=C(r_{m})+\sqrt{g(r_{m})(m^2+\frac{\ell_{\textmd{Ia}}^2}{r_{m}^2})}.
\label{E1}
\end{equation}

The Equation (\ref{Energy}) and (\ref{Angular Momentum}), for the
current case can yield:
$$\frac{dt}{d\tau}=\frac{E_{min}}{g(r_{m})},$$
$$\frac{d\phi}{d\tau}=\frac{\ell_{\textmd{Ia}}^2}{r_{m}^2}.$$

To calculate its period we need the rate of changing in the
coordinate time $t$ with respect to $\phi$ as following:
\begin{equation}
T=2\pi\frac{dt}{d\phi}. \label{period-general}
\end{equation}

By chain differentiation we have:
$$\frac{dt}{d\phi}=\frac{dt/d\tau}{d\phi/d\tau}=\frac{E_{min}/g(r_{m})}{\ell_{\textmd{Ia}}^2/r_{m}^2}.$$

Therefore for the period of stable orbits in Class Ia, (from Eq.
(\ref{E1})) we have:
\begin{equation}
T_{\textmd{Ia}}=2\pi\frac{r_{m}^2[C(r_{m})+\sqrt{g(r_{m})(m^2+\frac{\ell_{\textmd{Ia}}^2}{r_{m}^2})}]}{g(r_{m})\ell_{\textmd{Ia}}^2}.
\label{period-Ia}
\end{equation}

\subsection*{Appendices B}

\noindent{\bf Derivation of the precession of periastron in
planetary orbits}

We will find this precession here in the following way
\cite{Schutz}:\\

Substituting $u=\frac{1}{r}$ in Eq. (\ref{dr/dphi}) yields:
\end{multicols}
\ruleup
\begin{equation}
(\frac{du}{d\phi})^2=-q_0^2u^4+2m_0u^3-u^2(1+\frac{q_0^2m^2}{L^2})+\frac{2m_0m^2}{L^2}u+\frac{(E-qq_0u)^2-m^2}{L^2}.
\label{du-1}
\end{equation}
\\
\ruledown \vspace{0.5cm}

\begin{multicols}{2}

We consider an approximately stable orbit, with deviations on the
trajectory. This helps us to compare the trajectories with the
known circular orbits and simplify the calculations to derive the
precession. By defining:
$$z=u-u_m,$$

in which $z$ is the deviation from circularity. We substitute this
in (\ref{du-1}), ignoring the terms $O(z^3)$ because we concern
about nearly circular orbits ($z\ll1$). Considering
$L=\ell_{\textmd{Ia}}$ and $E=E_{min}$ for the stable orbits in
Class Ia, yields:
\end{multicols}
\ruleup
\begin{equation}
\begin{array}{l}
(\frac{dz}{d\phi})^2=[-u_{m}^2-u_{m}^4q_0^2+\frac{E_{min}^2}{\ell_{\textmd{Ia}}^2}-\frac{m^2q_0^2u_{m}^2}{\ell_{\textmd{Ia}}^2}-\frac{2E_{min}qq_0u_{m}}{\ell_{\textmd{Ia}}^2}+\frac{2m^2m_0^2u_{m}}{\ell_{\textmd{Ia}}^2}+2m_0u_{m}^3+
\frac{q^2q_0^2u_{m}^2}{\ell_{\textmd{Ia}}^2}-\frac{m^2}{\ell_{\textmd{Ia}}^2}]\\\\
+z[\frac{2q^2q_0^2u_{m}}{\ell_{\textmd{Ia}}^2}-4q_0^2u_{m}^3-\frac{2m^2q_0^2u_{m}}{\ell_{\textmd{Ia}}^2}-2u_{m}+6m_0u_{m}^2-\frac{2E_{min}qq_0}{\ell_{\textmd{Ia}}^2}
+\frac{2m^2m_0}{\ell_{\textmd{Ia}}^2}]\\\\
z^2[-6q_0^2u_{m}^2+\frac{q^2q_0^2}{\ell_{\textmd{Ia}}^2}+6m_0u_{m}^2-1-\frac{m^2q_0^2}{\ell_{\textmd{Ia}}^2}].
\end{array}
\label{dz-1}
\end{equation}
\\
\ruledown \vspace{0.5cm}

\begin{multicols}{2}

Here, we must have a periodic function, to return the initial
$\phi$ and $u$ after a period. As usual we choose
$z=c_1+c_2\cos(\alpha\phi+c_3)$ in which $c_1$, $c_2$ and $c_3$
are constants. However, the coefficient $\alpha$ is not equal to
one, and in Eq. (\ref{dz-1}) it would be the square root of the
coefficient of $z^2$ (see \cite{Schutz}). Therefore:
\begin{equation}
\alpha=[-6q_0^2u_{m}^2+\frac{q^2q_0^2}{\ell_{\textmd{Ia}}^2}+6m_0u_{m}^2-1-\frac{m^2q_0^2}{\ell_{\textmd{Ia}}^2}]^{\frac{1}{2}}.
\label{lambda-1}
\end{equation}

For $\alpha\phi=2\pi$, we have a complete orbit, therefore the
change in $\phi$ from one periastron to the next is:
\begin{equation}
\Delta\phi=\frac{2\pi}{\alpha}=2\pi[-6q_0^2u_{m}^2+\frac{q^2q_0^2}{\ell_{\textmd{Ia}}^2}+6m_0u_{m}^2-1-\frac{m^2q_0^2}{\ell_{\textmd{Ia}}^2}]^{\frac{1}{2}}.
\label{precession of periastron}
\end{equation}

\end{multicols}

\vspace{-1mm}
\centerline{\rule{80mm}{0.1pt}}
\vspace{2mm}

\begin{multicols}{2}

\end{multicols}

\clearpage


\begin{thebibliography}{90}

\vspace{3mm}

\bibitem{Str} Straumann N , Bieri L. Discovering the expanding universe. Cambridge University Press, 2009

\bibitem{Zel} Zel'dovich Ya B, Novikov I D. Relativistic Astrophysics, Vol. I: Stars and
Relativity. University of Chicago Press, Chicago III, 1971

\bibitem{Guilfoyl} Guilfoyle B S. Gen. Rel. Grav., 1999, \textbf{31}: 1645

\bibitem{Zhai} ZHAI Xiang-hau, YAUN Ning-yi, LI Xin-zhou. Chin. Phys. Lett. 1999,
\textbf{16}(5): 321

\bibitem{Blaga} Blaga P, Mioc V. Europhys. Lett., 1992 \textbf{17}(3):
275

\bibitem{Pugliese} Pugliese D, Quevedo H, Ruffini R. Phys. Rev. D., 2011
\textbf{83}: 104052

\bibitem{Stuchlik} Stuchlik Z.  Bull. Astron. Inst. Czechosl. 1983 \textbf{34}: 129

\bibitem{Murta} Casti\~{n}eiras Jorge, Crispino Lu\'{i}s C B, Rodrigo
Murta Rodrigo et al. Braz. J. Phys., 2005 \textbf{35}(4B)

\bibitem{Prasanna} Prasanna A R, Vishveshwara C V. Pramana, 1978 \textbf{11}(4):
359

\bibitem{Chandrasakhar}  Chandrasekhar S. The Mathematical Theory of Black Holes. Oxford University Press, New York, 1983

\bibitem{lam1} Hackmann E, L\"{a}mmerzahl C. Phys. Rev. Lett., 2008
\textbf{100}: 171101

\bibitem{lam2} Hackmann E, Kagramanova V, Kunz J et al. Phys. Rev. D. 2008 \textbf{78}:
124018

\bibitem{Lam3} Hackmann E, Kagramanova V, Kunz J et al. Europhys.
Lett.2009 \textbf{88}: 30008

\bibitem{lam4} Hackmann E, L\"{a}mmerzahl C, Kagramanova V et al. Phys. Rev. D. 2010 \textbf{81}:
044020

\bibitem{lam5} Hackmann E, Hartmann B, L\"{a}mmerzahl C et al. Phys. Rev. D. 2010 \textbf{81}:
064016

\bibitem{lam6} Kagramanova V, Kunz J, Hackmann E et al. Phys. Rev. D. 2010 \textbf{81}:
124044

\bibitem{lam7} Hackmann E, Hartmann B, L\"{a}mmerzahl C et al. Phys. Rev.
D.2010 \textbf{82}: 044024

\bibitem{lam8} Hackmann E, L\"{a}mmerzahl C. Phys. Rev. D. 2012 \textbf{85}:
044049

\bibitem{Wald} Wald Robert M. General Relativity. The University of Chicago Press, Ltd., London, 1984

\bibitem{Wheeler} Misner C W, Thorne K S, Wheeler J A. Gravitation.
Freeman, 1973

\bibitem{Greiner} Greiner W. Classical Mechanics: System of Particles and Hamiltonian Dynamics. Springer-Verlag, New York, Inc., 2003

\bibitem{Weyl} Weyl H. Ann. Phys. (Berlin)
1917 \textbf{359}: 117

\bibitem{Lemos} Lemos Jos\`{e} P S, Zanchin Vilson T. Phys. Rev. D. 2010 \textbf{81}:
124016

\bibitem{Majumdar} Majumdar S D. Phys. Rev. 1947 \textbf{72}: 390

\bibitem{Papapetrou} Papapetrou A. Proc. R. Irish Acad. A, 1947 \textbf{51}:
191

\bibitem{Schutz} Schutz Bernard F. A First Course in General Relativity. Second Edition. Cambridge University
Press, 2009

\end{thebibliography}
\end{document}